\documentclass[%
reprint,
amsmath,amssymb,aps,
]{revtex4-2}

\usepackage{graphicx}
\usepackage{dcolumn}%
\usepackage{bm}
\usepackage{hyperref}

\begin{document}
\title{Gravity-induced mixing, regeneration and CP violation}
\author{J-M Rax}
\email{jean-marcel.rax@universite-paris-saclay.fr}
\affiliation{Universit\'{e} de Paris-Saclay\\
IJCLab-Facult\'{e} des Sciences d'Orsay\\
91405 Orsay France}
\date{\today}

\begin{abstract}
The analysis of neutral mesons oscillations perturbed by Earth's gravity
reveals a small coupling between flavor oscillations and quarks {\it
zitterbewegung}. As a result of this small gravity-induced coupling
between two very low energy oscillations, the mass eigenstates are no longer
the Charge-Parity eigenstates. The three types of observed CP violation
effects result from this gravity-induced mixing. The associated violation
parameters are predicted in full agreement with the experimental data. Thus
gravity-induced CP violation offers the accurate framework to understand the
origin of CP symmetry violation in neutral mesons experiments. This origin renews
in depth the problem of the required large-scale asymmetric baryogenesis in 
the early universe. Furthermore, the issues of the small time reversal violation 
in neutral kaons experiments and of the strong CP problem are fully clarified within 
this new framework of analysis taking into account gravity. 
\end{abstract}

\maketitle

\section{Introduction}

Different types of Charge-Parity violation (CPV) observables associated with
neutral mesons ($K^{0}/\overline{K}^{0}$, $B^{0}/\overline{B}^{0}$,... noted 
$M^{0}/\overline{M}^{0}$) experiments have been identified and measured
since the first observations of the anomalous decay of long-lived kaons,
more than sixty years ago \cite{1}. The results of these CPV experiments are
interpreted within the Standard Model (SM) canonical framework, through the
adjustment between ({\it i}) a complex phase in the quarks mixing matrix \cite{2,3},
and ({\it ii}) the measurements of {\it phase-convention-independent} (PCI)
CPV observables, assuming CPT invariance.

For three quark families an {\it ad hoc} CPV phase can be included in the
Cabibbo-Kobayashi-Maskawa matrix (CKM), and is included according to the
Kobayashi-Maskawa (KM) prescription to reproduce the experimental PCI-CPV
observables \cite{2}. Despite the success of this KM interpretation, the
physical origin of CPV remains an open issue. Moreover, the various issues
associated with the {\it strong CP problem} remain a puzzling set of open
problems, as well as the {\it asymmetric baryogenesis} during the early
stages of the cosmic evolution \cite{4}.

The starting-point of this work is the conviction that: ({\it i}) it is
deeply unsatisfactory to be obliged to interpret such a fundamental, small,
symmetry breaking as a highly specific new law of physics restricted to
flavored neutral mesons, and that ({\it ii}) identifying the origin of CPV
should help clarify three fundamental issues: the small {\it time reversal
violation} (TV) in kaons experiments, the {\it strong CP problem} and the 
{\it asymmetric baryogenesis}.

This study explores the impact of Earth gravity on flavored neutral meson
dynamics and demonstrates that gravity-induced{\it \ }CPV (GICPV), resulting
from gravity induced anomalous mixing and regeneration, provides a pertinent
framework to explain the well-documented CPV experiments. Experiments
dedicated to CPV are classified according to three main types [({\it i)}, (%
{\it ii)} and ({\it iii)}]. ({\it i}) Indirect CPV in the mixing \cite{5,6},
for $K^{0}/\overline{K}^{0}$ experiments it is characterized by the complex
parameter $\varepsilon $. ({\it ii}) Direct CPV in decays into one final
state, for $K^{0}/\overline{K}^{0}$ experiments it is characterized by \ $%
\varepsilon ^{\prime }/\varepsilon $ \cite{6,7}. ({\it iii}) CPV in
interference between decays with and without mixing, the observable for such 
$B^{0}/\overline{B}^{0}$ experiments are $S_{f}=\sin 2\beta $ and $C_{f}$ 
\cite{6,7}. That GICPV explains the main puzzling features of CPV is
summarized in the table below where GICPV predictions are presented in the
second column, and the most recent experimental values, from {\it PDG 2024} 
\cite{7}, in the third one.

\begin{equation}
\begin{tabular}{ccc}
\hline\hline
{\it CPV observables} & {\it GICPV} & {\it Experim.-PDG2024} \\ \hline\hline
$\left. \mathop{\rm Re}\varepsilon \right| _{K^{0}/\overline{K}^{0}}\times
10^{3}$ & $1.66$ & $1.66\pm 0.02$ \\ \hline
$\left. \mathop{\rm Im}\varepsilon \right| _{K^{0}/\overline{K}^{0}}\times
10^{3}$ & $1.57$ & $1.57\pm 0.02$ \\ \hline
$\left. \mathop{\rm Re}\left( \varepsilon ^{\prime }/\varepsilon \right)
\right| _{K^{0}/\overline{K}^{0}}\times 10^{3}$ & $1.66$ & $1.66\pm 0.23$ \\ 
\hline
$\left. \mathop{\rm Im}\left( \varepsilon ^{\prime }/\varepsilon \right)
\right| _{K^{0}/\overline{K}^{0}}\times 10^{3}$ & $0$ & $\left( -2.\pm
5.\right) {}^{\circ }$ \\ \hline
$\left. S_{f}\right| _{\phi K_{S}-\psi K^{0}}\times 10$ & $6.1$ & $5.8\pm
1.2 $ - $7.0\pm 0.1$ \\ \hline
$-\left. C_{f}\right| _{\phi K_{S}-\psi K^{0}}\times 10^{2}$ & $0$ & $9.\pm
12.$\ - $0.8\pm 1.7$ \\ \hline\hline
\end{tabular}
\label{t1}
\end{equation}
The equality between the experimental direct violation parameter $%
\mathop{\rm Re}\left( \varepsilon ^{\prime }/\varepsilon \right) $, on line
three/column three, and the experimental indirect violation parameter $%
\mathop{\rm Re}\varepsilon $, on line one/column three, is considered,
within the KM-CPV framework, as a {\it meaningless numerical coincidence}.
However, GICPV predicts that these two parameters must be equal, $%
\mathop{\rm Re}\left( \varepsilon ^{\prime }/\varepsilon \right) =$ $%
\mathop{\rm Re}\varepsilon $, and they are. The left value on the last two
lines of the third column reports the experimental results for the decay
products $f=\left( \phi ,K_{S}\right) $ and the right value for $f=\left(
\psi ,K^{0}\right)$ \cite{7}.

Gravity-induced CPV might appear quite foreign to the usual elementary
particles models expected to interpret CPV, but it is very successful in the
interpretation of the experimental results (\ref{t1}), and does not rely on
any speculation on {\it new physics }as it rely solely on quantum mechanics
in the vicinity of a massive spherical object described within the framework
of weak field general relativity. In doing so, GICPV set up a new bridge
between gravitation and the SM. This bridge does not adjust free parameters
to experimental results, but simply takes into account the internal
structure of the mesons and the effect of gravity on weak interaction
mixing. Moreover, GICPV sheds new light on TV in kaons experiments and
clears up the common confusion between time reversal violation and
irreversible evolution.

Flavor eigenstates, $M^{0}$ or $\overline{M}^{0}$, are stationary di-quarks
bound states ($K^{0}/\overline{K}^{0}\sim $ $d\overline{s}/\overline{d}s$, $%
B^{0}/\overline{B}^{0}\sim $ $d\overline{b}/\overline{d}b$) made of one
quark $q^{\prime }$ or $\overline{q^{\prime }}$ and one other quark $%
\overline{q}$ or $q$, combined into a spin-singlet state. Weak interaction
is responsible of flavor oscillations $M^{0}\rightleftharpoons \overline{M}%
^{0}$ between particles and antiparticles. As a result, the mass
eigenstates are linear superpositions of the particles with the
antiparticles. Without CPV, these mass eigenstates are also CP eigenstates
and are noted $\left| M_{1/2}\right\rangle =$ ($\left| M^{0}\right\rangle $ $%
\pm $ $\left| \overline{M}^{0}\right\rangle $)/$\sqrt{2}$. A CP eigenstate, $%
M_{1}$ or $M_{2}$, is a combination of 4 quarks structured in 2 linear
superpositions and 2 bound states. The 2 linear superpositions involve $q$
with $\overline{q}$ and $q^{\prime }$ with $\overline{q}^{\prime }$, the 2
bound states involve $q$ with $\overline{q}^{\prime }$ and $q^{\prime }$
with $\overline{q}$. Such symmetric and antisymmetric linear superposition
of a spin 1/2 fermion with its antifermion are well known to be eigenstates
of the {\it Dirac velocity} operator with eigenvalues $\pm c$ \cite{8,9},
this property ({\it zitterbewegung}) is the source of a subtle coupling with
gravity.

It is very important to note that this universal, non intuitive, {\it %
zitterbewegung} behavior of fermion-antifermion is completely independent of
the mass and charge of the fermion; and of the shape and strength of the
interaction involved in the Dirac Hamiltonian describing quarks confinement
inside the mesons. The universal {\it zitterbewegung velocity }$\pm c${\it \ 
} is responsible of a small vertical fluctuation resulting in a small
gravity-induced coupling with the flavor oscillations. As demonstrated in
this paper, this coupling explains all the CPV effects observed in $K^{0}/%
\overline{K}^{0}$ and $B^{0}/\overline{B}^{0}$ experiments (\ref{t1}).

Neutral mesons oscillations ($K^{0}\rightleftharpoons \overline{K}^{0}$, $%
B^{0}\rightleftharpoons \overline{B}^{0}$) are very low energy oscillations:
[$10^{-6}$, $10^{-4}$] eV (in the following, the quantities are evaluated
with two different orders of magnitude, one for $K^{0}$/$\overline{K}^{0}$,
the other for $B^{0}/\overline{B}^{0}$). The gravitational effect on Earth
is a low energy coupling: [$10^{-9}$, $10^{-10}$] eV. Therefore, there is no
need to rely on quantum field theory. The usual Lee-Oehme-Yang (LOY) two
states Hamiltonian provides the pertinent framework to describe very low
energy flavor oscillations slightly perturbed by Earth's gravity \cite{10,11}%
.

The LOY Hamiltonian will be completed with a {\it zitterbewegung }position
operator obtained through a careful analysis of the Dirac spinors states
associated with the quarks structure of the mesons. A simple heuristic model
of the coupling between the two low energy quantum oscillations ({\it %
zitterbewegung velocity }$+c$ $\rightleftharpoons $ $-c$ and $M^{0}$ $%
\rightleftharpoons $ $\overline{M}^{0}$) leading to CPV, is presented in the
last section.

The GICPV mechanism is to be analyzed in two steps: ({\it i}) with a model
preserving unitarity ($d\left\langle M_{1/2}\right. $ $\left|
M_{1/2}\right\rangle /d\tau $ $=$ $0$); and then ({\it ii}) with the
possibility of irreversible decays ($d\left\langle M_{1/2}\right. \left|
M_{1/2}\right\rangle /d\tau <0$). \ First, a stable meson is considered to
describe the small gravity induced CP mixing resulting from the interplay
between the{\it \ zitterbewegung }oscillations{\it \ }$\left( \pm c\right) $%
{\it \ }and meson-antimeson {\it flavor oscillations (}$\Delta S=2$, $\Delta
B=2...${\it ). }Then, the finite lifetime of the meson is introduced and the
balance between the {\it gravity induced mixing/regeneration,} on the one
hand, and {\it decay }$M\rightarrow f${\it , }on the other hand, provides
the complete mechanism of GICPV and the right values of the measured
parameters in {\it PDG 2024} \cite{7}.

The unitary model ({\it i}) is an eigenvectors problem needed to identify a
symmetry violation, and the dissipative model ({\it ii}) can be viewed as an
initial value problem describing real experimental conditions.
Irreversibility is associated with the various decays $M\rightarrow f$ \ as
the classical phase space (position $\otimes $ momentum) available to $f$ \
is restricted in momentum but unbounded in space so that the associated
Poincar\'{e} recurrence time needed to observe  $f\rightarrow M$ \ is
extremely large. This irreversible evolution must not be confused with a
fundamental TV. The two steps GICPV mechanism clarify this issue. It will be
shown that if CPT invariance is assumed, irreversibility might be
interpreted as a TV, however it remains a CPT modeling ignoring the two
steps GICPV mechanism. This usual confusion between (statistical)
irreversibility, such as decay or radiation processes, and (dynamical) time
reversal violation will be clarified here with the two steps mechanism.

Table (\ref{t1}) provides a fairly convincing argument in support of GICPV.
As a consequence of the accuracy of these quantitative predictions, it is
conjectured that, far from any massive object, the CKM matrix might be free
from any CPV phase: CPV effects appear to be gravity induced near
astrophysical massive objects, or during the very early stages of the cosmic
evolution.

The new gravitational coupling, identified here, not only elucidates the
issues of the origin of CPV and clarify the confusion between
irreversibility and TV, it opens also very interesting perspectives to set
up cosmological models displaying an asymmetric baryogenesis compatible with
the present state of our universe \cite{4}. Indeed, if the experimental
observations of CPV are incorporated into the Standard Model in the form of
a weak {\it Universal symmetry breaking} (KM-CPV), and if such a model is
used within the framework of a canonical cosmological model, this approach
fails, by several orders of magnitude, to explain how our matter-dominated
universe came into being during its early evolution. During the very early
stages of the cosmic evolution, gravity was far more larger than today on
Earth, GICPV is a linear function of the gravitational field, this opens up
a promising avenue to resolve the present contradiction between the observed
very small CPV and the large CPV needed for a pertinent model of
our matter-dominated universe.

The study presented below complements previous studies \cite{12,13}, in
particular \cite{12} was based on two coupled Klein-Gordon equations
describing mesons evolutions on a Schwarzschild metric. The results given by
the weak field model below are similar to those of this previous fully
Einsteinian model, GICPV is thus model independent.

An effective SM Lagrangian on Earth can be constructed starting from a CPV
free CKM matrix, then including a CPV phase adjusted to recover $\varepsilon
_{GICPV}$ and $\varepsilon ^{\prime }/\varepsilon _{GICPV}$ ; {\it id est}
adjusted in fact to the actual KM phase. Indeed, the GICPV parameters, $%
\varepsilon _{GICPV}$ and $\varepsilon ^{\prime }/\varepsilon _{GICPV}$, are
in complete agreement with {\it PDG 2024} data, $\varepsilon _{PDG2024}$ and 
$\varepsilon ^{\prime }/\varepsilon _{PDG2024}$, used to parametrize the CKM
matrix. However, this effective SM Lagrangian on Earth obviates the problem
of the origin of CPV, thus creating misleading open issues such as the {\it %
strong CP problem,} the very small {\it asymmetric baryogenesis }and the
small{\it \ time reversal violation}. It is important to note that besides
the canonical parameters, $\varepsilon $, $\varepsilon ^{\prime
}/\varepsilon $ and $\beta $, analyzed in this paper, all others CPV effects
observed and considered {\it in agreements with the SM-KM model} will also
agree with this {\it GICPV effective Lagrangian} model of Earth's influence
on a (initially CPV free) CKM mixing matrix.

In the next section the various quantum states and Hilbert spaces needed to
set up GICPV are presented. Then, section three is devoted to a review of
the LOY Hamiltonian without CPV and without gravity. This Hamiltonian is
extended, in section four, to take into account the impact of Earth's
gravity. {\it Zitterbewegung} associated with the quarks content of the
mesons is analyzed in sections five. Section six and seven are devoted to
the analysis of the structure of the observed mesons mass eigenstates on
Earth, first as an eigenvector problem and then as an initial value problem
taking into account the finite lifetime. The confusion between
irreversibility and T violation is clarified by the two steps mechanism
presented in these two sections. Then, in the three following sections, the
GICPV theory of types ({\it i)}, ({\it ii)} and ({\it iii) }CPV experiments
is developed.

A simple heuristic model of GICPV is presented in the last section to
illustrate the basic mechanism at work and the perspectives opened by these
new results complete this last section. All the experimental numerical
values used to evaluate the expressions are taken from the recent reference: 
{\it PDG 2024} \cite{7}. Appendix A is devoted to a review of the
definitions of $\varepsilon $, $\varepsilon ^{\prime }$ and $\beta $ when
CPT is assumed. Appendix B provides the complete calculations associated
with section five.

\section{States, stability, irreversibility, phase convention}

In the following, $M^{0}/\overline{M}^{0}$ will stand either for $K^{0}/%
\overline{K}^{0}$ $\sim \left( d\overline{s}\right) /\left( \overline{d}%
s\right) $, or for $B^{0}/\overline{B}^{0}\sim \left( d\overline{b}\right)
/\left( \overline{d}b\right) $, or for any flavored neutral meson/antimeson $%
\left( q^{\prime }\overline{q}\right) /\left( \overline{q^{\prime }}q\right) 
$ pair with lifetimes long enough to allow type ({\it i)}, ({\it ii)} or (%
{\it iii) }experiments with clear CPV evidences.

As different models (eigenvalue {\it versus} initial value problems,
conservative {\it versus} dissipative dynamics) will be used, different
states in the LOY Hilbert space are to be defined. Moreover, different
Hilbert spaces will be used: a simple two level one, the usual LOY Hilbert
space ${\bf E}_{LOY}$, and a larger one $\widetilde{{\bf E}}_{Q}$ where the
details of the 4 quarks ($q$, $\overline{q}$, $q^{\prime }$, $\overline{q}%
^{\prime }$) structure can be described. Beside \ ${\bf E}_{LOY}$ and $%
\widetilde{{\bf E}}_{Q}$, a third type of Hilbert space will be used: ${\bf E%
}_{\bot }$ the space of wave functions along the vertical direction
perpendicular to the surface of Earth, ${\bf E}_{\bot }$ is simply the space
of {\it square-integrable complex functions} of a real variable. In all
these Hilbert spaces the identity operator is noted $\widehat{\text{I}}$ ( $%
\widehat{}$ indicates operators).

In ${\bf E}_{LOY}$ the notations for the different stable and unstable
states ({\it i-v}) are presented below.

({\it i}) The flavor eigenstates are noted $\left| M^{0}\right\rangle $ and $%
\left| \overline{M}^{0}\right\rangle $, these two states provide an
orthonormal basis in ${\bf E}_{LOY}$.

({\it ii}) The CP eigenstates $\left| M_{1}\right\rangle $ and $\left|
M_{2}\right\rangle $ are defined as

\begin{equation}
\left| M_{1/2}\right\rangle =\left( \left| M^{0}\right\rangle \pm \left| 
\overline{M}^{0}\right\rangle \right) /\sqrt{2}\text{,}  \label{1}
\end{equation}
these two states provide an orthonormal basis in ${\bf E}_{LOY}$.

({\it iii}) The stable mass eigenstates on Earth are $\left| M_{1}^{\oplus
}\right\rangle $ and $\left| M_{2}^{\oplus }\right\rangle $. With these states
 gravity is
taken into account and the lifetimes of the mesons is assumed infinite (stable
particles $d\left\langle M_{1/2}^{\oplus }\right. \left| M_{1/2}^{\oplus
}\right\rangle /d\tau =0$). As demonstrated in section six, these stable
mass eigenstates display a small gravity-induced CP mixing and are no longer
CP eigenstates,

\begin{equation}
\left| M_{2/1}^{\oplus }\right\rangle =\left| M_{2/1}\right\rangle \pm
j\kappa \left| M_{1/2}\right\rangle \text{.}  \label{a2}
\end{equation}
The small GICPV parameter $\kappa \ll 1$ is identified in section six (as $%
\kappa ^{2}$ $\sim $ O$\left[ 10^{-6}\text{, }10^{-12}\right] $ it is
neglected in the normalization of (\ref{a2})).

({\it iv}) Experiments are initial value problems starting from the mesons
production. Because of decays $M\rightarrow f$ \ to a set of final states $%
\left\{ f\right\} $, the observed mass eigenstates on Earth, $\left|
M_{H}^{\oplus }\right\rangle $ and $\left| M_{L}^{\oplus }\right\rangle $,
are different from $\left| M_{2}^{\oplus }\right\rangle $ and $\left|
M_{1}^{\oplus }\right\rangle $ (unstable particles $d\left\langle
M_{H/L}^{\oplus }\right. \left| M_{H/L}^{\oplus }\right\rangle /d\tau $ $<0$%
) . These heavy\ $\left( H\right) $ and light $\left( L\right) $ mass
eigenstates noted $\left| M_{H}^{\oplus }\right\rangle $ and $\left|
M_{L}^{\oplus }\right\rangle $ are the observed physical states in CPV
experiments.

({\it v}) Finally, the notation $\left| M_{H}\right\rangle $ and $\left|
M_{L}\right\rangle $ will be used for the usual KM-CPV mass eigenstates
parametrized under the assumption of CPT invariance and reviewed in Appendix
A.

Quarks are spin 1/2 fermions evolving in an appropriate Hilbert space where
various symmetry groups can be represented: $\left| q\right\rangle $ $\sim
\left| \theta _{\text{flavor}}\right. $ $\zeta _{\text{color}}$ $\sigma _{%
\text{spin}}$ $\left. \psi _{\text{space}}....\right\rangle $. Beside ${\bf E%
}_{LOY}$, to reveal the {\it zitterbewegung} coupling, a larger Hilbert
space is needed: $\widetilde{{\bf E}}_{Q}$.

The 4 quarks $\left( q,\overline{q},q^{\prime },\overline{q}^{\prime
}\right) $ will be described in the Hilbert space $\widetilde{{\bf E}}_{Q}$
where the notation $\left| \widetilde{M}_{0}\right\rangle \in \widetilde{%
{\bf E}}_{Q}$ and $\left| \widetilde{\overline{M}}_{0}\right\rangle \in 
\widetilde{{\bf E}}_{Q}$ will be used for the di-quarks states representing $%
M^{0}$ and $\overline{M}^{0}$.

The states $\left| \widetilde{M}_{0}\right\rangle $ ($\left| \widetilde{%
\overline{M}}_{0}\right\rangle $) $\in \widetilde{{\bf E}}_{Q}$ and $\left|
M^{0}\right\rangle $ ($\left| \overline{M}^{0}\right\rangle $) $\in $ ${\bf E%
}_{LOY}$ describe the very same physical object, but with a different level
of details, fewer details are available in ${\bf E}_{LOY}$. The set of
observables in ${\bf E}_{LOY}$ is thus smaller than the set of observables
in $\widetilde{{\bf E}}_{Q}$. Despite the reduction of the description, the
matrix elements, with $\left| M^{0}\right\rangle $ and $\left| \overline{M}%
^{0}\right\rangle $, of a given well defined physical observable, meaningful
in both $\widetilde{{\bf E}}_{Q}$ and ${\bf E}_{LOY}$, are equal to the
matrix elements, with $\left| \widetilde{M}_{0}\right\rangle $ and $\left| 
\widetilde{\overline{M}}_{0}\right\rangle $, of the same observable in $%
\widetilde{{\bf E}}_{Q}$. However, because of the reduction, some
observables in $\widetilde{{\bf E}}_{Q}$ are meaningless in ${\bf E}_{LOY}$.

Without decay, but with Earth's gravity, the stables mesons states, $%
M_{1}^{\oplus }$ and $M_{2}^{\oplus }$, identified in section six, display a
small CP violation and the time reversal symmetry T is preserved. Then, when
decays are switched on, in section seven, both CP and time reversibility are
broken with the {\it observed} mass eigenstates $M_{H}^{\oplus }$ and $%
M_{L}^{\oplus }$. A CPT invariant framework can be used to interpret the
structure of $\left| M_{H}^{\oplus }\right\rangle $ and $\left|
M_{L}^{\oplus }\right\rangle $ as a CPV with a TV and thus ignore the
underlaying GICPV two steps mechanism at the cost of a confusion between
irreversibility and TV which are physically different.

CPT theorem is demonstrated within the framework of three main hypothesis:
Lorentz group invariance, spin-statistics relations and local field theory.
On Earth, the first hypothesis is not fulfilled as a curved metric $\left[
g_{\mu \nu }\right] $ $\neq $ {\it diag}$\left[ -1,1,1,1\right] $ is used in
this study.

Thus, when Earth influence is taken into account, and mesons are assumed
stable, GICPV does not contradict the CPT theorem which no longer holds true
as CP invariance is broken by gravity and T invariance remains unaltered.
With KM-CPV an over-interpretation of the irreversibility of spontaneous
decays as a time reversal T violation leads to consider a CPT invariant
framework to describe CPV experiments and overlooks the nature of the stable
particle on Earth $M_{1}^{\oplus }$ and $M_{2}^{\oplus }$. Nevertheless,
reducing CPV to such a CPT invariant description, and introducing a KM phase
in the SM provides an operational framework of interpretation, but it hides
the two steps GICPV mechanism and assume a deeply unsatisfactory small TV
that obviates the understanding of the origin of CPV.

Within the framework of GICPV, Earth's gravity is described as an external
field. The evolution of a mass eigenstate alone, as a linear superposition
of two flavor eigenstates, $M^{0}$ and $\overline{M}^{0}$, does not provide
the complete picture of the quantum system. Thus, it can not be considered
as a good candidate displaying discrete symmetries invariance such as P and
T. It is a common evidence that the symmetries of a global system are lost
when the observations are restricted to one of \ its subsystem.

Ultimately, there are neither CPT, nor CP, nor T, violations {\it stricto
sensu}, all fundamental discrete{\it \ }symmetries are restored for a global
two bodies, meson + Earth ( $M$ + $\oplus $), evolution of a stable quantum
state $\left| M\left( \tau \right) ,\oplus \left( \tau \right) \right\rangle 
$ describing both the a stable meson and a stable Earth. Then, when decays
are considered, the transition $\oplus $ + $M\rightarrow $ $\oplus $ + $f\ $%
\ provide an irreversible evolution (the Poincar\'{e} recurrence time to
observe $\oplus $ + $f\ \rightarrow $ $M$ + $\oplus $ is extremely large)
not to be confused with a TV.

A global picture, free of broken discrete symmetries and irreversibility,
can be given through the study of states of the type $\left| M,\right.
\oplus ,$ $f_{1},f_{2},\left. f_{3}...\right\rangle $ in an appropriate very
large Hilbert space $\left\{ \left| M,\oplus ,\left\{ f\right\}
\right\rangle \right\} $. The dynamics in this large Hilbert space does not
display C or P or T violations or irreversibility. However, rather than
considering the unitary evolution of a complete quantum states $\left|
M,\right. \oplus ,$ $f_{1},f_{2},\left. f_{3}...\right\rangle $, up to now,
the analysis of neutral mesons experiments used a subspace of the tensorial
products $\left| M\right\rangle $ $\left| \oplus \right\rangle $ $\left|
f_{1}\right\rangle \left| f_{2}\right\rangle $... where: ({\it i}) $\left|
\oplus \right\rangle $ is completely ignored; ({\it ii}) the coupling
amplitude $a\left( f_{i}\rightarrow M\right) $ are all set equal to zero to
provide the observed irreversibility; and ({\it iii}) the coupling amplitude 
$a\left( M\rightarrow f_{i}\right) $ are described with an effective
operator $\widehat{\gamma }$ under a Markovian approximation. Then, assuming
CPT, this incomplete framework leads to interpret the experimental results
restricted to $M$ as CP and T violations. With GICPV the analysis is also
restricted to $M$ but completed with an external Earth's field and a careful
study of the internal quarks {\it zitterbewegung}. Within the GICPV
framework, a complete clarification of the various issues associated with
CPV is achieved, these clarifications are summarized in the tables (\ref{t2}%
, \ref{t3}) in the last section.

A physical state is not represented by a single vector in an Hilbert space,
but by a set of vectors related by an unphysical global phase factor. The
various global {\it bra} and {\it ket} in a quantum model are defined up to
a phase factor. These arbitrary conventional phases, inherent to all {\it ab
initio} theoretical models and parametrizations, are to be eliminated when
it comes to interpret experiments and it is mandatory to define PCI
observables.

The relative phase between $\left| M^{0}\right\rangle $ and $\left| 
\overline{M}^{0}\right\rangle \,$ is arbitrary because it is not measurable
and will be fixed by convention. The CP operator in ${\bf E}_{LOY}$ \ as a
function of this arbitrary phase $\delta $ is given by 
\begin{equation}
\widehat{CP}=\exp -j\delta \left| M^{0}\right\rangle \left\langle \overline{M%
}^{0}\right| +\exp j\delta \left| \overline{M}^{0}\right\rangle \left\langle
M^{0}\right| .
\end{equation}
The phase convention used in this study \ is $\delta =0$ so that $\widehat{CP%
}\cdot \left| M^{0}\right\rangle $ $=$ $\left| \overline{M}^{0}\right\rangle 
$ and $\widehat{CP}\cdot \left| \overline{M}^{0}\right\rangle $ $=$ $\left|
M^{0}\right\rangle $. With this usual convention $\widehat{CP}\cdot \left|
M_{1/2}\right\rangle $ = $\pm \left| M_{1/2}\right\rangle $ and without CPV $%
\left| M_{H/L}\right\rangle $ = $\left| M_{2/1}\right\rangle $. With the
other popular convention $\delta $ $=$ $\pm \pi $, $\widehat{CP}\cdot \left|
M^{0}\right\rangle $ = $-\left| \overline{M}^{0}\right\rangle $ and $\left|
M_{H/L}\right\rangle $ = $\left| M_{1/2}\right\rangle $. However, for both
conventions, without CPV, the observable mass eigenstates fullfil $\widehat{%
CP}\cdot \left| M_{H}\right\rangle $ = $-\left| M_{H}\right\rangle $ and $%
\widehat{CP}\cdot \left| M_{L}\right\rangle $ = $+\left| M_{L}\right\rangle $%
.

\section{Oscillations and decay without CPV and gravity}

A meson state $\left| M\left( \tau \right) \right\rangle $ is a linear
superposition of two flavor eigenstates with amplitudes $u\left( \tau
\right) $ and $v\left( \tau \right) $ functions of the proper time $\tau $.
This superposition decays on a short time scale to a final state $\left|
f\right\rangle $. Because of the possibility of these decays, the Hilbert
space where $\left| M\left( \tau \right) \right\rangle $ evolves should be
extended to the set $\left\{ \left| f\right\rangle \right\} $, and the time
evolution should be described with an additional set of amplitudes $\left\{
w_{f}\left( \tau \right) \right\} $, 
\begin{equation}
\left| M\left( \tau \right) \right\rangle =u\left( \tau \right) \left|
M^{0}\right\rangle +v\left( \tau \right) \left| \overline{M}%
^{0}\right\rangle +\Sigma _{f}w_{f}\left( \tau \right) \left| f\right\rangle 
\text{.}  \label{13}
\end{equation}
Besides the flavor basis, the evolution on the CP basis (\ref{1}), with
amplitudes $u^{\prime }\left( \tau \right) $ and $v^{\prime }\left( \tau
\right) $, is given by 
\begin{equation}
\left| M\left( \tau \right) \right\rangle =u^{\prime }\left( \tau \right)
\left| M_{1}\right\rangle +v^{\prime }\left( \tau \right) \left|
M_{2}\right\rangle +\Sigma _{f}w_{f}\left( \tau \right) \left|
f\right\rangle \text{.}  \label{14}
\end{equation}
The amplitudes $\left\{ w_{f}\left( \tau \right) \right\} $ describe the
possibility of reversible transitions of the type $\left|
M_{1/2}\right\rangle \rightleftharpoons \left| f\right\rangle $. Because of
the very large phase space (position $\otimes $ momentum) explored in
position space by the outgoing final products $\left\{ \left| f\right\rangle
\right\} $, the restriction to the description of irreversible transitions $%
\left| M_{1/2}\right\rangle \rightarrow \left| f\right\rangle $, excluding
the possibility of $\left| f\right\rangle \rightarrow \left|
M_{1/2}\right\rangle $, is sufficient. Following Lee, Oehme and Yang, the
Weisskopf-Wigner (WW) approximation \cite{14} is thus used to describe the
coupling to $\left\{ \left| f\right\rangle \right\} $ as an irreversible
decay.

A non-Hermitian decay operator $j\widehat{\gamma }$ is introduced to
describe $\left| M_{1/2}\right\rangle \rightarrow \left\{ \left|
f\right\rangle \right\} $ irreversible decays. It is to be noted that
reversed transitions $\left| f\right\rangle \rightarrow \left|
M_{1/2}\right\rangle $, avoided within the framework of the WW
approximation, are in fact also experimentally prohibited due to the very
large phase space explored by the outgoing products $\left\{ \left|
f\right\rangle \right\} $. The decay process, both at the level of the
experiments and at the level of the theoretical WW description, is an
irreversible process which should not be confused with a T violation
attributed to a fundamental interaction. Irreversibility is statistical and
TV is dynamical. If irreversibility is ultimately interpreted as a T
violation to preserve CPT, then the usual view on CPV can be recovered, this
issue will be addressed in section seven.

Within the framework of the WW approximation, the time evolution of $\left|
M\right\rangle $ is restricted to a two states Hilbert space ${\bf E}_{LOY}$%
, at the cost of the loss of unitarity ($d\left\langle M\right. \left|
M\right\rangle /d\tau <0$) induced by $j\widehat{\gamma }$. \ 

Without CPV and gravity, the usual LOY Hamiltonian is the sum of three terms 
\cite{5,6},

\begin{equation}
j\hbar \frac{d\left| M\right\rangle }{d\tau }=\left[ \widehat{m}c^{2}-\frac{%
\widehat{\delta m}}{2}c^{2}-j\frac{\hbar \widehat{\gamma }}{2}\right] \cdot
\left| M\right\rangle \text{.}  \label{Z34}
\end{equation}
The Hermitian part $\widehat{H}_{M}$ = $\widehat{m}c^{2}$ $-$ $\widehat{%
\delta m}c^{2}/2$ preserves unitarity and describes stable particles
dynamics. Here $\tau $ is the proper time measured in the meson \ frame. The
mass operator $\widehat{m}$, the mixing operator $\widehat{\delta m}$,
describing $M^{0}\rightleftharpoons \overline{M}^{0}$ oscillations, and the
decay operator $\widehat{\gamma }$, describing $\left| M\right\rangle
\rightarrow \left\{ \left| f\right\rangle \right\} $ irreversible
transitions, are expressed as: $\widehat{m}=m\widehat{\text{I}}$ and 
\begin{eqnarray}
\widehat{\delta m} &=&\delta m\left| M^{0}\right\rangle \left\langle 
\overline{M}^{0}\right| +\delta m\left| \overline{M}^{0}\right\rangle
\left\langle M^{0}\right| \text{,}  \label{17} \\
\widehat{\gamma } &=&\Gamma \widehat{\text{I}}+\delta \Gamma \widehat{\delta
m}/\delta m\text{,}  \label{18}
\end{eqnarray}
where $\delta m>0$ is the mass splitting between the heavy $\left( m_{H}%
\text{ = }m+\delta m/2\right) $ and light $\left( m_{L}\text{ = }m-\delta
m/2\right) $ mass eigenstates ($\delta m/m$ $\sim $ [$10^{-14}$, $10^{-13}$%
]). The widths $\Gamma >0$, $\delta \Gamma $ are respectively the average
and the difference between the decay widths of the these two mass
eigenstates ($\delta \Gamma _{K^{0}/\overline{K}^{0}}<0$ and $\delta \Gamma
_{B^{0}/\overline{B}^{0}}\sim 0$). The experimental values of $\ m$, $\delta
m$, $\Gamma $, $\delta \Gamma $ for the various flavored neutral mesons can
be found in {\it PDG 2024}. \ On the CP basis, without CPV and gravity $%
\left| M_{H/L}\right\rangle $ = $\left| M_{2/1}\right\rangle $, and

\begin{eqnarray}
\widehat{\delta m} &=&-\delta m\left| M_{2}\right\rangle \left\langle
M_{2}\right| +\delta m\left| M_{1}\right\rangle \left\langle M_{1}\right| 
\text{,}  \label{20} \\
\widehat{\gamma } &=&\Gamma _{2}\left| M_{2}\right\rangle \left\langle
M_{2}\right| +\Gamma _{1}\left| M_{1}\right\rangle \left\langle M_{1}\right| 
\text{,}  \label{21}
\end{eqnarray}
where $\Gamma _{2}$ = $\Gamma +\delta \Gamma /2$ and $\Gamma _{1}$ = $\Gamma
-\delta \Gamma /2$. The mixing $\widehat{\delta m}$ and decay $\widehat{%
\gamma }$ terms involve first and second order weak interaction couplings.
It is important to note that in the mass term $m=\left\langle q^{\prime }%
\overline{q}\right| {\cal H}\left| q^{\prime }\overline{q}\right\rangle $ = $%
\left\langle \overline{q^{\prime }}q\right| {\cal H}\left| \overline{%
q^{\prime }}q\right\rangle $ (${\cal H}$ is the sum of all the elementary
particles interactions) both the potential energy (ensuring $q^{\prime }%
\overline{q}$ binding) of the quark structure and\ the kinetic energy of the
quarks motion, including the {\it zitterbewegung fluctuations, }are taken
into account.

\section{Oscillations and free fall on Earth without decay}

The CPV free LOY dynamics (\ref{Z34}) is to be extended to take into account
the impact of Earth. First, a brief review of the protocol used to set up
the fundamental Hamiltonian 
\begin{eqnarray}
\widehat{H}_{M} &=&\left( m+\delta m/2\right) c^{2}\left| M_{2}\right\rangle
\left\langle M_{2}\right|  \label{hm} \\
&&+\left( m-\delta m/2\right) c^{2}\left| M_{1}\right\rangle \left\langle
M_{1}\right| \text{,}  \nonumber
\end{eqnarray}
is proposed. Then, this Hamiltonian is completed with the small
gravity-induced energy term. The Hamiltonian $\widehat{H}_{M}$ (\ref{hm}) is
obtained from the particle energy-momentum dispersion relation, in an
appropriate frame, with the correspondence principle{\it \ (}$E\rightarrow
j\hbar \partial /\partial \tau ...${\it )}.

Consider a typical experiment with a collimated horizontal beam. In a given
frame, this beam is fully characterized by the energy-momentum four-vector
of each meson $\left( P^{\mu }\right) $ = $\left( P^{0}\text{, }{\bf P}%
\right) $. The inertial mass $m\pm \delta m/2$ is the invariant ensuring the
identity of the meson. In a frame where the metric tensor components are $%
g_{\mu \nu }$ and the energy-momentum components $P^{\mu }$, the energy $%
E_{\pm }$ and the mass $m\pm \delta m/2$ are defined by 
\begin{eqnarray}
E_{\pm } &\equiv &-cg_{0\nu }P^{\nu }\text{,}  \label{b21} \\
\left( m\pm \delta m/2\right) ^{2} &\equiv &-g_{\mu \nu }P^{\mu }P^{\nu
}/c^{2}\text{.}  \label{b22}
\end{eqnarray}
In the laboratory frame, the measurement of the energy gives $E_{\pm
}^{\prime }$ and the measurement of the linear momentum gives ${\bf P}%
^{\prime }$ $=$ $P_{\Vert }^{\prime }{\bf e}_{\Vert }$ + $p_{\bot }^{\prime }%
{\bf e}_{\bot }$, $P_{\Vert }^{\prime }$ is the horizontal momentum parallel
to the surface of earth, $p_{\bot }^{\prime }$ is the vertical momentum
perpendicular this surface and (${\bf e}_{\Vert }$, ${\bf e}_{\bot }$, ${\bf %
e}_{\Vert }\times {\bf e}_{\bot }$) is a local laboratory Cartesian basis.
For example, the typical momentum of a kaon beam is in the range [$10-100$]
GeV/c \ and the imperfections of the beam horizontal collimation are kept as
small as possible with an angle below the mrad such that $p_{\bot }^{\prime
}/P_{\Vert }^{\prime }$ \ remains in the range [$10^{-4}$ $-$ $10^{-6}$];
thus $P_{\Vert }^{\prime }\gg mc$ and $p_{\bot }^{\prime }$ $\ll mc$.
Neglecting the physical impact of Earth and assuming that the laboratory is
an inertial frame, the Cartesian components of the metric are given by 
\begin{equation}
\left[ g_{\mu \nu }\right] =\text{{\it diag}}\left[ -1,1,1,1\right] \text{.}
\label{b23}
\end{equation}
In the laboratory frame the dispersion relations (\ref{b21}, \ref{b22})
becomes: $E_{\pm }^{\prime }=cP^{^{\prime }0}$ and 
\begin{equation}
\left( P^{^{\prime }0}\right) ^{2}=\left( m\pm \delta m/2\right)
^{2}c^{2}+P_{\Vert }^{\prime 2}+p_{\bot }^{\prime 2}\text{.}  \label{b25}
\end{equation}
The experimental data $\left( E_{\pm }^{\prime },P_{\Vert }^{\prime
},p_{\bot }^{\prime }\right) $ and the final decay products are measured in
the laboratory frame. Then, to produce the CPV parameters reported in the
experimental literature, such as {\it PDG 2024} \cite{7}, these data are
transformed, through a Lorentz boost, to provide a new description ($E_{\pm
} $, $P_{\Vert }=0$, $p_{\bot }$) in a frame where the meson horizontal
momentum cancels ($P_{\Vert }^{\prime }$ $\gg mc$ $\gg p_{\bot }^{\prime }$) 
$\rightarrow $ ($P_{\Vert }$ $=0$, $p_{\bot }=p_{\bot }^{\prime }$ $\ll mc$%
). In this {\it horizontal co-moving frame}, the small vertical momentum $%
p_{\bot }$ $\ll mc$ remains non relativistic and the associated time $\tau $
is also the meson proper time as the meson rest frame and the horizontal
co-moving frame are related by a simple Galilean transform. In the
horizontal co-moving frame the energy-momentum relation is given by the
expansion of $\sqrt{\left( m\pm \delta m/2\right) ^{2}c^{2}+p_{\bot }^{2}}$ with 
respect to $p_{\bot }/mc\ll 1$, 
\begin{equation}
E_{\pm }=\left( m\pm \delta m/2\right) c^{2}+p_{\bot }^{2}/2m\text{.}
\label{23}
\end{equation}
Despite its smallness, the mass spliting $\delta m/m\sim $ [$10^{-14}$, $%
10^{-13}$] is important in $\widehat{H}_{M}$ to describe the oscillation $%
M^{0}\rightleftharpoons \overline{M}^{0}$, however, in the remaining small
kinetic energy $p_{\bot }^{2}/2\left( m\pm \delta m/2\right) $ $\ll $ $%
mc^{2} $, $\delta m$ can be neglected in front of $m$. Then, following the
correspondence principle providing the rules for forming the quantum
evolution equation{\it :} $E\rightarrow $ $j\hbar \partial /\partial \tau $
and $p_{\bot }\rightarrow \widehat{p}_{\bot }$, the dynamics of a stable
neutral meson $\left| N\right\rangle $ is 
\begin{equation}
j\hbar d\left| N\right\rangle /d\tau =\widehat{H}_{M}\cdot \left|
N\right\rangle +\widehat{p}_{\bot }^{2}/2m\cdot \left| N\right\rangle \text{.%
}  \label{b26}
\end{equation}
The Hamiltonian $\widehat{H}_{M}$ operates in ${\bf E}_{LOY}$ and $\widehat{p%
}_{\bot }$ operates in the vertical direction $\left( \xi _{\bot }\right) $
Hilbert space ${\bf E}_{\bot }$ where $\left[ \widehat{\xi }_{\bot },%
\widehat{p}_{\bot }\right] =j\hbar $. Equation (\ref{b26}) is solved with a
state $\left| M\right\rangle \in {\bf E}_{LOY}$ and a wave function $\left|
\Phi \right\rangle \in {\bf E}_{\bot }$, 
\begin{equation}
\left| N\left( \tau \right) \right\rangle \equiv \left| M\left( \tau \right)
\right\rangle \left| \Phi \left( \tau \right) \right\rangle \text{,}
\label{b27}
\end{equation}
allowing to separate the vertical space dynamics in ${\bf E}_{\bot }$ from
the internal flavor dynamic in ${\bf E}_{LOY}$. The small vertical drift and
spreading of the meson wave function is the solution of $j\hbar $ $d\left|
\Phi \right\rangle /d\tau $ $=$ $\widehat{p}_{\bot }^{2}/2m\cdot \left| \Phi
\right\rangle $, starting from the initial wave function $\left| \Phi \left(
\tau =0\right) \right\rangle $ ($\tau =0$ when the meson is produced). A
complete set of normalized plane waves $\left\{ \left| \Phi
_{_{n}}\right\rangle \right\} $ with $\widehat{p}_{\bot }^{2}\cdot $ $\left|
\Phi _{_{n}}\right\rangle $ $=$ $2mE_{n}\left| \Phi _{_{n}}\right\rangle $,
is introduced. For example, if the meson is localized in a large vertical
vacuum gap, displaying a width $L$ between two horizontal walls, then $%
\left\langle \xi _{\bot }\right. \left| \Phi _{_{n}}\right\rangle $ $\sim $ $%
\sin \left( n\pi \xi _{\bot }/L\right) /\sqrt{L}$ if $0$ $\leq $ $\xi _{\bot
}$ $\leq $ $L$,\ $\xi _{\bot }$ is the vertical position with respect to the
lower wall. The vertical dynamics is obtained with the expansion 
\begin{equation}
\left| \Phi \right\rangle =\sum_{n}\left| \Phi _{_{n}}\right\rangle
\left\langle \Phi _{_{n}}\right. \left| \Phi \left( \tau =0\right)
\right\rangle \exp -j\frac{E_{n}\tau }{\hbar }\text{.}  \label{b30}
\end{equation}
As a consequence $\left| M\right\rangle $ fulfils the expected LOY dynamics (%
\ref{Z34}), with $\widehat{\gamma }=\widehat{0}$, 
\begin{equation}
j\hbar d\left| M\right\rangle /d\tau =\widehat{H}_{M}\cdot \left|
M\right\rangle \text{,}  \label{b31}
\end{equation}
as a function of the meson proper time $\tau $.

Consider the very same experimental situation, an horizontal collimated
meson beam and a meson, $M_{1}$ or $M_{2}$, with an energy-momentum
four-vector $\left( P^{\mu }\right) $ = ($P^{^{\prime }0}$, $P_{\Vert
}^{\prime }{\bf e}_{\Vert }$ $+$ $p_{\bot }^{\prime }{\bf e}_{\bot }$)
measured in the laboratory frame. But, now, the impact of Earth gravity is
taken into account.

To set up a quantum model with gravity, a conservative view on gravitation
is adopted and the canonical Einsteinian general relativity near a
non-rotating massive spherical object is used under the assumption that the
distance from its center is far larger than the Schwarzschild radius. The
gravitational field on the surface of Earth is weak: $G_{N}M_{\oplus
}/R_{\oplus }c^{2}\sim 10^{-9}$. The definition of a zero altitude reference
level does not require a comparison of the various laboratories altitudes as 
$dg/dh_{\bot }\sim 10^{-7}g$ m$^{-1}$ ($g$ = $G_{N}M_{\oplus }/R_{\oplus
}^{2}$ = $9.81$ m/s$^{2}$). Beside the altitude, the variation with the
latitude is also small, of the order of $0.5\%$ between the poles and the
equator where $0.3\%$ of these $0.5\%$ are due to the centrifugal force. The
impact of Earth slow rotation $\Omega _{\oplus }R_{\oplus }$ $\sim 10^{-6}c$
is neglected on the time scale of the meson lifetime $\left( 1/\Gamma
\right) $ as the typical displacement $R_{\oplus }$ $\Omega _{\oplus
}/\Gamma $ $\sim $ $\left[ 10^{-7},10^{-10}\right] $ m. Thus, on the surface
of Earth, in the laboratory frame, the Cartesian components of \ the metric
tensor are given by the usual {\it weak field approximation }on the surface
of a{\it \ }non-rotating massive spherical object

\begin{equation}
\left[ g_{\mu \nu }\left( \phi \right) \right] =\text{{\it diag}}\left[
-1,1,1,1\right] -2\phi \times \text{{\it diag}}\left[ 1,1,1,1\right] \text{.}
\label{b40}
\end{equation}
In the laboratory frame $\phi $ is the normalized static Newtonian potential,

\begin{equation}
\phi \left( r\right) =-G_{N}M_{\oplus }/rc^{2}\sim 10^{-9}\text{,}
\label{b41}
\end{equation}
where $r$ is the radial distance between the meson and the Earth's center.
In the laboratory frame, the position of a meson, $l_{\Vert }{\bf e}_{\Vert
} $ $+$ $h_{\bot }{\bf e}_{\bot }$, above the surface of Earth (which is
taken, for simplicity, as the floor of the laboratory) is given by the
horizontal $\left( l_{\Vert }\right) $ and vertical $\left( h_{\bot }\right) 
$ coordinates on a Cartesian basis (${\bf e}_{\Vert }$, ${\bf e}_{\bot }$, 
${\bf e}_{\Vert }\times {\bf e}_{\bot }$) attached to a fixed point and $%
r^{2}=\left( R_{\oplus }+h_{\bot }\right) ^{2}+l_{\Vert }^{2}$ where both $
h_{\bot }\ll R_{\oplus }$ and $l_{\Vert }$ $\ll R_{\oplus }$. A lateral
horizontal drift along ${\bf e}_{\Vert }\times {\bf e}_{\bot }$ does change
the final conclusions. The energy in the
Laboratory frame is given by 
\begin{eqnarray}
E_{\pm }^{\prime } &=&-cg_{0\nu }P^{\prime \nu }=c\left( 1+2\phi \right)
P^{^{\prime }0}\text{,} \\
\left( m\pm \delta m/2\right) ^{2}c^{2} &=&-g_{\mu \nu }P^{\mu }P^{\nu
}=\left( P^{^{\prime }0}\right) ^{2}\left( 1+2\phi \right)  \label{b44} \\
&&-\left( P_{\Vert }^{\prime 2}+p_{\bot }^{\prime 2}\right) \left( 1-2\phi
\right) \text{.}  \nonumber
\end{eqnarray}
With the usual interpretation of CPV experiments, gravity is ignored $\left(
\phi \ll 1\right) $, and (\ref{b25}) is recovered. Moving to the meson
horizontal co-moving frame, the horizontal momentum cancels but not the
perpendicular one ($P_{\Vert }^{\prime }$ $\gg mc$ $\gg p_{\bot }^{\prime }$
) $\rightarrow $ ($P_{\Vert }$ $=0$, $p_{\bot }=p_{\bot }^{\prime }\ll mc$)
. In this frame the remaining small perpendicular momentum $p_{\bot
}=p_{\bot }^{\prime }$ is non relativistic, so the time is similar to the
proper time of the meson. Indeed, the vertical velocity of the meson rest frame with
respect to the horizontal co-moving frame is far smaller than $c$. This
relative motion is due to the imperfect horizontal beam collimation and to
the very small meson free fall.

In the horizontal co-moving frame, used in the interpretation of CPV
experiments, Earth is moving and the metric to be considered is $g_{\mu \nu
}\left( \phi ^{*}\right) $ where $\phi ^{*}$($\neq \phi $) is the potential
of a large spherical mass $M_{\oplus }$ moving at a constant velocity.

In the laboratory frame the meson starts at ($h_{\bot }=h_{\bot i}$, $%
l_{\Vert i}=0$) and, after a typical lifetime $1/\Gamma $, it decays away at
($h_{\bot f}\approx h_{\bot i}$, $l_{\Vert f}>l_{\Vert i}$). The length of
the horizontal path of an unstable meson in the laboratory frame $l_{\Vert
f} $ is a fraction of mm, up to several tens of meters or few hundreds of
meters. In the laboratory frame, the field $\phi \left( h_{\bot f},l_{\Vert
f}\right) $ experienced by the meson when it disappears can be compared to
the field experienced when it was created $\phi \left( h_{\bot i},l_{\Vert
i}\right) $%
\begin{equation}
\phi \left( h_{\bot f},l_{\Vert f}\right) /\phi \left( h_{\bot i},l_{\Vert
i}\right) \approx 1-l_{\Vert f}^{2}/2R_{\oplus }^{2}\text{.}
\end{equation}
For a large value $l_{\Vert f}\approx 50$ m: $\phi \left( h_{\bot
f},l_{\Vert f}\right) /\phi \left( h_{\bot i},l_{\Vert i}\right) $ $-$ $1$ $%
= $ O$\left[ 10^{-11}\right] $, so it can be safely assumed that, during its
whole life, the meson experiences the very same Earth field $\phi \left(
h_{\bot f},l_{\Vert f}\right) $ = $\phi \left( h_{\bot i},l_{\Vert i}\right) 
$. As a consequence, the field $\phi ^{*}$ in the horizontal co-moving{\it \ 
}frame, can be taken static during the meson lifetime and $\phi ^{*}=\phi $.
The momentum and the energy in this horizontally co-moving frame fullfil

\begin{equation}
\frac{E_{\pm }^{2}}{1+2\phi }=\left( m\pm \delta m/2\right)
^{2}c^{4}+p_{\bot }^{2}c^{2}\left( 1-2\phi \right) \text{.}  \label{b46}
\end{equation}
The remaining momentum $p_{\bot }$ describes the small, non relativistic,
vertical motion due ({\it i}) to the imperfect horizontal beam collimation
and ({\it ii}) to the small free fall, free fall starting just after the
meson production. The free fall momentum component starts from zero and
remains non relativistic during the short lifetime $\left( 1/\Gamma \right) $
of the meson: $mg/\Gamma \ll mc$ (the scaling $\hbar \Gamma \sim \delta
mc^{2}$, common for several neutral mesons, gives $p_{\bot }/mc$ $\sim $ $%
g/c\Gamma \sim $ $\left( \hbar g/c\right) /\delta mc^{2}$ $\sim $ [$10^{-17}$%
, $10^{-19}$]).

A Taylor expansion (\ref{b46}) with respect to the small
parameters $\phi \ll 1$ and $p_{\bot }^{2}/m^{2}c^{2}\ll 1$ gives

\begin{equation}
E_{\pm }=\left( m\pm \delta m/2\right) c^{2}\left( 1+\phi \right) +p_{\bot
}^{2}/2\left( m\pm \delta m/2\right) \text{.}  \label{b47}
\end{equation}
Despite its smallness, the mass spliting $\delta m/m\sim \left[ 10^{-14}%
\text{, }10^{-13}\right] $ is important in $\widehat{H}_{M}$ to ensure the
oscillation $M^{0}\rightleftharpoons \overline{M}^{0}$. However, in $\left(
m\pm \delta m/2\right) \phi $ and $p_{\bot }^{2}/\left( m\pm \delta
m/2\right) $, which are very small perturbations to this oscillation, $%
\delta m$ can neglected in front of $m$. The first two terms of the
expansion of $\phi \left( R_{\oplus }+h_{\bot }\right) $ with respect to the
instantaneous vertical position $h_{\bot }\ll R_{\oplus }$, is 
\begin{equation}
-\frac{G_{N}M_{\oplus }m}{R_{\oplus }+h_{\bot }}\approx -\frac{%
G_{N}M_{\oplus }m}{R_{\oplus }}+mgh_{\bot }\text{.}  \label{b48}
\end{equation}
Thus, the energy-mass-momentum relations to be considered for $M_{1}$ and $%
M_{2}$ are

\begin{equation}
E_{\pm }=\left( m\pm \frac{\delta m}{2}\right) c^{2}-G_{N}\frac{M_{\oplus }m%
}{R_{\oplus }}+mgh_{\bot }+\frac{p_{\bot }^{2}}{2m}\text{.}  \label{b49}
\end{equation}
To set up a full picture of the experiments, in addition to the free fall
and the small initial vertical velocity, the possibility that the meson
altitude $h_{\bot }$ might slightly fluctuate, ({\it i}) independently of $%
\phi $, ({\it ii}) at very high frequency $\left( c/\lambda _{C}\sim
mc^{2}/\hbar \gg \delta mc^{2}/\hbar \right) $, ({\it iii}) around an {\it %
average} level $l_{\bot }$, ({\it iv}) with a very small amplitude $x$ of
the order of the Compton wavelength ($x\sim \lambda _{C}\sim [10^{-16}$, $%
10^{-17}]$ m $\ll l_{\bot }$), must be considered. This possibility results
from the fact that \ flavored neutral mesons mass eigenstates are not
structureless point particles, but display: ({\it i}) a finite size of the
order of their Compton wavelength $\lambda _{C}$ and ({\it ii}) an internal
dynamical structure made of quarks, $q$, $\overline{q}^{\prime }$, $%
\overline{q}$ and $q^{\prime }$, that might display the customary spin 1/2
fermion {\it zitterbewegung }behavior \cite{8,9}. To anticipate this
non-intuitive quantum behavior of spin 1/2 fermions and anti-fermions, the
instantaneous position of the center of mass/barycenter of the quark
structure is to be separated into ({\it i}) a slowly evolving average
barycenter position (free fall + small initial vertical velocity, $l_{\bot }$%
) plus ({\it ii}) a small fluctuation around this average ({\it %
zitterbewegung}, $x$), \ 
\begin{equation}
h_{\bot }=\left\langle h_{\bot }\right\rangle +\left[ h_{\bot }-\left\langle
h_{\bot }\right\rangle \right] =l_{\bot }+x\text{.}  \label{aver}
\end{equation}
These small $\left( \lambda _{C}\right) $, high frequency $\left(
mc^{2}/\hbar \right) $, vertical fluctuations $x\left( \tau \right) $,
resulting from the internal dynamics, are independent of $\phi $ and are
identified in the next section. To take into account the occurrence of these 
{\it zitterbewegung }fluctuations, the final energy-mass-momentum relation
used to apply the correspondence principle becomes

\begin{eqnarray}
E_{\pm } &=&\left( m\pm \frac{\delta m}{2}\right) c^{2}-\frac{G_{N}M_{\oplus
}m}{R_{\oplus }}  \label{b50} \\
&&+\frac{p_{\bot }^{2}}{2m}+mgl_{\bot }+mgx\text{.}  \nonumber
\end{eqnarray}
As expected, this expression is just the sum of a small classical kinetic
energy plus a small classical (average plus fluctuations) potential energy
completed by a large relativistic inertial mass energy. It is important to
note that, beside the gravity-induced potential energy $mgx$ of the small
fluctuations, the kinetic energy associated with the quarks {\it %
zitterbewegung} internal motion is already taken into account in the $m\pm
\delta m=\left\langle \widetilde{M}_{2/1}\right| {\cal H}\left| \widetilde{M}%
_{2/1}\right\rangle /c^{2}$ inertial mass term of the dispersion relation (%
\ref{b50}).

We are now in position to write down the equation describing neutral mesons
flavor oscillations on Earth as a function of the meson proper time.
Following the correspondence principle{\it ,} $E\rightarrow $ $j\hbar
\partial /\partial \tau $, $l_{\bot }\rightarrow \widehat{l}_{\bot }$, $%
p_{\bot }\rightarrow \widehat{p}_{\bot }=-j\hbar \partial /\partial l_{\bot
} $, $x\left( \tau \right) \rightarrow \widehat{x}\left( \tau \right) $, the
evolution of a meson state $\left| N\right\rangle $ $\in $ ${\bf E}%
_{LOY}\otimes {\bf E}_{\bot }$ is given by 
\begin{eqnarray}
j\hbar \frac{d\left| N\right\rangle }{d\tau } &=&\left[ \ \widehat{H}_{M}-%
\frac{mG_{N}M_{\oplus }}{R_{\oplus }}\widehat{\text{I}}\right] \cdot \left|
N\right\rangle  \label{31} \\
&&+\left[ \frac{\widehat{p}_{\bot }^{2}}{2m}+mg\widehat{l}_{\bot }+mg%
\widehat{x}\left( \tau \right) \right] \cdot \left| N\right\rangle \text{.} 
\nonumber
\end{eqnarray}
The meson altitude operator $\widehat{l}_{\bot }$ and the vertical momentum
operator $\widehat{p}_{\bot }$ are canonically conjugated, $\left[ \widehat{l%
}_{\bot },\widehat{p}_{\bot }\right] $ = $j\hbar $. The relation (\ref{31})
provides a Schr\"{o}dinger representation of the quantum dynamics where $d%
\widehat{l}_{\bot }/d\tau $ = $d\widehat{p}_{\bot }/d\tau $ = $\widehat{0}$,
and, as will be demonstrated in the next section, within the framework of
this Schr\"{o}dinger representation, $d\widehat{x}/d\tau $ $\neq $ $\widehat{%
0}$. The operators $\widehat{l}_{\bot }$ and $\widehat{p}_{\bot }$ operate
in ${\bf E}_{\bot }$. The Hamiltonian $\ \widehat{H}_{M}$ operates in ${\bf E%
}_{LOY}$.

The physical nature of the small{\it \ vertical zitterbewegung }operator $%
\widehat{x}\left( \tau \right) $ is elucidated in the next sections and 
Appendix B through a reduction from its representation in $\widetilde{{\bf E}}_{Q}$,
where the quarks structure and dynamics can be described, to its
representation in ${\bf E}_{LOY}$.

\section{Dirac velocity operator and zitterbewegung fluctuations}

In the next section it will be demonstrated that a precise knowledge of $%
\widehat{x}\left( \tau \right) $ in (\ref{31}) is not required to set up and
understand GICPV. Indeed, the dominant coupling between Earth and flavored
mesons ultimately involves the associated {\it zitterbewegung velocity
fluctuations }operator $d\widehat{x}/d\tau $. This section focuses on the
analysis of $d\widehat{x}/d\tau ${\it .}

A quark, $q$ or $\overline{q}$, is a confined spin 1/2 fermion evolving in
an appropriate Hilbert space where various symmetry groups can be
represented: $\left| q\right\rangle $ $\sim \left| \theta _{\text{flavor}%
}\right. $ $\zeta _{\text{color}}$ $\sigma _{\text{spin}}$ $\left. \psi _{%
\text{space}}....\right\rangle $. The energy exchanged during flavor
oscillations ($\delta mc^{2}\sim $ [$10^{-6}$, $10^{-4}$] eV) and the energy
involved in the gravity-induced coupling ($mg\hbar /\delta mc\sim $ [$%
10^{-9} $, $10^{-10}$] eV) are both very small. To describe such extremely
low energy processes, quantum field theory is not needed and the set of
symmetry groups necessary to build a relevant quark model can be restricted.
The useful quark variables are : the spin, charge and mass variables to
define the low energy identity of a quark and ${\bf X}$ = $\left(
X,Y,Z\right) $ defining its position; $q$ and $\overline{q}$ are thus described
here as simple {\it Poincar\'{e} elementary particles}.

Dirac's equation, within the framework of its Feynman's interpretation, is
thus sufficient. The structure of $M_{0}$, or $\overline{M}_{0}$, is rather
simple, this is a combination of 2 different quarks, structured as a spin
singlet bound state, $q$ with $\overline{q}^{\prime }$, or $\overline{q}$
with $q^{\prime }$. Therefore, it appears that the tensorial combinations of
\ two 4 components Dirac spinors $\left| q/\overline{q}\right\rangle \left|
q^{\prime }/\overline{q}^{\prime }\right\rangle $ provides a necessary and
sufficient framework. Indeed, a single $4$ components spinor $\left| q/%
\overline{q}\right\rangle $ describes both a fermion $q$ and its
anti-fermion $\overline{q}$, thus only two Dirac Hilbert spaces will be
used. The tensorial product of these two Dirac spinors spaces is $\widetilde{%
{\bf E}}_{Q}$.

Appendix B provides an explicit construction of the spinor representation of
the spin singlet meson states $\left| \widetilde{M}_{0}\right\rangle $ $\in 
\widetilde{{\bf E}}_{Q}$ and $\left| \widetilde{\overline{M}}%
_{0}\right\rangle $ $\in \widetilde{{\bf E}}_{Q}$. These states $\left| 
\widetilde{M}_{0}\right\rangle $ and $\left| \widetilde{\overline{M}}%
_{0}\right\rangle $ describe the same particle and antiparticle as the state 
$\left| M^{0}\right\rangle \in {\bf E}_{LOY}$ and $\left| \overline{M}%
^{0}\right\rangle \in {\bf E}_{LOY}$, but with far more details about the
quark structure. However they share few common well defined properties such
as the {\it zitterbewegung }velocity fluctuations around an average position
which will be expressed by the operators $d\widehat{x}_{Q}/d\tau $ in $%
\widetilde{{\bf E}}_{Q}$ and $d\widehat{x}/d\tau $ in ${\bf E}_{LOY}$.

Two Dirac Hamiltonian, $\widehat{H}_{D}$ and $\widehat{H}_{D}^{\prime }$,
are required to set up the appropriate framework needed to study quark
confinement and low energy (from $10^{-9}$-$10^{-10}$ eV to $10^{-6}$-$%
10^{-4}$ eV ) dynamics. The evolutions of the\ two $4$ components spinors $%
\left| q/\overline{q}\right\rangle $ and $\left| q^{\prime }/\overline{q}%
^{\prime }\right\rangle $ are given by Dirac's equations 
\begin{eqnarray}
j\hbar d\left| q/\overline{q}\right\rangle /d\tau &=&\widehat{H}_{D}\cdot
\left| q/\overline{q}\right\rangle \text{,}  \label{ZS1} \\
j\hbar d\left| q^{\prime }/\overline{q}^{\prime }\right\rangle /d\tau &=&%
\widehat{H}_{D}^{\prime }\cdot \left| q^{\prime }/\overline{q}^{\prime
}\right\rangle \text{.}  \label{ZS2}
\end{eqnarray}
The (unknown) Dirac Hamiltonian $\widehat{H}_{D}$ describes the effective
confinement resulting from the interaction of $q/\overline{q}$ with $%
\overline{q}^{\prime }$/$q^{\prime }$. The (unknown) Dirac Hamiltonian $%
\widehat{H}_{D}^{\prime }$ describes the interaction of $\overline{q}%
^{\prime }$/$q^{\prime }$ with $q/\overline{q}$. The structures of these
Hamiltonians are canonical: 
\begin{eqnarray}
\widehat{H}_{D}\left( \widehat{{\bf X}},\widehat{{\bf P}}\text{,}\tau
\right) &=&c\widehat{\alpha }\cdot \widehat{{\bf P}}+\widehat{\beta }%
m_{q}c^{2}+F\left( \widehat{{\bf X}},\tau \right) \text{,} \\
\widehat{H}_{D}^{\prime }\left( \widehat{{\bf X}}^{\prime },\widehat{{\bf P}}%
^{\prime }\text{,}\tau \right) &=&c\widehat{\alpha }^{\prime }\cdot \widehat{%
{\bf P}}^{\prime }+\widehat{\beta }^{\prime }m_{q^{\prime }}c^{2}+G\left( 
\widehat{{\bf X}}^{\prime },\tau \right) \text{,}
\end{eqnarray}
where $F$ and $G$ are the (unknown) self-consistent confinement fields.
These self-consistent confinement fields, $F\left( \widehat{{\bf X}},\tau
\right) $ and $G\left( \widehat{{\bf X}}^{\prime },\tau \right) $, are
functions of the other quark, for example through the positions $%
\left\langle q^{\prime }/\overline{q}^{\prime }\right| \widehat{{\bf X}}%
^{\prime }\left| q^{\prime }/\overline{q}^{\prime }\right\rangle $ in $F$
and $\left\langle q/\overline{q}\right| \widehat{{\bf X}}\left| q/\overline{q%
}\right\rangle $ in $G$, so that (\ref{ZS1}) and (\ref{ZS2}) are coupled.
The position operator $\widehat{{\bf X}}=\left( \widehat{X},\widehat{Y},%
\widehat{Z}\right) $ operates on $4$ components spinors of the type $\left|
q/\overline{q}\right\rangle $. The position operator $\widehat{{\bf X}}%
^{\prime }=\left( \widehat{X}^{\prime },\widehat{Y}^{\prime },\widehat{Z}%
^{\prime }\right) $ operates on $4$ components spinors of the type $\left|
q^{\prime }/\overline{q}^{\prime }\right\rangle $. The usual Dirac-Pauli
representation of the Dirac algebra ($\widehat{\alpha ^{i}}\cdot \widehat{%
\alpha ^{j}}$ \ $+$ $\widehat{\alpha ^{j}}\cdot \widehat{\alpha ^{i}}$ $=$ $%
2\delta ^{ij}\widehat{\text{I}}...$) \cite{8,9}, will be used in this study.
The {\it Dirac velocity operator} $\widehat{\alpha }\equiv $ ($\widehat{%
\alpha }_{X},$ $\widehat{\alpha }_{Y},\widehat{\alpha }_{Z}$) is expressed
in terms of the usual Pauli spin operators ($\widehat{\sigma }_{X},$ $%
\widehat{\sigma }_{Y},$ $\widehat{\sigma }_{Z}$) as $\widehat{\alpha }_{i}$ $%
=$ $\widehat{\sigma }_{X}\otimes \widehat{\sigma }_{i}$ $\ $and $\widehat{%
\alpha }_{i}^{\prime }$ $=$ $\widehat{\sigma }_{X}^{\prime }\otimes \widehat{%
\sigma }_{i}^{\prime }$. \ Rather than Schr\"{o}dinger's representation (\ref
{ZS1}, \ref{ZS2}), Heisenberg's representation of the quark dynamics will be
used here. The quark states are thus stationary ($d\left| q/\overline{q}%
\right\rangle /d\tau $ = $d\left| q^{\prime }/\overline{q}^{\prime
}\right\rangle /d\tau $ = $0$) and the operators ($\widehat{{\bf X}},%
\widehat{{\bf P}}...$) become time dependant ($\widehat{{\bf X}}\left( \tau
\right) ,\widehat{{\bf P}}\left( \tau \right) ...$). The time dependant
position operators $\widehat{{\bf X}}\left( \tau \right) $ and $\widehat{%
{\bf X}}^{\prime }\left( \tau \right) $ are solutions of Heisenberg's
equations 
\begin{equation}
d\widehat{{\bf X}}/d\tau =j\left[ \widehat{H}_{D},\widehat{{\bf X}}\right]
/\hbar \text{, }d\widehat{{\bf X}}^{\prime }/d\tau =j\left[ \widehat{H}%
_{D}^{\prime },\widehat{{\bf X}}^{\prime }\right] /\hbar \text{.}
\end{equation}
Since $\widehat{{\bf X}}$ and $\widehat{{\bf X}}^{\prime }$ commute with all
terms in $\widehat{H}_{D}$ and $\widehat{H}_{D}^{\prime }$ except $\widehat{%
\alpha }\cdot \widehat{{\bf P}}$ and $\widehat{\alpha }^{\prime }\cdot 
\widehat{{\bf P}}^{\prime }$ the velocities are given by 
\begin{eqnarray}
\frac{d\widehat{{\bf X}}}{d\tau } &=&c\widehat{\alpha }\text{,}  \label{zs3}
\\
\frac{d\widehat{{\bf X}}^{\prime }}{d\tau } &=&c\widehat{\alpha }^{\prime }%
\text{.}  \label{zs4}
\end{eqnarray}
The eigenvalues of $\widehat{\alpha }$ are $\pm 1$ and $\widehat{\alpha }%
_{X}^{2}$ = $\widehat{\alpha }_{Y}^{2}$ = $\widehat{\alpha }_{Z}^{2}$ = $%
\widehat{\text{I}}$.

To simplify, the same notations have been used for $\widehat{{\bf X}}$ and $%
\widehat{{\bf X}}\left( \tau \right) $ despite the fact that they have
different structures ($\widehat{{\bf X}}\left( \tau \right) $ = $\widehat{%
{\bf U}}^{-1}\left( \tau \right) \cdot \widehat{{\bf X}}\cdot \widehat{{\bf U%
}}\left( \tau \right) $). The {\it zitterbewegung }behavior displayed by (%
\ref{zs3}, \ref{zs4}): 
\begin{eqnarray}
\left( d\widehat{X}/d\tau \right) ^{2} &=&\left( d\widehat{Y}/d\tau \right)
^{2}=\left( d\widehat{Z}/d\tau \right) ^{2}=c^{2}\widehat{\text{I}}\text{,}
\label{zs5} \\
\left( d\widehat{X}^{\prime }/d\tau \right) ^{2} &=&\left( d\widehat{Y}%
^{\prime }/d\tau \right) ^{2}=\left( d\widehat{Z}^{\prime }/d\tau \right)
^{2}=c^{2}\widehat{\text{I}}\text{,}  \label{zs6}
\end{eqnarray}
holds true for any, free or bound, spin 1/2 fermions. The relations (\ref
{zs3}, \ref{zs4}, \ref{zs5}, \ref{zs6}) are independent of ({\it i}) the
mass of the fermion-antifermion, ({\it ii}) the charge of the
fermion-antifermion, ({\it iii}) the shape and ({\it iv}) the strength of
the self-consistent confinement potential $F$ and $G$. This {\it %
zitterbewegung }behavior is completely independent of the quark confinement
model ($\widehat{H}_{D}$, $\widehat{H}_{D}^{\prime }$), and is universal.

The universal {\it zitterbewegung} behavior is usually presented as
counter-intuitive in standard textbook. It can be viewed as a quiver ({\it %
zitter}) behavior ({\it bewegung}) at an instantaneous velocity equal to $%
\pm c$. The {\it zig-zag} model of the electron proposed in \cite{15} offers
a pictorial representation of the relations (\ref{zs5}, \ref{zs6}).
Moreover, the model of reference \cite{15} provides an interpretation of P
violation in weak interaction in terms of preferential W or Z bosons
coupling with left and right helicity. Zitterbewegung provides also an
operational framework to interpret the Darwin term in the hydrogen fine
structure problem as a correction due to the electron position fluctuation
of the order of the Compton wavelength. These are the very same positions
fluctuations and their interplay with $M^{0}\rightleftharpoons \overline{M}%
^{0}$ oscillations which provide the origin of CPV.

In Appendix B a large class of di-quark tensorial states encompassing the
particular singlet type states, $\left| \widetilde{M}_{0}\right\rangle $ $%
\sim $ $\left| q\uparrow \right\rangle \left| \overline{q}^{\prime }{\bf %
\downarrow }\right\rangle -$ $\left| q{\bf \downarrow }\right\rangle \left| 
\overline{q}^{\prime }\uparrow \right\rangle $ $\in \widetilde{{\bf E}}_{Q}$
and $\left| \widetilde{\overline{M}}_{0}\right\rangle $ $\sim $ $\exp j%
\overline{\delta }$ ($\left| \overline{q},\uparrow \right\rangle \left|
q^{\prime }{\bf \downarrow }\right\rangle $ $-$ $\left| \overline{q}{\bf %
\downarrow }\right\rangle \left| q^{\prime }\uparrow \right\rangle $) $\in 
\widetilde{{\bf E}}_{Q}$, are explicitly constructed. In this same appendix
the combination $\widehat{{\bf X}}_{Q}\left( \tau \right) $ of the positions 
$\widehat{{\bf X}}\left( \tau \right) $ and $\widehat{{\bf X}}^{\prime
}\left( \tau \right) $, defining the instantaneous position of the center of
mass/barycenter of two quarks in a meson $\left| \widetilde{M}%
_{0}\right\rangle $ or $\left| \widetilde{\overline{M}}_{0}\right\rangle $,
is identified. The barycentric/center of mass position operator $\widehat{%
{\bf X}}_{Q}$ operates on $\left| \widetilde{M}_{0}\right\rangle $ and $%
\left| \widetilde{\overline{M}}_{0}\right\rangle $ in $\widetilde{{\bf E}}%
_{Q}$. Following the definition (\ref{aver}), the{\it \ }fluctuations
operator around the barycenter \ is 
\begin{equation}
\widehat{{\bf x}}_{Q}\left( \tau \right) =\widehat{{\bf X}}_{Q}\left( \tau
\right) -\left\langle \widehat{{\bf X}}_{Q}\left( \tau \right) \right\rangle
_{\tau }\text{,}  \label{XX}
\end{equation}
where $\left\langle \widehat{{\bf X}}_{Q}\left( \tau \right) \right\rangle
_{\tau }$ is a time average eliminating the high frequency content of $%
\widehat{{\bf X}}_{Q}\left( \tau \right) $. The associated vertical{\it \ }%
velocity fluctuations operator in $\widetilde{{\bf E}}_{Q}$ is thus simply 
\begin{equation}
d\widehat{{\bf x}}_{Q}/d\tau =d\widehat{{\bf X}}_{Q}/d\tau \text{,}
\label{xx}
\end{equation}
as $d\left\langle \widehat{X}_{Q}\right\rangle _{\tau }/d\tau $ $=$ $%
\widehat{0}$. In Appendix B, starting from (\ref{zs3}, \ref{zs4}), after
some calculations, the following matrix elements of $d\widehat{{\bf x}}%
_{Q}/d\tau $ are obtained 
\begin{equation}
\left\langle \widetilde{M}_{0}\right| d\widehat{{\bf x}}_{Q}/d\tau \left| 
\widetilde{M}_{0}\right\rangle =\left\langle \widetilde{\overline{M}}%
_{0}\right| d\widehat{{\bf x}}_{Q}/d\tau \left| \widetilde{\overline{M}}%
_{0}\right\rangle ={\bf 0}\text{,}
\end{equation}
for the diagonal part. For the non diagonal part, the {\it zitterbewegung}
behavior reappears and leads to 
\begin{eqnarray}
\left\langle \widetilde{M}_{0}\right| d\widehat{x}_{Q}/d\tau \left| 
\widetilde{\overline{M}}_{0}\right\rangle &=&\pm c\text{,}  \label{ult1} \\
\left\langle \widetilde{\overline{M}}_{0}\right| d\widehat{x}_{Q}/d\tau
\left| \widetilde{M}_{0}\right\rangle &=&\pm c\text{,}  \label{ult2}
\end{eqnarray}
The results are similar along the two others cartesian directions $d\widehat{%
y}_{Q}/d\tau $ and $d\widehat{z}_{Q}/d\tau $. The states ($\left| \widetilde{%
M}_{0}\right\rangle $, $\left| \widetilde{\overline{M}}_{0}\right\rangle $)
in $\widetilde{{\bf E}}_{Q}$ contain far more details than the states ($%
\left| M^{0}\right\rangle $,$\left| \overline{M}^{0}\right\rangle $) \ in $%
{\bf E}_{LOY}$. Some observables in $\widetilde{{\bf E}}_{Q}$ are
meaningless in ${\bf E}_{LOY}\otimes {\bf E}_{\bot }$, but observables such
as the instantaneous barycenter vertical position are meaningful in both
Hilbert spaces $\widetilde{{\bf E}}_{Q}$ and ${\bf E}_{LOY}\otimes {\bf E}%
_{\bot }$. The states $\left| \widetilde{M}_{0}\right\rangle $ or $\left| 
\widetilde{\overline{M}}_{0}\right\rangle $ describe the very same physical
object as $\left| M^{0}\right\rangle $ or $\left| \overline{M}%
^{0}\right\rangle $ and the {\it vertical } {\it zitterbewegung }%
fluctuations operator $d\widehat{x}_{Q}/d\tau $ and $d\widehat{x}/d\tau $
describe the very same fluctuations around an average position (\ref{aver}, 
\ref{XX}) thus the matrix elements of $d\widehat{x}/d\tau $ are given by 
\begin{eqnarray}
\left\langle M^{0}\right| d\widehat{x}/d\tau \left| M^{0}\right\rangle
&=&\left\langle \overline{M}^{0}\right| d\widehat{x}/d\tau \left| \overline{M%
}^{0}\right\rangle =0\text{,} \\
\left\langle M^{0}\right| d\widehat{x}/d\tau \left| \overline{M}%
^{0}\right\rangle &=&\pm c\text{,}  \label{ult3} \\
\left\langle \overline{M}^{0}\right| d\widehat{x}/d\tau \left|
M^{0}\right\rangle &=&\pm c\text{.}  \label{ult4}
\end{eqnarray}
It is important to note that a reduction from $\widetilde{{\bf E}}_{Q}$ to $%
{\bf E}_{LOY}\otimes {\bf E}_{\bot }$ was expected, but $d\widehat{{\bf X}}%
_{Q}/d\tau $ does not operate on the quarks space variables $X,Y,Z$ and $%
X^{\prime },Y^{\prime },Z^{\prime }$ (see Appendix B) because $d\widehat{%
{\bf X}}/d\tau $ and $d\widehat{{\bf X}}^{\prime }/d\tau $ fullfil (\ref{zs3}%
, \ref{zs4}), so $d\widehat{x}/d\tau $ does not operate in ${\bf E}_{\bot }$
(or equivalently operate as $\widehat{\text{I}}$ in ${\bf E}_{\bot }$) and
the final reduction is restricted simply to ${\bf E}_{LOY}$. Note also that
the phase convention adopted in $\widetilde{{\bf E}}_{Q}$ is similar to the
phase convention adopted in ${\bf E}_{LOY}$ and is either $\exp j\delta =+1$
or $\exp j\delta =-1$ (Appendix B).

With these results (\ref{ult3}, \ref{ult4}) there are four possibilities
to construct $d\widehat{x}/d\tau $. The reduction from a detailed
description ($\widetilde{{\bf E}}_{Q}$) to a simplified one (${\bf E}_{LOY}$%
) is not just a matter of eliminating redundant degrees of freedom while
preserving the same type of algebraic structure in both Hilbert spaces, care
must also be exercised on the physical structures of the larger and smaller
Hilbert spaces involved in such a reduction. In a 4 components spinor
Hilbert space, describing coupled spin 1/2 fermions and antifermions, the
mass of the fermion appears to be positive and the mass of the antifermion
appears to be negative \cite{8,9}; although, in the 2 component LOY model (%
\ref{hm}), the mass $m$ of the antiparticle $\overline{M}^{0}$ is positive
and equal to the mass $m$ of the particle $M^{0}$. This difference between $%
\widetilde{{\bf E}}_{Q}$ and ${\bf E}_{LOY}$ on the relative sign of the
particle-antiparticle mass is to be resolved to achieve the construction of $%
d\widehat{x}/d\tau $ with \ the four possibilities offered by (\ref{ult3}, 
\ref{ult4}). The clue to resolve this issue is to consider the usual Feynman
interpretation of the solutions of Dirac's equations regarding
particle-antiparticle kinematics. The Dirac equation negative mass puzzle
was resolved through the Feynman-Stueckelberg interpretation of an
antiparticle (with negative mass) as a particle (with positive mass)
propagating backward in time. This usual Feynman prescription (antiparticle
are particle propagating backward in time) tells us that the representation
of a velocity $d\widehat{x}/d\tau $ on the ${\bf E}_{LOY}$
particle/antiparticle basis ($M^{0},\overline{M}^{0}$) must display opposite
sign when the flavor are reversed. With (\ref{ult3}, \ref{ult4}), the PCI
formulation of the Feynman prescription within the framework of the reduced
LOY model imply that if 
\begin{equation}
\frac{d\widehat{x}}{d\tau }\cdot \left| M^{0}\right\rangle =\pm c\text{ }%
\widehat{CP}\cdot \left| M^{0}\right\rangle \text{,}
\end{equation}
then, the antiparticle behavior is similar, but backward in time, and the
associated relation is 
\begin{equation}
\frac{d\widehat{x}}{d\tau }\cdot \left| \overline{M}^{0}\right\rangle =\mp c%
\text{ }\widehat{CP}\cdot \left| \overline{M}^{0}\right\rangle \text{,}
\end{equation}
with a positive mass antiparticle such as $\overline{M}^{0}$ in the LOY
model. This sign prescription agree with the discussion on parity in
Appendix B, and leads to 
\begin{equation}
\frac{d\widehat{x}}{d\tau }=c\left| M^{0}\right\rangle \left\langle 
\overline{M}^{0}\right| -c\left| \overline{M}^{0}\right\rangle \left\langle
M^{0}\right| \text{,}  \label{ff}
\end{equation}
the {\it zitterbewegung }velocity operator to be used in (\ref{31}). The
choice of the opposite signs for both flavor eigenstates projectors
correspond to the other phase convention $\exp j\delta =-1$ discussed at the
end of section two. Observables are Hermitian operators, but non-observables
are not necessarily described by Hermitian operators, in particular $d%
\widehat{x}_{Q}/d\tau $, $d\widehat{X}_{Q}/d\tau $ and $d\widehat{x}/d\tau $
are not observables because their definitions imply two successive
measurements of the position with $d\tau \rightarrow 0$ such that the
uncertainty on the energy $\sim \hbar /d\tau $ becomes infinite. This type
of uncertainty in momentum space favors large values of momentum associated
with velocities very close to the speed of light. Following simply an
analysis of parity, the very same final result (\ref{ff}) is obtained at the
end of Appendix B. The Feynman prescription on particles-antiparticles
kinematics provides a point of view different from \ the parity argument
used at the end of Appendix B, thus the construction of $d\widehat{x}/d\tau $
(\ref{ff}) is robust and model independent.

\section{Stable mass eigenstates on Earth: $\left| M_{1}^{\oplus
}\right\rangle $ and $\left| M_{2}^{\oplus }\right\rangle $}

The operators, $\widehat{m}c^{2}$, $\widehat{\delta m}c^{2}$ and $mg\widehat{%
x}$ \ in (\ref{31}) do not display the same status as the operator $\widehat{%
\gamma }$ in (\ref{Z34}). The effective operator $\widehat{\gamma }$ is
associated with the WW description of $\left| M_{1/2}\right\rangle
\rightarrow \left| f\right\rangle $ transitions excluding the possibility of 
$\left| f\right\rangle \rightarrow \left| M_{1/2}\right\rangle $ and thus
providing an irreversible time evolution. If Earth's gravity impacts
flavored neutral mesons dynamics, this influence must be explored at the
fundamental level with a model preserving {\it a priori} unitarity,
reversibility and T invariance. The impact of the WW description of
irreversible transitions $\left| M_{1/2}\right\rangle \rightarrow \left|
f\right\rangle $ is to be considered after a clear understanding of the
unitary evolution on Earth. This two steps analysis not only provides the
right value of the CPV parameters, it also clarify the common confusion
between (statistical) irreversibility and (dynamical) time reversal
violation. If the reversibility of a time evolution is preserved in a
dynamical model, time reversal can be assumed without any ambiguity. The
analysis below will follow two steps. First, assuming $\widehat{\gamma }=%
\widehat{0}$, the small gravity induced CP mixing resulting from the
interplay between {\it \ zitterbewegung }and {\it flavor oscillations }will
be identified. This gives the stable mass eigenstates $\left| M_{1}^{\oplus
}\right\rangle $ and $\left| M_{2}^{\oplus }\right\rangle $ identified in
this section. Then, with $\widehat{\gamma }\neq \widehat{0}$, the observed
mass eigenstates $\left| M_{H}^{\oplus }\right\rangle \neq \left|
M_{2}^{\oplus }\right\rangle $ and $\left| M_{L}^{\oplus }\right\rangle \neq
\left| M_{1}^{\oplus }\right\rangle $ will result from the balance between
the {\it gravity induced mixing/regeneration} and {\it mesons} {\it decay }%
for{\it \ }$K^{0}/\overline{K}^{0}$ and a combination of {\it gravity
induced mixing }and {\it mesons} {\it decay }for{\it \ }$B^{0}/\overline{B}%
^{0}$.

Following this two steps program, the dynamics of a stable meson on Earth,
in the horizontal co-moving frame is given by (\ref{31}) 
\begin{eqnarray}
j\hbar \frac{d\left| N\right\rangle }{d\tau } &=&c^{2}\left[ m\widehat{\text{%
I}}-\frac{\widehat{\delta m}}{2}-\frac{mG_{N}M_{\oplus }}{R_{\oplus }c^{2}}%
\widehat{\text{I}}\right] \cdot \left| N\right\rangle  \label{52} \\
&&+\left[ mg\widehat{x}\left( \tau \right) +mg\widehat{l}_{\bot }+\frac{%
\widehat{p}_{\bot }^{2}}{2m}\right] \cdot \left| N\right\rangle \text{.} 
\nonumber
\end{eqnarray}
The inertial mass $m$ and the bulk gravity term $mG_{N}M_{\oplus }/R_{\oplus
}\sim 10^{-9}mc^{2}$ are eliminated with the change of variables 
\begin{equation}
\left| N\right\rangle =\left| M^{\prime }\right\rangle \exp -jmc^{2}\left(
1-G_{N}M_{\oplus }/c^{2}R_{\oplus }\right) \tau /\hbar \text{,}  \label{W1}
\end{equation}
where $2G_{N}M_{\oplus }/c^{2}$ $\sim 8.8$ mm $\ll R_{\oplus }$ is the
Schwarzschild radius. The resulting evolution of $\left| M^{\prime
}\right\rangle $ is given by 
\begin{eqnarray}
j\hbar \frac{d\left| M^{\prime }\right\rangle }{d\tau } &=&-\frac{\widehat{%
\delta m}}{2}c^{2}\cdot \left| M^{\prime }\right\rangle  \label{Z67} \\
&&+\left[ mg\widehat{x}\left( \tau \right) +mg\widehat{l}_{\bot }+\frac{%
\widehat{p}_{\bot }^{2}}{2m}\right] \cdot \left| M^{\prime }\right\rangle 
\text{.}  \nonumber
\end{eqnarray}
The Hilbert space where $\left| M^{\prime }\right\rangle $ evolves is the
tensorial product ${\bf E}_{LOY}\otimes {\bf E}_{\bot }$. The operator $%
\widehat{\delta m}$ operates in ${\bf E}_{LOY}$. The\ operators $\widehat{l}%
_{\bot }$ and $\widehat{p}_{\bot }$, with $\left[ \widehat{l}_{\bot },%
\widehat{p}_{\bot }\right] =j\hbar $, operate in ${\bf E}_{\bot }$. The
fluctuating position operator $\widehat{x}\left( \tau \right) $ operates, 
{\it a priori}, both in ${\bf E}_{LOY}$ and in ${\bf E}_{\bot }$.

In (\ref{Z67}) the energy operator $mg\widehat{x}\left( \tau \right) $
describes the fluctuating vertical dynamics complementing the average one $mg%
\widehat{l}_{\bot }$. However, the {\it average} potential energy $mgl_{\bot
}\sim $ [$10^{-4}$, $10^{-3}$] $\delta mc^{2}$ (with $l_{\bot }\sim 10$. cm)
is far more larger than the fluctuating potential energy $mgx\sim $ [$%
10^{-17}$, $10^{-19}$] $\delta mc^{2}$ (with $x\sim \lambda _{C}$). It would
be appropriate to ignore the action of $mg\widehat{x}\left( \tau \right) $,
in front of the action of $mg\widehat{l}_{\bot }$ in ${\bf E}_{\bot }$, but,
it would be inappropriate to completely ignore the action of $mg\widehat{x}%
\left( \tau \right) $, if any, in ${\bf E}_{LOY}$. So, this remaining action
of $mg\widehat{x}\left( \tau \right) $ on the flavor degrees of freedom, if
any, will not be ignored to keep track of \ Earth's influence.

The non-relativistic vertical drift/free fall/broadening of the wave
function, and the flavor mixing can be studied with the tensorial product 
\begin{equation}
\left| M^{\prime }\left( \tau \right) \right\rangle \equiv \left| M\left(
\tau \right) \right\rangle \left| \Psi \left( \tau \right) \right\rangle 
\text{,}  \label{Z09}
\end{equation}
where $\left| M\right\rangle $ $\in {\bf E}_{LOY}$ and $\left| \Psi
\right\rangle \in {\bf E}_{\bot }$. The four operators, $\widehat{\delta m}$%
, $\widehat{x}$, $\widehat{l}_{\bot }$ and $\widehat{p}_{\bot }$, on the
right hand side of (\ref{Z67}), and the {\it zitterbewegung} velocity
fluctuations operator $d\widehat{x}/d\tau $ (\ref{ff}), operate on different
degrees of freedom which are summarized below.

The two operators $\widehat{\delta m}$ (\ref{20})\ and $d\widehat{x}/d\tau $
(\ref{ff}) operate exclusively in ${\bf E}_{LOY}$ where $\left|
M_{2/1}\right\rangle $ is an orthonormal basis, 
\begin{eqnarray}
\widehat{\delta m}\cdot \left( \left| M_{2/1}\right\rangle \left| \Psi
\right\rangle \right) &=&\left( \mp \delta m\left| M_{2/1}\right\rangle
\right) \left| \Psi \right\rangle \text{ , }  \label{d23} \\
d\widehat{x}/d\tau \cdot \left( \left| M_{2/1}\right\rangle \left| \Psi
\right\rangle \right) &=&\left( \mp c\left| M_{1/2}\right\rangle \right)
\left| \Psi \right\rangle \text{.}  \label{d24}
\end{eqnarray}

The two operators, $\widehat{l}_{\bot }$ and $\widehat{p}_{\bot }$, are just
a set of conjugated position and momentum operators used to observe the
average meson position or momentum, without the {\it zitterbewegung}
fluctuations, along the vertical direction, in the horizontal co-moving
frame. The actions of $\widehat{l}_{\bot }$ and $\widehat{p}_{\bot }$ are
exclusively in ${\bf E}_{\bot }$, 
\begin{eqnarray}
\widehat{l}_{\bot }\cdot \left( \left| M_{2/1}\right\rangle \left| \Psi
\right\rangle \right) &=&\left| M_{2/1}\right\rangle \left( \widehat{l}%
_{\bot }\cdot \left| \Psi \right\rangle \right) \text{, }  \label{D45} \\
\widehat{p}_{\bot }\cdot \left( \left| M_{2/1}\right\rangle \left| \Psi
\right\rangle \right) &=&\left| M_{2/1}\right\rangle \left( \widehat{p}%
_{\bot }\cdot \left| \Psi \right\rangle \right) \text{.}  \label{D46}
\end{eqnarray}

A meson is a composite object whose complete internal structure and dynamics
is not resolved within the framework of equation (\ref{Z67}). Nevertheless,
this internal dynamics is not fully neglected with the possibility of small
vertical fluctuations $\widehat{x}\left( \tau \right) $. As a consequence of
the ordering $mgl_{\bot }=mc^{2}\left( gl_{\bot }/c^{2}\right) $ $\sim $ [$%
10^{-8}$, $10^{-7}$] eV $\ll $ $mgx\sim \hbar g/c\sim $ $10^{-23}$ eV ($%
l_{\bot }\sim 10.$ cm, $x\sim \lambda _{C}$), the effect of the operator $%
\widehat{x}$ on $\left| \Psi \right\rangle $, if any, can be neglected in
front the action of $\widehat{l}_{\bot }$ on $\left| \Psi \right\rangle $%
\begin{eqnarray}
\widehat{x}\cdot \left( \left| M_{2/1}\right\rangle \left| \Psi
\right\rangle \right) &=&\left( \widehat{x}\cdot \left| M_{2/1}\right\rangle
\right) \left| \Psi \right\rangle  \label{dry} \\
&&+\left| M_{2/1}\right\rangle \text{O}\left[ 10^{-15}\right] \widehat{l}%
_{\bot }\cdot \left| \Psi \right\rangle \text{.}  \nonumber
\end{eqnarray}
The (unknown) action of $\widehat{x}$ on $\left| M_{2/1}\right\rangle $ in $%
{\bf E}_{LOY}$, must not be ignored. This summary of the actions of the
various operators allows to separate equation (\ref{Z67}). The set of
normalized states $\left\{ \left| \Psi _{_{n}}\right\rangle \right\} $ ($%
\left\langle \Psi _{_{n}}\right. \left| \Psi _{_{n}}\right\rangle =1$),
solutions of the Schr\"{o}dinger eigenvalue problem in ${\bf E}_{\bot }$: 
\begin{equation}
\left( \widehat{p}_{\bot }^{2}/2m+mg\widehat{l}_{\bot }\right) \cdot \left|
\Psi _{_{n}}\right\rangle =E_{n}\left| \Psi _{_{n}}\right\rangle \text{,}
\label{D98}
\end{equation}
(with, for example, the following boundary conditions: $\left\langle l_{\bot
}=0\right. \left| \Psi _{_{n}}\right\rangle $ $=$ $\left\langle l_{\bot
}=+\infty \right. \left| \Psi _{_{n}}\right\rangle =0$, to ensure that (\ref
{D98}) is a well posed problem) provides a complete orthonormal basis to
expand $\left| \Psi \right\rangle $. With this basis an initial state $%
\left| \Psi \left( \tau =0\right) \right\rangle $ evolves at $\tau $ as 
\begin{equation}
\left| \Psi \right\rangle =\sum_{n}\left| \Psi _{_{n}}\right\rangle
\left\langle \Psi _{_{n}}\right. \left| \Psi \left( \tau =0\right)
\right\rangle \exp -j\frac{E_{n}\tau }{\hbar }\text{.}  \label{D64}
\end{equation}
The initial wave function $\left\langle l_{\bot }\right. \left| \Psi \left(
\tau =0\right) \right\rangle $ is centered at the meson production altitude.
The small vertical drift/free fall/broadening described by $\left| \Psi
\left( \tau \right) \right\rangle $ (\ref{D64}) fulfils 
\begin{equation}
j\hbar d\left| \Psi \right\rangle /d\tau =\left[ mg\widehat{l}_{\bot }+%
\widehat{p}_{\bot }^{2}/2m\right] \cdot \left| \Psi \right\rangle \text{.}
\label{Z098}
\end{equation}
The evolution of $\left| M\right\rangle \left| \Psi \right\rangle $ is
described by (\ref{Z67}), thus (\ref{Z098}) and (\ref{dry}) imply that $%
\left| M\right\rangle $ must be the solution of 
\begin{equation}
j\hbar \frac{d\left| M\right\rangle }{d\tau }=\left[ -\frac{\widehat{\delta m%
}}{2}c^{2}+mg\widehat{x}\left( \tau \right) \right] \cdot \left|
M\right\rangle \text{.}  \label{H76}
\end{equation}

For the sake of completeness, an explicit expression of the free-fall energy
eigenstates $\left\{ \left| \Psi _{_{n}}\right\rangle \right\} $ is given
below. The solutions of (\ref{D98}) \ are given by the set of Airy functions 
\begin{equation}
\left\langle l_{\bot }\right. \left| \Psi _{_{n}}\right\rangle =\text{Ai}%
\left( \frac{l_{\bot }-E_{n}/mg}{\left( \hbar ^{2}/2m^{2}g\right) ^{1/3}}%
\right)
\end{equation}
if $l_{\bot }>0$ and $\left\langle l_{\bot }\right. \left| \Psi
_{_{n}}\right\rangle =0$ if $l_{\bot }\leq 0$, where, to simplify, the
normalization factor has been omitted.

The probability amplitude is exponentially small along the upward vertical
direction, $\left\langle l_{\bot }\right. \left| \Psi _{_{n}}\right\rangle
\rightarrow 0$ when $l_{\bot }\rightarrow +\infty $. \ The spectrum of
allowed energies $E_{1,2...n}$ $\ =$ $\left( m\hbar ^{2}g^{2}/2\right)
^{1/3} $ $\times $ [$2.3,$ $4.0,...\left( 3\pi n/2\right) ^{2/3}...$] is
given by the boundary conditions on the ground Ai$\left( -E_{n}/\left(
m\hbar ^{2}g^{2}/2\right) ^{1/3}\right) =0$.

The strong energy ordering, $mgx\sim \hbar g/c\sim 10^{-23}$ eV $\ll $ $%
\delta mc^{2}\sim $ [$10^{-6}$, $10^{-4}$] eV, leads to the identification
of $\ \hbar g/\delta mc^{3}$ $\sim $ [$10^{-19}$, $10^{-17}$] as the small
parameter associated with the interplay between gravity and weak mixing.
Given the smallness of this parameter, the evolution of $\left|
M\right\rangle $ in (\ref{H76}) is largely dominated by weak mixing $%
\widehat{\delta m}$. Neglecting O[$\left( \hbar g/\delta mc^{3}\right)
^{2}]< $ O$\left[ 10^{-34}\right] $ in front of O[$\hbar g/\delta mc^{3}]$,
the state $-\widehat{\delta m}c^{2}\cdot \left| M\right\rangle $ $=$ $\
2j\hbar d\left| M\right\rangle /d\tau $ can be substituted into the term $mg%
\widehat{x}\cdot \left| M\right\rangle $ of (\ref{H76}) to obtain 
\begin{equation}
j\hbar \frac{d\left| M\right\rangle }{d\tau }=-\left[ \frac{\widehat{\delta m%
}}{2}c^{2}+2j\frac{mg\hbar }{c^{2}}\ \ \widehat{x}\cdot \widehat{\delta m}%
^{-1}\cdot \frac{d}{d\tau }\right] \left| M\right\rangle \text{.}
\label{zdde}
\end{equation}
Then, using the relation $\widehat{\delta m}^{-1}=\widehat{\delta m}/\delta
m^{2}$ and defining the fundamental GICPV coupling parameter 
\begin{equation}
\kappa =\frac{2mg\hbar }{\delta m^{2}c^{3}}\text{,}  \label{ZD35}
\end{equation}
($\kappa \sim $ [$10^{-3}$, $10^{-6}$]), equation (\ref{zdde}) can be
rewritten 
\begin{eqnarray}
&&j\hbar \frac{d}{d\tau }\left[ \left( \widehat{\text{I}}+\kappa \frac{c}{%
\hbar }\ \ \widehat{x}\cdot \widehat{\delta m}\right) \cdot \left|
M\right\rangle \right]  \label{ult58} \\
&=&-\frac{\delta m}{2}c^{2}\left( \widehat{\text{I}}-2j\frac{\kappa }{c}\ 
\frac{d\widehat{x}}{d\tau }\right) \cdot \frac{\widehat{\delta m}}{\delta m}%
\cdot \left| M\right\rangle \text{.}  \nonumber
\end{eqnarray}
The operator $\widehat{x}\cdot \widehat{\delta m}$ is O$\left[ \lambda
_{C}\delta m\right] $, thus the operator $\kappa c\ \ \widehat{x}\cdot 
\widehat{\delta m}/\hbar $, on the left hand side of (\ref{ult58}), is O$%
\left[ \kappa \left( \delta m/m\right) \right] $ in front of $\ \widehat{%
\text{I}}$. The operator $\left( \kappa /c\right) d\widehat{x}/d\tau $, on
the right hand side of (\ref{ult58}), is O$\left[ \kappa \right] $ in front
of $\widehat{\text{I}}$. As $\delta m/m\sim $[$10^{-14}$, $10^{-13}$], to
keep only the dominant gravity-induced perturbation, the $d\widehat{x}/d\tau 
$ term will remain on the right hand side and the $\widehat{x}$ term, on the
left hand side, will be neglected. The evolution of \ the meson state $%
\left| M\right\rangle $, taking into account Earth's influence, becomes 
\begin{equation}
j\hbar \frac{d\left| M\right\rangle }{d\tau }=\left[ -\frac{\widehat{\delta m%
}}{2}c^{2}+j\kappa c\left( \frac{d\widehat{x}}{d\tau }\cdot \widehat{\delta m%
}\right) \right] \cdot \left| M\right\rangle \text{.}  \label{D77}
\end{equation}
The evolution (\ref{D77}) reveals an interplay between the flavor
oscillation $\left( +\delta m\rightleftharpoons -\delta m\right) $ and the 
{\it zitterbewegung velocity } $\left( +c\rightleftharpoons -c\right) $. On
the CP basis, the action of the two operators involved in (\ref{D77}) are
given by 
\begin{eqnarray}
\widehat{\delta m}\cdot \left| M_{2/1}\right\rangle &=&\mp \delta m\left|
M_{2/1}\right\rangle \text{,}  \label{d56} \\
\ d\widehat{x}/d\tau \cdot \widehat{\delta m}\cdot \left|
M_{2/1}\right\rangle &=&\delta mc\left| M_{1/2}\right\rangle \text{.}
\label{d57}
\end{eqnarray}
With the help of these two relations it is straightforward to identify the
energy eigenstates of (\ref{D77}) fulfilling the eigenvector equation 
\begin{equation}
\left[ -\frac{\widehat{\delta m}}{2}+j\frac{\kappa }{c}\frac{d\widehat{x}}{%
d\tau }\cdot \widehat{\delta m}\right] \cdot \left| M_{2/1}^{\oplus
}\right\rangle =\pm \frac{\delta m}{2}\text{ }\left| M_{2/1}^{\oplus
}\right\rangle \text{.}  \label{equa}
\end{equation}
Neglecting O$\left[ \kappa ^{2}\right] \left| M_{2/1}\right\rangle $ terms
in front of O$\left[ 1\right] \left| M_{2/1}\right\rangle $ terms, the
solutions are 
\begin{eqnarray}
\left| M_{2}^{\oplus }\right\rangle &=&\left| M_{2}\right\rangle +j\kappa
\left| M_{1}\right\rangle \text{,}  \label{Z799} \\
\left| M_{1}^{\oplus }\right\rangle &=&\left| M_{1}\right\rangle -j\kappa
\left| M_{2}\right\rangle \text{,}  \label{Z999}
\end{eqnarray}
with the normalization $\sqrt{1+\kappa ^{2}}$ $=1+$ O$\left[
10^{-6},10^{-12}\right] $ $\approx 1$. These relations define the stable
mesons mass eigenstates on Earth $\left| M_{2/1}^{\oplus }\right\rangle $ as
a linear combination of the CP eigenstates $\left| M_{2/1}\right\rangle $
resulting from the unitary evolution perturbed by gravity. On the flavor
basis 
\begin{equation}
\left| M_{2/1}^{\oplus }\right\rangle =\frac{1\pm j\kappa }{\sqrt{2}}\left|
M^{0}\right\rangle \mp \frac{1\mp j\kappa }{\sqrt{2}}\left| \overline{M}%
^{0}\right\rangle \text{.}
\end{equation}
\ It is to be noted that this gravity-induced mixing (\ref{Z799}, \ref{Z999}%
) does not violate T and violates CP. In the next sections reversibility
will be lost as irreversible decays are taken into account. This offers the
opportunity to set up an apparent CPT invariant description with CPV and TV
for \ $K^{0}/\overline{K}^{0}$ at the cost of a confusion between
irreversibility and TV. Considering this CPT invariant description obtained
for $K^{0}/\overline{K}^{0}$ as fundamental overlooks the problem of the
origin of CPV: the two steps GICPV mechanism without TV.

\section{Unstable mass eigenstates on Earth: $\left| M_{H}^{\oplus
}\right\rangle $ and $\left| M_{L}^{\oplus }\right\rangle $}

Mesons are unstable and the stable eigenstates $\left| M_{2}^{\oplus
}\right\rangle $ and $\left| M_{1}^{\oplus }\right\rangle $ defined by (\ref
{Z799}, \ref{Z999}), can not be observed. The mesons short lifetimes result
in the {\it observed eigenstates} $\left| M_{H}^{\oplus }\right\rangle \neq
\left| M_{2}^{\oplus }\right\rangle $ and $\left| M_{L}^{\oplus
}\right\rangle \neq $ $\left| M_{1}^{\oplus }\right\rangle $. The
theoretical identification of the eigenstates $\left| M_{2}^{\oplus
}\right\rangle $ and $\left| M_{1}^{\oplus }\right\rangle $ is useful to
understand the impact of gravity at the fundamental level: CPV without TV.
However, \ experiments are initial value problems, with an initial state
preparation and a final state measurement, and are not eigenvalues problems.
A selected set of measurements is conducted after the sequence
preparation/selection/evolution/decay. The result of these measurements is
interpreted with a model based, for example with the KM-CPV CPT invariant
eigenstates review in Appendix A.

To describe the interplay between ({\it i}) the irreversible decay and ({\it %
ii}) the dynamics of gravity induced mixing, two different cases are to
considered: ({\it i}) $\Gamma _{1}\gg \Gamma _{2}$ and ({\it ii}) $\Gamma
_{1}\sim \Gamma _{2}$, the former apply to $K^{0}/\overline{K}^{0}$ and the
later to $B^{0}/\overline{B}^{0}$. The confusion between irreversibility and
TV when CPT is assumed is particularly clear for the $K^{0}/\overline{K}^{0}$
as $\Gamma _{1}\gg \Gamma _{2}$.

\underline{{\bf Case 1:} $\Gamma _{1}\gg \Gamma _{2}$}. In this subsection
the $M_{2}$ lifetime is far larger than the $M_{1}$ one, thus $M_{2}$ is
assumed stable and 
\begin{equation}
\widehat{\gamma }\cdot \left| M_{1}\right\rangle =\Gamma _{1}\left|
M_{1}\right\rangle \text{, }\widehat{\gamma }\cdot \left| M_{2}\right\rangle
=0\text{.}  \label{decay}
\end{equation}
An initial value problems is considered: the initial preparation of the
system is a meson state $M_{2}^{\oplus }$. This situation is typical of type
({\it i}) $K^{0}/\overline{K}^{0}$ indirect CPV experiments.

The $M_{1}$ content of the $M_{2}^{\oplus }$ mass eigenstate (\ref{Z799})
disappears at a rate $\Gamma _{1}$ so that the remaining $M_{2}$ content is
no longer a mass eigenstate $M_{2}^{\oplus }$, but, as it is observed it
must remain a mass eigenstate. To remain so, the operator $\left( d\widehat{x%
}/d\tau \right) \cdot \widehat{\delta m}$ regenerates the small $M_{1}$
content, destroyed at a rate $\Gamma _{1}$ through decay. Indeed, $\left( d%
\widehat{x}/d\tau \right) \cdot \widehat{\delta m}$ \ transforms $\left|
M_{2}\right\rangle $ into $\left| M_{1}\right\rangle $ (\ref{d57}). The
dynamics induced by Earth's gravity can be viewed as a {\it gravity induced
regeneration} $M_{2}\rightarrow M_{1}$ (\ref{d57}), competing with a fast
decay $M_{1}\rightarrow f$ (\ref{decay}). The dissipative version of
equation (\ref{D77}) is 
\begin{equation}
j\hbar \frac{d\left| M\right\rangle }{d\tau }=-\left[ \frac{\widehat{\delta m%
}}{2}c^{2}+j\hbar \frac{\widehat{\gamma }}{2}-j\kappa c\frac{d\widehat{x}}{%
d\tau }\cdot \widehat{\delta m}\right] \cdot \left| M\right\rangle \text{.}
\label{th4}
\end{equation}
Two terms on the right hand side of (\ref{th4}) are not Hermitian, they can
compensate each other and provide a steady state. This source-sink process: $%
M_{2}\stackrel{\text{Gravity}}{\rightarrow }$ $M_{1}\stackrel{\text{Decay}}{%
\rightrightarrows }$ $f\,$ sustains the $M_{1}$ content of $M_{H}^{\oplus }$%
. The $M_{2}$ component of $M_{H}^{\oplus }$ is assumed stable and the
depletion of its amplitude due to the {\it gravitational regeneration} of $%
M_{1}$ is negligible. The observed heavy state $\left| M_{H}^{\oplus
}\right\rangle $, steady state solution of this source-sink problem, behave
as a superposition of the CP eigenstates and, assuming a CPT invariant form, a
mixing of the type 
\begin{eqnarray}
\left| M_{H}^{\oplus }\right\rangle &=&\left| M_{2}\right\rangle +\epsilon
\left| M_{1}\right\rangle \text{,}  \label{j1} \\
\left| M_{L}^{\oplus }\right\rangle &=&\left| M_{1}\right\rangle +\epsilon
\left| M_{2}\right\rangle \text{,}  \label{j2}
\end{eqnarray}
can be considered, where $\epsilon \sim $ O$\left[ \kappa \right] $ is\ the
unknown GICPV observed CPV mixing parameter to be calculated. This CPT
invariant form assumption for the dissipative CP violating dynamics leads to assume TV, but,
there are no TV if we consider the full dynamical picture with the GICPV two
steps mechanism or the appropriate large Hilbert space \{$\left| M,\right.
\oplus ,$ $f_{1},f_{2},\left. f_{3}...\right\rangle $\} without the
irreversibility introduced by the WW approximation.

Looking for a steady-state solution of (\ref{th4}) such that $d\left\langle
M_{H}^{\oplus }\right. \left| M_{H}^{\oplus }\right\rangle /d\tau $ $=$ $0$
under the combined influence of gravity and decay $\left| M_{H}^{\oplus
}\right\rangle $ (\ref{j1}) must fulfill the relation

\begin{equation}
\left. j\kappa c\frac{d\widehat{x}}{d\tau }\cdot \widehat{\delta m}\cdot
\left| M_{H}^{\oplus }\right\rangle \right| _{\text{gravity}}=\left. j\hbar 
\frac{\widehat{\gamma }}{2}\cdot \left| M_{H}^{\oplus }\right\rangle \right|
_{\text{decay}}\text{.}
\end{equation}
Neglecting a very small second order O$\left[ \kappa \epsilon \right] $ $%
\sim $ O$\left[ \kappa ^{2}\right] \sim \left[ 10^{-6},10^{-12}\right] $
term, the observed state $\left| M_{H}^{\oplus }\right\rangle $ (\ref{j1})
is a steady-state if and only if 
\begin{equation}
\epsilon =\kappa \frac{2\delta mc^{2}}{\hbar \Gamma _{1}}\text{.}  \label{j3}
\end{equation}
The very same result was obtained with a different method in \cite{12,13},
where a gravity induced regeneration rate was introduced and calculated, so
this result (\ref{j3}) is method independent. As a result of the
irreversible evolution, gravity induced mixing can be interpreted as a CPV
with CPT conservation, that is to say with TV. However irreversibility has
nothing to do with TV and the confusion is clarified here by the two step
mechanism: first unitary CPV with $\left| M_{1/2}^{\oplus }\right\rangle $
and then the impact of decay, with $\Gamma _{1}\gg \Gamma _{2}$, leading to $%
\left| M_{H/L}^{\oplus }\right\rangle $ (\ref{j1}) under the assumption of
an apparent CPT invariance which is physically meaningless in a dissipative
problem such as (\ref{th4}) moreover on a curved space-time such as (\ref
{b40}).

For kaons experiments, it is important to note that the state (\ref{j1}) is
observed experimentally with $\epsilon $ given by (\ref{j3}), but the state (%
\ref{j2}) is a CPT invariant form difficult to observe
experimentally because of the ordering $\Gamma _{1}\gg \Gamma _{2}$.

\underline{{\bf Case 2: }$\Gamma _{1}\approx \Gamma _{2}$ $=\Gamma $}. In
this subsection $\Gamma _{1}\approx \Gamma _{2}$, thus $\widehat{\gamma }%
=\Gamma \widehat{\text{I}}$. The Hamiltonian taking into account both the
decay and the impact of gravity is obtained on the basis of a derivation
similar to the one leading to (\ref{equa}), but here with $\widehat{\gamma }%
\neq \widehat{0}$. Defining the parameters $\chi $ $=$ $\hbar \Gamma /\delta
mc^{2}$, $\zeta $ $=$ $\arctan \chi $ and $\kappa ^{\prime }=$ $\kappa /%
\sqrt{1+\chi ^{2}}$ associated with the interplay between mixing, decay and
gravity and using the relation 
\begin{equation}
\widehat{\delta m}/\delta m-j\chi \widehat{\text{I}}=\left( 1+\chi
^{2}\right) \left( \widehat{\delta m}/\delta m+j\chi \widehat{\text{I}}%
\right) ^{-1}\text{,}
\end{equation}
the equation describing the evolution of an unstable meson state under the
influence of weak interaction and gravity when $\Gamma _{1}\approx \Gamma
_{2}$ $=\Gamma $ becomes 
\begin{eqnarray}
j\hbar \frac{d\left| M\right\rangle }{d\tau } &=&-\frac{\delta m}{2}%
c^{2}\left( \frac{\widehat{\delta m}}{\delta m}+j\chi \widehat{\text{I}}%
\right) \cdot \left| M\right\rangle \\
&&+j\delta m\frac{\kappa c}{\left( 1+\chi ^{2}\right) }\frac{d\widehat{x}}{%
d\tau }\cdot \left( \frac{\widehat{\delta m}}{\delta m}-j\chi \widehat{\text{%
I}}\right) \cdot \left| M\right\rangle \text{,}  \nonumber
\end{eqnarray}
rather than (\ref{D77}). After some algebra, the initial value solutions
associated with experiments can be described as a superposition of two
solutions

\begin{equation}
\left| M\right\rangle =\sum_{HL}\left| M_{H/L}^{\oplus }\right\rangle \left.
\left\langle M_{H/L}^{\oplus }\right. \left| M\right\rangle \right| _{\tau
=0}\exp \mp j\frac{\delta mc^{2}}{2\hbar }\tau -\frac{\Gamma }{2}\tau
\end{equation}
where the observables states $\left| M_{H}^{\oplus }\right\rangle $ and $%
\left| M_{L}^{\oplus }\right\rangle $ are given by

\begin{eqnarray}
\left| M_{H}^{\oplus }\right\rangle &=&\left| M_{2}\right\rangle +j\kappa
^{^{\prime }}\exp \left( j\zeta \right) \left| M_{1}\right\rangle \text{,}
\label{BBB1} \\
\left| M_{L}^{\oplus }\right\rangle &=&\left| M_{1}\right\rangle -j\kappa
^{^{\prime }}\exp \left( -j\zeta \right) \left| M_{2}\right\rangle \text{.}
\label{BBB2}
\end{eqnarray}
This result displays what can be called a {\it dissipative dressing} of the
fundamental relation (\ref{Z799}, \ref{Z999}) with an additional phase
factor $\exp \left( \pm j\zeta \right) $ which will be the clue to
understand the results of type ({\it iii}) CPV experiments with $B^{0}/%
\overline{B}^{0}$.

\section{Type ({\it i}) $K^{0}/\overline{K}^{0}$experiments: $\varepsilon $}

The usual notations for kaons physics will be used in this section: the
light kaon $L$ is the short lived kaon $S$ and the heavy kaon $H$ is the
long lived kaon $L$. The ordering $\Gamma _{L}=\Gamma _{2}\ll \Gamma
_{1}=\Gamma _{S}\approx 578\times \Gamma _{L}$) allows to use the relations (%
\ref{j1}, \ref{j2}, \ref{j3}). Thus, assuming a stable $K_{L}$, the GICPV
mixing parameter (\ref{j3}) is given by

\begin{equation}
\epsilon =\frac{\left\langle K_{L}^{\oplus }\right. \left| K_{S}^{\oplus
}\right\rangle }{2}=\frac{2\delta m_{K}c^{2}}{\hbar \Gamma _{S}}\frac{%
2m_{K}g\hbar }{\delta m_{K}^{2}c^{3}}=1.66\times 10^{-3}\text{,}  \label{eps}
\end{equation}
in remarkable agreement with the experimental value reported in {\it PDG
2024 }\cite{7}

\begin{equation}
\mathop{\rm Re}\left. \left\langle K_{L}^{\oplus }\right. \left|
K_{S}^{\oplus }\right\rangle \right| _{\exp }/2=\left( 1.66\pm 0.02\right)
\times 10^{-3}\text{.}
\end{equation}
This result was obtained from the source-sink mechanism: $\left|
K_{2}\right\rangle \stackrel{\text{Gravity}}{\rightarrow }\left|
K_{1}\right\rangle \stackrel{\text{Decay}}{\rightrightarrows }\left|
f\right\rangle $ with a stable $K_{2}$. A complete view on the decays, with
an unstable $K_{2}$: $\left| K_{2}\right\rangle \stackrel{\text{Decay}}{%
\rightarrow }\left| f^{\prime }\right\rangle $, reveals a additional {\it %
phenomenological dissipative }imaginary part{\it \ }attached to the above
indirect CPV parameter.

Consider now the full experimental situation with both $\Gamma _{2}$ $=$ $%
\Gamma _{L}\neq 0$ 
\mbox{$<$}%
$\Gamma _{1}=\Gamma _{S}$ decay rates defined as: $\Gamma _{S}$ = $\Sigma
_{f}\left| \left\langle f\right| {\cal T}\left| K_{S}^{\oplus }\right\rangle
\right| ^{2}$ and $\Gamma _{L}$ = $\Sigma _{f}\left| \left\langle f\right| 
{\cal T}\left| K_{L}^{\oplus }\right\rangle \right| ^{2}$. Bell and
Steinberger have demonstrated a general relation based on global unitarity
starting from the evaluation of $\ d\left\langle M\right. \left|
M\right\rangle /d\tau $ at $\tau =0$ \cite{16}. For $K_{S}$, the sum $\Sigma
_{f}$ \ over the final states is dominated (99.9\%) by $K_{S}^{\oplus
}\rightarrow 2\pi $ decays, more precisely by the $K_{S}^{\oplus }$ $%
\rightarrow I_{0}$ decays (95\%) to the isospin-zero combination of $\left|
\pi ^{+}\pi ^{-}\right\rangle $ with $\left| \pi ^{0}\pi ^{0}\right\rangle $ and  
Bell-Steinberger \cite{16} derived the PCI relation:

\begin{equation}
\left( j\frac{\delta m_{K}c^{2}}{\hbar }+\frac{\Gamma _{S}}{2}\right)
\left\langle K_{S}^{\oplus }\right. \left| K_{L}^{\oplus }\right\rangle
=\left\langle I_{0}\right| {\cal T}\left| K_{L}^{\oplus }\right\rangle
\left\langle I_{0}\right| {\cal T}\left| K_{S}^{\oplus }\right\rangle ^{*}%
\text{.}  \label{a42}
\end{equation}
The restriction of $\sum_{f}\left| f\right\rangle $ to $\left|
I_{0}\right\rangle $ reduces the $K_{S}^{\oplus }$ width to $\Gamma _{S}$ = $%
\left\langle I_{0}\right| {\cal T}\left| K_{S}^{\oplus }\right\rangle
\left\langle I_{0}\right| {\cal T}\left| K_{S}^{\oplus }\right\rangle ^{*}$
so that

\begin{equation}
\left\langle I_{0}\right| {\cal T}\left| K_{L}^{\oplus }\right\rangle
\left\langle I_{0}\right| {\cal T}\left| K_{S}^{\oplus }\right\rangle
^{*}=\Gamma _{S}\left\langle I_{0}\right| {\cal T}\left| K_{L}^{\oplus
}\right\rangle /\left\langle I_{0}\right| {\cal T}\left| K_{S}^{\oplus
}\right\rangle \text{.}  \label{a43}
\end{equation}
This PCI relation is then substituted in Bell-Steinberger's relation (\ref
{a42}) to obtain the final expression 
\begin{equation}
\frac{\left\langle I_{0}\right| {\cal T}\left| K_{L}^{\oplus }\right\rangle 
}{\left\langle I_{0}\right| {\cal T}\left| K_{S}^{\oplus }\right\rangle }=%
\frac{\left\langle K_{S}^{\oplus }\right. \left| K_{L}^{\oplus
}\right\rangle }{2}\left( 1+j\frac{2\delta m_{K}c^{2}}{\hbar \Gamma _{S}}%
\right) \text{.}  \label{a44}
\end{equation}
It is important to note that this relation is PCI. The left hand side of (%
\ref{a44}) is just the canonical definition of the complex indirect CPV
parameter $\varepsilon $. The relation (\ref{eps}) gives $\left\langle
K_{S}^{\oplus }\right. \left| K_{L}^{\oplus }\right\rangle $ so that $%
\varepsilon _{GICPV}$ is given by 
\begin{equation}
\varepsilon _{GICPV}=\epsilon \left( 1+j\frac{2\delta m_{K}c^{2}}{\hbar
\Gamma _{S}}\right) \text{.}
\end{equation}
Comparing the predicted and experimental values resulting from (\ref{a44})
provides an extremely compelling support for GICPV as 
\begin{equation}
\mathop{\rm Re}\varepsilon _{GICPV}=\left( \frac{2\delta m_{K}c^{2}}{\hbar
\Gamma _{S}}\right) \frac{2m_{K}g\hbar }{\delta m_{K}^{2}c^{3}}=1.66\times
10^{-3}\text{,}
\end{equation}
is in excellent agreement with the value reported in \cite{7}:

\begin{equation}
\mathop{\rm Re}\varepsilon _{PDG2024}=\left( 1.66\pm 0.02\right) \times
10^{-3}\text{.}
\end{equation}
The imaginary part deduced from (\ref{a44}) provides also an extremely
compelling support for GICPV as 
\begin{equation}
\mathop{\rm Im}\varepsilon _{GICPV}=\left( \frac{2\delta m_{K}c^{2}}{\hbar
\Gamma _{S}}\right) ^{2}\frac{2m_{K}g\hbar }{\delta m_{K}^{2}c^{3}}%
=1.57\times 10^{-3}\text{,}
\end{equation}
is in excellent agreement with the value reported in {\it PDG2024}:

\begin{equation}
\mathop{\rm Im}\varepsilon _{PDG2024}=\left( 1.57\pm 0.02\right) \times
10^{-3}\text{.}
\end{equation}
It is very important to note that the fundamental parameter describing CPV
is associated with the unitary evolution perturbed by Earth's gravity
leading to the mixing (\ref{Z799}, \ref{Z999}). The measurement of the
complex CPV parameter $\varepsilon _{PDG2024}$ is in agreement with $%
\varepsilon _{GICPV}$ as a result of a {\it dissipative dressing} of the
fundamental gravity-induced mixing (\ref{Z799}, \ref{Z999}), {\it %
dissipative dressing} resulting from the finite lifetime of both the $K_{L}$
and the $K_{S}$ mesons. One consequence of this {\it dissipative dressing}
is to consider a violation of the time reversal symmetry T resulting from a
confusion with an irreversible evolution. This {\it dissipative dressing} is
not {\it stricto sensu} a CPV effects but is inherent to experiments with
unstable particles, and is thus observable. 

The results of this section are summarized in the following table 
\begin{equation}
\begin{tabular}{ccc}
\hline\hline
{\it \ Observables} & {\it GICPV} & {\it PDG2024} \\ \hline\hline
$\mathop{\rm Re}\varepsilon \times 10^{3}$ & $1.66$ & $1.66\pm 0.02$ \\ 
\hline
$\mathop{\rm Im}\varepsilon \times 10^{3}$ & $1.57$ & $1.57\pm 0.02$ \\ 
\hline\hline
\end{tabular}
\end{equation}

It can be stated, with a high degree of plausibility, that the origin of
type ({\it i}) CPV effects observed in  kaons experiments is Earth's gravity.

\section{Type ({\it ii}) $K^{0}/\overline{K}^{0}$ experiments: $\varepsilon
^{\prime }/\varepsilon $}

The findings of the previous sections can be summarized as follows. ({\it i}%
) Earth's gravity induces a mixing of the (unobservables) stable ($\Gamma
_{1}=\Gamma _{2}=0$) CP eigenstates such that these states are given by $%
\left| K_{2/1}^{\oplus }\right\rangle $ (\ref{Z999}, \ref{Z799}) with (\ref
{ZD35}). ({\it ii}) When the finite lifetime of $K_{S}$ is taken into
account ($\Gamma _{1}\neq 0$, $\Gamma _{2}=0$), the mass eigenstate $\left|
K_{S/L}^{\oplus }\right\rangle $ are given by (\ref{j1}) with (\ref{j3}, \ref
{eps}) and look like CP and T violating states preserving CPT. ({\it iii})
When the finite lifetimes of both $K_{S}$ and $K_{L}$ are taken into account
($\Gamma _{1}\neq 0$, $\Gamma _{2}\neq 0$), the observed mass eigenstates $%
\left| K_{S/L}\right\rangle $ are obtained in remarkable agreement with the
observed eigenstates reported in {\it PDG2024}. The eigenstates, $\left|
K_{S/L}^{\oplus }\right\rangle $ and $\left| K_{S/L}\right\rangle $ display
what can be called a {\it dissipative dressing} of the fundamental
gravity-induced stables states $\left| K_{2/1}^{\oplus }\right\rangle $. On
the CP basis this set of mass eigenstates are: 
\begin{eqnarray}
\left| K_{2/1}^{\oplus }\right\rangle  &=&\left| K_{2/1}\right\rangle \pm
j\kappa \left| K_{1/2}\right\rangle \text{,}  \label{fund} \\
\left| K_{S/L}^{\oplus }\right\rangle  &=&\left| K_{1/2}\right\rangle
+\epsilon \left| K_{2/1}\right\rangle \text{,} \\
\left| K_{S/L}\right\rangle  &=&\left| K_{1/2}\right\rangle +\varepsilon
_{GICPV}\left| K_{2/1}\right\rangle \text{,}  \label{fund2}
\end{eqnarray}
where $\varepsilon _{GICPV}=\varepsilon _{PDG2024}=\varepsilon $. 

In Appendix A the issue of the {\it phase-convention-independent } (PCI)
definitions of the experimental CPV parameters is reviewed. In particular,
the usual (non-PCI) definition of the indirect violation parameter $%
\varepsilon ^{\prime }$ is 
\begin{equation}
\eta _{00}=\frac{\left\langle \pi ^{0}\pi ^{0}\right| {\cal T}\left|
K_{L}\right\rangle }{\left\langle \pi ^{0}\pi ^{0}\right| {\cal T}\left|
K_{S}\right\rangle }\equiv \varepsilon \left( 1-2\frac{\varepsilon ^{\prime }%
}{\varepsilon }\right) \text{.}  \label{zert}
\end{equation}
The isospin decomposition of $\left\langle \pi \pi \right| $ zero angular
momentum states \ (involving isospin $I=0$ and $I=$ $2$ states, without
contribution from $I=1$) and the strong ordering between the $I=2$ and $I=0$
amplitudes ($\left\langle I=2\right| {\cal T}\left| K_{S}\right\rangle $ $%
\ll $ $\left\langle I=0\right| {\cal T}\left| K_{S}\right\rangle $) allow to
relate $\eta _{00}$, $\eta _{+-}$ \ $=$ $\left\langle \pi ^{+}\pi
^{-}\right| {\cal T}\left| K_{L}\right\rangle $ $/\left\langle \pi ^{+}\pi
^{-}\right| {\cal T}\left| K_{S}\right\rangle $ (see Appendix A) and $%
\varepsilon $ (defined as $\varepsilon $ $=$ $\left\langle I=0\right| {\cal T%
}\left| K_{L}\right\rangle /$ $\left\langle I=0\right| {\cal T}\left|
K_{S}\right\rangle $) \cite{5,6,7}: 
\begin{equation}
3\frac{\left\langle I=0\right| {\cal T}\left| K_{L}\right\rangle }{%
\left\langle I=0\right| {\cal T}\left| K_{S}\right\rangle }=2\frac{%
\left\langle \pi ^{+}\pi ^{-}\right| {\cal T}\left| K_{L}\right\rangle }{%
\left\langle \pi ^{+}\pi ^{-}\right| {\cal T}\left| K_{S}\right\rangle }+%
\frac{\left\langle \pi ^{0}\pi ^{0}\right| {\cal T}\left| K_{L}\right\rangle 
}{\left\langle \pi ^{0}\pi ^{0}\right| {\cal T}\left| K_{S}\right\rangle }%
\text{.}  \label{11}
\end{equation}
It is important to note that the relation (\ref{11}) is PCI, although each
term is not PCI. This relation (\ref{11}) leads to the two conventional
definitions of $\varepsilon ^{\prime }$ as ({\it i}) $\eta _{00}=\varepsilon
-2\varepsilon ^{\prime }$ or as ({\it ii}) $\eta _{+-}=\varepsilon
+\varepsilon ^{\prime }$. So, in principle, the measurement of a single $%
\eta _{f}$ provides $\varepsilon ^{\prime }$ by comparison with the value of 
$\varepsilon $ obtained from the $K_{L}$ semi-leptonic charge asymmetry. But 
$\varepsilon ^{\prime }$ turns out to be very small and this experimental
approach cannot reach the required precision. This experimental obstacle
imply a measurement of both $\eta _{+-}$ and $\eta _{00}$. From the
theoretical point of view, the analysis can be restricted to $\eta _{00}$ to
predict $\varepsilon ^{\prime }$ and then $\eta _{+-}$ is deduced from the
PCI isospin relation (\ref{11}).

To interpret the measurements of the direct violation parameter $\varepsilon
^{\prime }$ \cite{17,18,19}, the arbitrary conventional phases inherent to
quantum theoretical models are to be eliminated to define PCI observables.
The usual definition of $\eta _{00}$ given by (\ref{zert}) is invariant
under rephasing of the pions state $\left\langle \pi ^{0}\pi ^{0}\right| $,
but not with respect to a rephasing of the kaons mass eigenstates $\left|
K_{L/S}\right\rangle $. A PCI quantity is defined through the multiplication
of $\eta _{00}$ with the {\it rephasing factor} (RF) 
\begin{equation}
\varphi _{00}=\frac{\left\langle K^{0}\right. \left| K_{S}\right\rangle }{%
\left\langle K^{0}\right. \left| K_{L}\right\rangle }=\frac{\left\langle
K_{L}\right. \left| K^{0}\right\rangle }{\left\langle K_{S}\right. \left|
K^{0}\right\rangle }\text{.}  \label{rf}
\end{equation}
For $\left| K_{S/L}\right\rangle $, it can be checked that $\varphi _{00}=1$
and the usual definition of $\varepsilon ^{\prime }$ remains unchanged but
is now PCI when written as 
\begin{equation}
\eta _{00}\times \varphi _{00}=\eta _{00}\equiv \varepsilon -2\varepsilon
^{\prime }\text{.}  \label{defep}
\end{equation}
But at the fundamental level, in order to avoid any dissipative phase, the $%
\left| K_{1/2}^{\oplus }\right\rangle $ states (\ref{fund}) must be used to
define the RF $\varphi _{00}$, 
\begin{equation}
\left. \varphi _{00}^{\oplus }\right| _{\Gamma _{S}=0,\Gamma _{L}=0}=\frac{%
\left\langle K^{0}\right. \left| K_{1}^{\oplus }\right\rangle }{\left\langle
K^{0}\right. \left| K_{2}^{\oplus }\right\rangle }=1-\left\langle
K_{1}^{\oplus }\right. \left| K_{2}^{\oplus }\right\rangle \text{,}
\end{equation}
where O$\left[ 10^{-6}\right] $ and higher orders terms are neglected. The
interaction between a $\pi ^{0}$ state and a neutral kaon state, $K^{0}$ or $%
\overline{K}^{0}$, can not differentiate the $K^{0}$ from the $\overline{K}%
^{0}$ (a final state phase can be absorbed by a proper phase convention
between $K^{0}$ and $\overline{K}^{0}$). Thus the amplitude of $%
K^{0}\rightarrow \pi ^{0}\pi ^{0}$ can be taken to be equal to the amplitude
of $\ \overline{K}^{0}\rightarrow \pi ^{0}\pi ^{0}$. \ This assumption is
coherent with the previous section where it is demonstrated that, what is
attributed to a weak interaction CPV at the fundamental level (KM-CPV) is in
fact just a low energy gravity-induced mixing. Thus, within the framework of
GICPV, CP commute with the weak interaction Hamiltonian and the particle and
antiparticle decay amplitudes are equal. The $\eta _{00}^{\oplus }$
parameter describes pion decays, so the $K_{S/L}$ fully dressed states given
by (\ref{fund2}) is to be used in the expression 
\begin{equation}
\eta _{00}^{\oplus }=\frac{\left\langle \pi ^{0}\pi ^{0}\right| {\cal T}%
\left| K_{L}\right\rangle }{\left\langle \pi ^{0}\pi ^{0}\right| {\cal T}%
\left| K_{S}\right\rangle }=\frac{\left\langle K_{S}\right. \left|
K_{L}\right\rangle }{2}=\frac{\left\langle K_{S}^{\oplus }\right. \left|
K_{L}^{\oplus }\right\rangle }{2}\text{.}
\end{equation}
Thus, within the framework of GICPV the measurement of the PCI amplitude
ratio $\eta _{00}$ (\ref{defep}) gives in fact $\eta _{00}^{\oplus }\varphi
_{00}^{\oplus }$ where 
\begin{equation}
\eta _{00}^{\oplus }\left. \varphi _{00}^{\oplus }\right| _{\Gamma
_{S}=0,\Gamma _{L}=0}=\frac{\left\langle K_{S}\right. \left|
K_{L}\right\rangle }{2}\left[ 1-2\frac{\left\langle K_{1}^{\oplus }\right.
\left| K_{2}^{\oplus }\right\rangle }{2}\right] \text{.}  \label{a55}
\end{equation}
We have seen in the previous section that a measurement of $\left\langle
K_{1}^{\oplus }\right. \left| K_{2}^{\oplus }\right\rangle $ gives in fact $%
\left\langle K_{S}\right. \left| K_{L}\right\rangle $ $=$ $\left\langle
K_{S}^{\oplus }\right. \left| K_{L}^{\oplus }\right\rangle $ because of the 
{\it dissipative dressing}. Decays provides a {\it dissipative dressing}
transforming the {\it unobservable} ideal eigenstates (\ref{fund}) into the 
{\it observed} eigenstates (\ref{fund2}). Thus, the measured PCI ratio is
given by 
\begin{equation}
\eta _{00}^{\oplus }\left. \varphi _{00}^{\oplus }\right| _{\Gamma _{S}\neq
0,\Gamma _{L}\neq 0}=\frac{\left\langle K_{S}\right. \left|
K_{L}\right\rangle }{2}\left[ 1-2\frac{\left\langle K_{S}\right. \left|
K_{L}\right\rangle }{2}\right] \text{.}  \label{a56}
\end{equation}
The comparison of (\ref{a56}) with the definition (\ref{defep}) gives 
\begin{equation}
\mathop{\rm Re}\left( \varepsilon ^{\prime }/\varepsilon \right)
_{GICPV}=\left( \frac{2\delta m_{K}c^{2}}{\hbar \Gamma _{S}}\right) \frac{%
2m_{K}g\hbar }{\delta m_{K}^{2}c^{3}}=1.66\times 10^{-3}\text{.}  \label{a57}
\end{equation}
This value is in agreement with the recent experimental value: 
\begin{equation}
\mathop{\rm Re}\left( \varepsilon ^{\prime }/\varepsilon \right) _{\text{%
{\it PDG2024}}}=\left( 1.66\pm 0.23\right) \times 10^{-3}\text{.}  \label{b8}
\end{equation}
The puzzling experimental relation between direct and indirect CPV, 
\begin{equation}
\mathop{\rm Re}\left( \varepsilon ^{\prime }/\varepsilon \right) _{\text{%
{\it PDG2024}}}\simeq \mathop{\rm Re}\left( \varepsilon \right) _{\text{{\it %
PDG2024}}}\text{,}
\end{equation}
was considered, up to now, as a {\it numerical coincidence.} Within the
framework of GICPV it finds here a simple explanation in term of a PCI
amplitudes and RF. Recent experimental values \cite{7} of $\mathop{\rm Im}%
\left( \varepsilon ^{\prime }/\varepsilon \right) _{\text{{\it PDG2024}}%
}=\left( -2.\pm 5.\right) {{}^{\circ }\times 10}^{-3}$ display a very large
incertitude and these measurements are considered compatible with $%
\mathop{\rm Im}\left( \varepsilon ^{\prime }/\varepsilon \right) $ $=$ $0$.

We have used a PCI definition to analyze the measurement of $\eta _{00}$ and
the relation 
\begin{equation}
2\text{ }\frac{\left\langle \pi ^{+}\pi ^{-}\right| {\cal T}\left|
K_{L}\right\rangle }{\left\langle \pi ^{+}\pi ^{-}\right| {\cal T}\left|
K_{S}\right\rangle }+\frac{\left\langle \pi ^{0}\pi ^{0}\right| {\cal T}%
\left| K_{L}\right\rangle }{\left\langle \pi ^{0}\pi ^{0}\right| {\cal T}%
\left| K_{S}\right\rangle }=3\text{ }\varepsilon \text{,}  \label{iso}
\end{equation}
based on the isospin decomposition (\ref{11}), is also a PCI relation, so an
additional PCI definition of $\eta _{+-}$ is to be avoided if this relation
is used. The PCI relation (\ref{iso}) and the PCI value $\eta _{00}=\eta
_{00}^{\oplus }\varphi _{00}^{\oplus }$ (\ref{a57}) imply $\eta
_{+-}=\varepsilon \left( 1+\varepsilon ^{\prime }/\varepsilon \right) $ in
agreement with the experiments. 

The results of this section are summarized in the following table 
\begin{equation}
\begin{tabular}{ccc}
\hline\hline
{\it \ Observables} & {\it GICPV} & {\it PDG2024} \\ \hline\hline
$\mathop{\rm Re}\varepsilon ^{\prime }/\varepsilon \times 10^{3}$ & $1.66$ & 
$1.66\pm 0.23$ \\ \hline
$\mathop{\rm Im}\varepsilon ^{\prime }/\varepsilon \times 10^{3}$ & $0$ & $%
\left( -2.\pm 5.\right) {{}^{\circ }}$ \\ \hline\hline
\end{tabular}
\end{equation}

It can be stated, with a high degree of plausibility, that the origin of
type ({\it ii}) CPV effects observed in  kaons experiments is Earth's gravity.

\section{Type ({\it iii}) $B^{0}/\overline{B}^{0}$ experiments: $S_{f}$, $
C_{f}$}

During the last decades of the twenty century, CPV evidences where
restricted to ({\it i}) $K$ mesons experiments and ({\it ii}) the baryons
asymmetry of the universe. In 2001 a clear identification of CPV with {\it %
B-factories\ }was reported \cite{20,21}. The lifetime of the CP eigenstate $%
B_{1}$ is equal to the lifetime of $B_{2}$. This ordering allows to use the
results given by (\ref{BBB1},\ref{BBB2}) 
\begin{equation}
\left| B_{H/L}^{\oplus }\right\rangle =\left| B_{2/1}\right\rangle \pm
j\kappa ^{^{\prime }}\exp \left( \pm j\zeta \right) \left|
B_{1/2}\right\rangle \text{.}  \label{bbb2}
\end{equation}
The mass and width ordering associated are given by : $\delta
m_{B}/m_{B}\sim 10^{-19}$, $\delta m_{B}c^{2}/\hbar \Gamma _{B}\sim 0.77$,
and the GICPV parameters are 
\begin{eqnarray}
\text{ }\zeta &=&\arctan \left( \hbar \Gamma _{B}/\delta m_{B}c^{2}\right) 
\text{,} \\
\text{ }\kappa ^{^{\prime }} &=&\left( 2m_{B}g\hbar /\delta
m_{B}^{2}c^{3}\right) /\sqrt{1+\left( \hbar \Gamma _{B}/\delta
m_{B}c^{2}\right) ^{2}}\text{,}
\end{eqnarray}
with $\kappa ^{^{\prime }}$ $\sim O\left[ 10^{-6}\right] $. Using the flavor
basis $\left| B^{0}\right\rangle $ and $\left| \overline{B}^{0}\right\rangle 
$, rather than the CP basis $\left| B_{1}\right\rangle $ and $\left|
B_{2}\right\rangle $, the mass eigenstates (\ref{bbb2}) are 
\begin{eqnarray}
\left| B_{L/H}^{\oplus }\right\rangle &=&\left[ 1\mp j\kappa ^{^{\prime
}}\exp \left( \mp j\zeta \right) \right] \left| B^{0}\right\rangle /\sqrt{2}
\label{a72} \\
&&\pm \left[ 1\pm j\kappa ^{^{\prime }}\exp \left( \mp j\zeta \right)
\right] \left| \overline{B}^{0}\right\rangle /\sqrt{2}\text{.}  \nonumber
\end{eqnarray}
This GICPV result requires two real numbers, $\zeta $ and $\kappa ^{\prime }$%
; although, to interpret type ({\it iii}) interference experiments, the
standard KM-CPV parametrization ((\ref{9}) in Appendix A), under the
assumption of CPT invariance, is based on a single angle $\beta $ and
display a different structure, 
\begin{equation}
\left| B_{L/H}\right\rangle =\exp +j\beta \left| B^{0}\right\rangle /\sqrt{2}%
\pm \exp -j\beta \left| \overline{B}^{0}\right\rangle /\sqrt{2}\text{,}
\label{a75a}
\end{equation}
Despite the difference between (\ref{a72}) and (\ref{a75a}), the
experimental results can be understood and explained within the framework of
GICPV. The situation here is similar to the one encountered in the previous
section devoted to the study of $\varepsilon ^{\prime }$: if CPT is assumed,
then the rephasing factors $\varphi =1$, and the interpretation of the
experimental measurements is based on the hypothesis of direct violation and
imply a CPV at the fundamental level. However, if Earth's gravity effects
are taken into account $\varphi \neq 1$ and the very same PCI measured
quantities agree with the experiments without any additional assumption at
the fundamental level.

The analysis below will use two different approaches to interpret the
measurement of $\ \beta $, each providing the same final result. The two
issues addressed below are: ({\it i}) the invariance under rephasing of the
flavor eigenstates to define a PCI observable, ({\it ii}) the invariance
under rephasing of the mass eigenstates to define a PCI observable.

In order to compare the usual eigenstates CPT parametrization based on the
angle $\beta $ (\ref{a75a}) with the gravity-induced mass eigenstates
involving ($\zeta $ , $\kappa ^{^{\prime }}$) (\ref{a72}), $\beta $ is to be
defined through a {\it gedanken} experiment providing $\exp j\beta $ as a
PCI expression. To do so, consider $\rho _{B}$ a simple, symmetric and
complete, combination of the four components of the mass eigenstates
providing a measure of $\beta $ 
\begin{equation}
\rho _{B}=\frac{\left\langle B^{0}\right. \left| B_{L}\right\rangle }{%
\left\langle \overline{B}^{0}\right. \left| B_{L}\right\rangle }\frac{%
\left\langle B^{0}\right. \left| B_{H}\right\rangle }{\left\langle \overline{%
B}^{0}\right. \left| B_{H}\right\rangle }=-\exp 4j\beta \text{.}  \label{a86}
\end{equation}
This definition of $\beta $ through $\rho _{B}$ takes into account all
flavor and mass eigenstates but suffers from a lack of phase compensation
with respect to the flavor eigenstates. All measured observables must be PCI
quantities. The RF needed to provide a PCI observable associated with $\rho
_{B}$ is $\varphi _{B}$: 
\begin{equation}
\varphi _{B}=\frac{\left\langle \overline{B}^{0}\right. \left|
B_{2}\right\rangle \left\langle B_{2}\right. \left| B_{H}\right\rangle }{%
\left\langle B^{0}\right. \left| B_{1}\right\rangle \left\langle
B_{1}\right. \left| B_{L}\right\rangle }\frac{\left\langle \overline{B}%
^{0}\right. \left| B_{2}\right\rangle \left\langle B_{2}\right. \left|
B_{L}\right\rangle }{\left\langle B^{0}\right. \left| B_{1}\right\rangle
\left\langle B_{1}\right. \left| B_{H}\right\rangle }\text{,}  \label{a87}
\end{equation}
where $B_{1/2}$ projectors have been used because $B_{1/2}$ are CP
eigenstates. It can be checked that the product $\rho _{B}\varphi _{B}$ is
PCI and thus can be measured in a {\it gedanken} experiment which does not
need to be described here.

If the usual CPT parametrization (\ref{a75a}) is used , this RF $\varphi _{B}
$ changes nothing because it is equal to one 
\begin{eqnarray}
\varphi _{B} &=&1\text{,}  \label{a89} \\
\rho _{B} &=&-\exp 4j\beta \text{.}
\end{eqnarray}

If CPV is gravity induced, $\left| B_{H}^{\oplus }\right\rangle $ and $%
\left| B_{L}^{\oplus }\right\rangle $ from (\ref{a72}) replace $\left|
B_{H/L}\right\rangle $ in (\ref{a86}, \ref{a87}) and the very same PCI
observable $\rho _{B}\varphi _{B}$ is the product of the following factors 
\begin{eqnarray}
\varphi _{B}^{\oplus } &=&\exp \left( -2j\zeta +j\pi \right) \text{,}
\label{a91} \\
\rho _{B}^{\oplus } &=&-1+\text{O}\left[ 10^{-12}\right] \text{.}
\end{eqnarray}
The measurement of the PCI observable $\rho _{B}\varphi _{B}$ on Earth gives
the result $\rho _{B}^{\oplus }\rho _{B}^{\oplus }$ within the GICPV
framework and $\varphi _{B}^{\oplus }\rho _{B}^{\oplus }$ $=$ $\varphi
_{B}\rho _{B}$ imply 
\begin{equation}
\sin 2\beta =\cos \zeta =0.61\text{.}
\end{equation}
The conclusion of this $\beta $ {\it gedanken experiment} measurement, with
two frameworks of interpretation, is $\sin 2\beta =0.61$, in rather good
agreement with the experiments. The measurement of the angle $\beta $ is one
of the major subjects at the forefront of the studies related to the physics
of the SM \cite{22,23,24}. \ 

Rather than a {\it gedanken experiment}, consider now a second approach and
the decay to one final appropriate CP eigenstate $\left| f\right\rangle $
such that (see Appendix A) 
\begin{equation}
\left. \lambda _{f}\right| _{CPT}=\frac{\left\langle \overline{B}^{0}\right.
\left| B_{L}\right\rangle }{\left\langle B^{0}\right. \left|
B_{L}\right\rangle }\frac{\left\langle f\right| {\cal T}\left| \overline{B}%
^{0}\right\rangle }{\left\langle f\right| {\cal T}\left| B^{0}\right\rangle }%
\text{.}
\end{equation}
It is more appropriate to consider a \ parameter $\widetilde{\lambda }_{f}$
constructed with the amplitude ratio $\left\langle \overline{B}^{0}\right.
\left| B_{L}\right\rangle /\left\langle B^{0}\right. \left|
B_{H}\right\rangle $, 
\begin{equation}
\widetilde{\lambda }_{f}=\frac{\left\langle \overline{B}^{0}\right. \left|
B_{L}\right\rangle }{\left\langle B^{0}\right. \left| B_{H}\right\rangle }%
\frac{\left\langle f\right| {\cal T}\left| \overline{B}^{0}\right\rangle }{%
\left\langle f\right| {\cal T}\left| B^{0}\right\rangle }=\left. \lambda
_{f}\right| _{CPT}\text{,}  \label{a79}
\end{equation}
which is better suited to characterize the dynamics as it takes into account
all the eigenstates: the two flavor eigenstates and the two mass
eigenstates, with the same lifetime, involved in experiments. This $%
\widetilde{\lambda }_{f}$ parameter is not PCI with respect to the mass
eigenstates. The symmetric RF $\varphi _{B}^{\prime }$ is

\begin{equation}
\varphi _{B}^{\prime }=\sqrt{\frac{\left\langle B_{1}\right. \left|
B_{H}\right\rangle }{\left\langle B_{1}\right. \left| B_{L}\right\rangle }%
\frac{\left\langle B_{2}\right. \left| B_{H}\right\rangle }{\left\langle
B_{2}\right. \left| B_{L}\right\rangle }}=1\text{,}  \label{a80}
\end{equation}
where $B_{1/2}$ are CP eigenstates like $f$. \ The experimental amplitude
ratio is given by the PCI product $\widetilde{\lambda }_{f}\varphi _{B}$

\begin{eqnarray}
\widetilde{\lambda }_{f}\varphi _{B}^{\prime } &=&\frac{\left\langle 
\overline{B}^{0}\right. \left| B_{L}\right\rangle }{\left\langle
B^{0}\right. \left| B_{H}\right\rangle }\frac{\left\langle f\right| {\cal T}%
\left| \overline{B}^{0}\right\rangle }{\left\langle f\right| {\cal T}\left|
B^{0}\right\rangle }\varphi _{B}^{\prime }  \label{a81} \\
&=&\frac{\left\langle f\right| {\cal T}\left| \overline{B}^{0}\right\rangle 
}{\left\langle f\right| {\cal T}\left| B^{0}\right\rangle }\exp -2j\beta
=\left. \lambda _{f}\right| _{CPT}\text{.}  \nonumber
\end{eqnarray}
When the same RF $\varphi _{B}^{\prime \oplus }$ is calculated within the
framework of GICPV with this gives

\begin{equation}
\varphi _{B}^{\prime \oplus }=\sqrt{\frac{\left\langle B_{1}\right. \left|
B_{H}^{\oplus }\right\rangle }{\left\langle B_{1}\right. \left|
B_{L}^{\oplus }\right\rangle }\frac{\left\langle B_{2}\right. \left|
B_{H}^{\oplus }\right\rangle }{\left\langle B_{2}\right. \left|
B_{L}^{\oplus }\right\rangle }}=\exp -j\left( \frac{\pi }{2}-\zeta \right) 
\text{.}
\end{equation}
The PCI product $\widetilde{\lambda }_{f}^{\oplus }\varphi _{B}^{\prime
\oplus }$ is given by

\begin{eqnarray}
\widetilde{\lambda }_{f}^{\oplus }\varphi _{B}^{\prime \oplus } &=&\frac{%
\left\langle \overline{B}^{0}\right. \left| B_{L}^{\oplus }\right\rangle }{%
\left\langle B^{0}\right. \left| B_{H}^{\oplus }\right\rangle }\frac{%
\left\langle f\right| {\cal T}\left| \overline{B}^{0}\right\rangle }{%
\left\langle f\right| {\cal T}\left| B^{0}\right\rangle }\varphi
_{B}^{\prime \oplus }  \label{a84} \\
&=&\frac{\left\langle f\right| {\cal T}\left| \overline{B}^{0}\right\rangle 
}{\left\langle f\right| {\cal T}\left| B^{0}\right\rangle }\exp -j\left( 
\frac{\pi }{2}-\zeta \right) \text{.}  \nonumber
\end{eqnarray}
where O$\left[ 10^{-6}\right] $ terms are neglected in front of one. To
compare the interpretations based on the usual SM-CKM eigenstates $\left|
B_{L/H}\right\rangle $ (\ref{a75a}) with the gravity induced mass
eigenstates $\left| B_{L/H}^{\oplus }\right\rangle $ (\ref{a72}), $\beta $
is to be defined such that $2\beta =$ $\arctan \left( 0.77\right) $. The
ratio $\left\langle f\right| {\cal T}\left| \overline{B}^{0}\right\rangle
/\left\langle f\right| {\cal T}\left| B^{0}\right\rangle $ is assumed real
and equal to one (with GICPV CP commute with the weak interaction
Hamiltonian), as a result the experiments dedicated to the measure of $%
\lambda _{f}$ give 
\begin{equation}
S_{f}=\frac{2\mathop{\rm Im}\widetilde{\lambda }_{f}^{\oplus }\varphi
_{B}^{\prime \oplus }}{1+\left| \widetilde{\lambda }_{f}^{\oplus }\varphi
_{B}^{\prime \oplus }\right| ^{2}}=\cos \zeta \text{, }C_{f}=\frac{1-\left| 
\widetilde{\lambda }_{f}^{\oplus }\varphi _{B}^{\prime \oplus }\right| ^{2}}{%
1+\left| \widetilde{\lambda }_{f}^{\oplus }\varphi _{B}^{\prime \oplus
}\right| ^{2}}=0\text{,}  \label{sf2}
\end{equation}
(see Appendix A). The decays $B_{0}\rightarrow \phi K_{S}$ and $%
B_{0}\rightarrow J/\psi K^{0}$ have been studied in depth and, according to
the data reported in \cite{7}, the present status of the associated $S_{f}$
and $C_{f}$ values is $S_{f}=0.58\pm 0.12\text{, }S_{f}=0.701\pm 0.01$ and $%
C_{f}=\left( -9.\pm 12.\right) \times 10^{-2}$, $C_{f}$ $=\left( -0.8\pm
1.7\right) \times 10^{-2}$. Other neutral final states, such as $\psi K^{*0}$
and $K^{0}\pi ^{0}$, give $S_{f}$ $=0.60\pm 0.24\pm 0.08$ and $S_{f}$ $=$ $%
0.64\pm 0.13$, also in good agreement with the GICPV result if $\left\langle
f\right| {\cal T}\left| \overline{B}^{0}\right\rangle $ $\approx $ $%
\left\langle f\right| {\cal T}\left| B^{0}\right\rangle $. 

These results are summarized in the table 
\begin{equation}
\begin{tabular}{ccc}
\hline\hline
{\it \ Observables} & {\it GICPV} & {\it PDG2024} \\ \hline\hline
$\left. S_{f}\right| _{\phi K_{S}^{0}}$ & $0.61$ & $0.58\pm 0.12$ \\ \hline
$\left. S_{f}\right| _{J/\psi K^{0}}$ & $0.61$ & $0.70\pm 0.01$ \\ 
\hline\hline
\end{tabular}
\end{equation}
The difficulty to evaluate $\arg \left\langle f\right| {\cal T}\left| 
\overline{B}^{0}\right\rangle /\left\langle f\right| {\cal T}\left|
B^{0}\right\rangle $ is one source of the dispersion, note also that the
sign of $\left\langle f\right| \widehat{CP}\left| f\right\rangle $ is to be
considered and the fact that a single parameter CPT modeling (\ref{a75a}) is
assumed rather than a two parameters modeling (\ref{a72}) is probably also a
source of dispersion. A clear understanding of the $\sin 2\beta $
distribution around $0.6-0.7$ requires to drop (\ref{a75a}) and to adopt the
GICPV mass eigenstates (\ref{a72}).

\section{Conclusion and perspectives}

The interplay between ({\it i}) flavor oscillations ($\delta m$), ({\it ii})
zitterbewegung fluctuations ($\kappa $) and ({\it iii}) irreversible decays (%
$\gamma $) requires a careful analysis of the various regimes. The ordering
and coupling associated with the regimes studied in the previous sections
are summarized below. 
\begin{equation}
\begin{tabular}{ccc}
\hline\hline
{\it Ordering} & {\it Coupling} & {\it Symmetry} \\ \hline\hline
$\kappa =0$, $\widehat{\gamma }=\widehat{0}$ & $M_{2}$ \ \ \ {\it \ }$M_{1}$
& CP, T \\ \hline
$\kappa =0$, $\widehat{\gamma }\neq \widehat{0}$ & $f^{\prime }\stackrel{%
\Gamma _{2}}{\leftarrow }M_{2}$ \ \ {\it \ }$M_{1}\stackrel{\Gamma _{1}}{%
\rightarrow }f$ & CP, T \\ \hline
$\kappa \neq 0$, $\widehat{\gamma }=\widehat{0}$ & $M_{2}\stackrel{\kappa }{%
\leftrightarrows }M_{1}$ & T \\ \hline
{\it Revers. trans.} & $f^{\prime }\stackrel{w}{\leftrightarrows }M_{2}%
\stackrel{\kappa }{\leftrightarrows }M_{1}\stackrel{w}{\leftrightarrows }f$
& T \\ \hline
$\Gamma _{1}\gg \Gamma _{2}\approx 0$ & $M_{2}\stackrel{\kappa }{\rightarrow 
}M_{1}\stackrel{\Gamma _{1}}{\rightrightarrows }f$ & {\it CPT*} \\ \hline
$\Gamma _{1}\gg \Gamma _{2}\neq 0$ & $f^{\prime }\stackrel{\Gamma _{2}}{%
\leftarrow }M_{2}\stackrel{\kappa }{\leftrightarrows }M_{1}\stackrel{\Gamma
_{1}}{\rightrightarrows }f$ & {\it CPT*} \\ \hline
$\Gamma _{1}=\Gamma _{2}=\Gamma $ & $f^{\prime }\stackrel{\Gamma }{%
\leftleftarrows }M_{2}\stackrel{\kappa }{\leftrightarrows }M_{1}\stackrel{%
\Gamma }{\rightrightarrows }f$ & {\it CPT*} \\ \hline\hline
\end{tabular}
\end{equation}
The first and second lines are associated with a simple LOY dynamics where
weak interaction is responsible of the oscillations $M^{0}\stackrel{\delta m%
}{\leftrightarrows }\overline{M}^{0}$ such that the mass eigenstates $M_{2}$-%
$M_{1}$ are decoupled. The third line is associated with the unitary GICPV
model of section six ($\left| M_{2/1}^{\oplus }\right\rangle $). The fourth
line describes an hypothetical model with reversible transitions to the {\it %
final} states. This model was not considered in this paper. Section seven
was devoted to the study of the ordering of line five ($\left|
M_{H/L}^{\oplus }\right\rangle $) and seven. The case of line six was
studied in section eight for kaons ($\left| K_{S/L}\right\rangle $). The
star in the last column indicates that the parametrization of the mass
eigenstates might look like CPT invariant, but, in fact, as Lorentz
invariance is broken by the metric $g_{\mu \nu }\neq $ diag [$-1,1,$ $1,1$],
this apparent CPT invariance no longer holds true. 

To gain a deeper
understanding that the combination of gravity and zitterbewegung is reponsible
for the CPV mixing, the following simple heuristic model can be developed. 

Consider the neutral kaons CPV free dynamics described by $\widehat{H}_{M}$
and a perturbation $\widehat{V}$ $\ll \widehat{H}_{M}$\ that might slightly
mix the CP eigenstates $\left| M_{1/2}\right\rangle $. The perturbed
eigenstates, $\left| M_{1/2}^{V}\right\rangle $, are solutions of $\left( 
\widehat{H}_{M}+\widehat{V}\right) \cdot \left| M_{1/2}^{V}\right\rangle $ = 
$E_{1/2}^{V}\left| M_{1/2}^{V}\right\rangle $. These mixed eigenstates are
given by first order perturbation theory: 
\begin{equation}
\left| M_{1/2}^{V}\right\rangle =\left| M_{1/2}\right\rangle +\frac{%
\left\langle M_{2/1}\right| \widehat{V}\left| M_{1/2}\right\rangle }{%
E_{1}-E_{2}}\left| M_{2/1}\right\rangle \text{.}  \label{mm}
\end{equation}
With a {\it \ }small zitterbewegung vertical fluctuation $\widehat{x}$, such
that $\widehat{V}\sim mg\widehat{x}$ , the expected, if any, matrix element $%
\left\langle M_{2/1}\right| \widehat{V}\left| M_{1/2}\right\rangle $ $\sim
mg\lambda _{C}=\hbar g/c\sim 10^{-23}$ eV is rather small in front of $%
E_{1}-E_{2}\sim \delta mc^{2}\sim 10^{-6}$ eV. The CPV mixing parameter $%
\left\langle M_{2/1}\right| \widehat{V}\left| M_{1/2}\right\rangle $ $%
/\left( E_{1}-E_{2}\right) $ $\ \sim $ $\hbar g/\delta mc^{3}$ $\sim 10^{-17}
$ \ is far smaller than the experimental one. This small parameter $\hbar
g/\delta mc^{3}$ was identified in Ref. \cite{25} by Fischbach who reaches
the same negative conclusion. Fischbach also noted that the smallness of the
parameter $\hbar g/\delta mc^{3}$ $\sim 10^{-17}$ can be compensated by the
large ratio $m/\delta m\sim 10^{-14}$ to provide the right order of
magnitude $\varepsilon \sim 10^{-3}$ for the $K^{0}/\overline{K}^{0}$ case.
This puzzling remark was later revaluated by Chardin to explore the
hypothesis of antigravity as the origin of CPV \cite{26,27}.

The origin of CPV is to be traced to the internal structure and internal
dynamics of a meson and requires a detailed analysis revealing a subtle
interplay between quarks zitterbewegung ($+c\rightleftharpoons $ $-c$) and
flavor oscillations ($M^{0}\rightleftharpoons \overline{M}^{0}$). Neglecting
decay, the balance between ({\it i}) flavor oscillations ($F$) and ({\it ii}%
) zitterbewegung fluctuations/oscillations ($Z$) in the Earth field, is
described by equation (\ref{H76}) 
\begin{equation}
j\hbar \frac{d\left| M\right\rangle }{d\tau }=-\left. c^{2}\frac{\widehat{%
\delta m}}{2}\cdot \left| M\right\rangle \right| _{F}+\left. mg\widehat{x}%
\cdot \left| M\right\rangle \right| _{Z}\text{.}  \label{dyna}
\end{equation}
The matrix elements of $\ \widehat{x}$ are unknown but those of the Dirac
velocity$\ d\widehat{x}/dt$ \ are given by (\ref{ff}): 
\begin{equation}
d\widehat{x}/d\tau \cdot \left| \overline{M}^{0}\right\rangle =c\left|
M^{0}\right\rangle \text{ , }d\widehat{x}/d\tau \cdot \left|
M^{0}\right\rangle =-c\left| \overline{M}^{0}\right\rangle \text{,}
\label{cos1}
\end{equation}
where the $\pm $ signs reflects the Feynman-Stueckelberg interpretation of
an antiparticle as a particle propagating backward in time. Flavor
oscillations $M^{0}\rightleftharpoons \overline{M}^{0}$ take place at a
frequency\ $\delta mc^{2}/2\hbar $. A meson initially in a flavor eigenstate 
$\left| M\left( \tau =0\right) \right\rangle =\left| M^{0}\right\rangle $
oscillates as 
\begin{equation}
\left. \left| M\left( \tau \right) \right\rangle \right| _{F}=\cos \frac{%
\delta mc^{2}\tau }{2\hbar }\left| M^{0}\right\rangle +j\sin \frac{\delta
mc^{2}\tau }{2\hbar }\left| \overline{M}^{0}\right\rangle \text{.}
\label{sin}
\end{equation}
As $\widehat{\delta m}$ is dominant and drives the main dynamics in (\ref
{dyna}), the {\it zitterbewegung }Dirac velocity follows the changing flavor
according to the relations (\ref{cos1}) and (\ref{sin}), this gives 
\begin{equation}
\left\langle M\right| \left. \frac{d\widehat{x}}{d\tau }\right| _{Z}\left|
M\right\rangle =cj\sin \left( \delta mc^{2}\tau /\hbar \right) \text{.}
\label{sin2}
\end{equation}
A simple picture of this interplay between $Z$ oscillations and $F$
oscillations (\ref{sin2}) can be drawn as follows. The unobservable vertical
z{\it itterbewegung velocity} (\ref{sin2}) \ exchanges a small power $%
dE/d\tau $ with the vertical force $mg$, 
\begin{equation}
\frac{dE}{d\tau }=mg\left\langle M\right| \left. \frac{d\widehat{x}}{d\tau }%
\right| _{Z}\left| M\right\rangle =jmgc\sin \left( \delta mc^{2}\tau /\hbar
\right) \text{.}
\end{equation}
The energy exchanged through this $F/Z$ interplay is $\delta E_{F/Z}$ $\sim $
$\left( mg\hbar /\delta mc\right) $ $\cos \left( \delta mc^{2}\tau /\hbar
\right) $. The time average stored energy $\left\langle \delta
E_{F/Z}\right\rangle =0$ but the quadratic average $\left\langle \delta
E_{F/Z}^{2}\right\rangle \neq 0$.

In the theory of coupled oscillators it is quite usual to observe secular
effects such that $\left\langle \delta E^{2}\right\rangle \neq 0$ despite
the fact that $\left\langle \delta E\right\rangle =0$. One typical example
is the adiabatic secular shift in Hamiltonian dynamics \cite{28} and an
other example is the ponderomotive effects in electrodynamics \cite{29}. A
third example, in QED, is the Darwin term in the hydrogen fine structure
Hamiltonian analyzed as a correction due to the electron {\it zitterbewegung}
fluctuation $\left\langle \delta x^{2}\right\rangle $ \cite{8,9}.

The secular energy shift $\sqrt{\left\langle \delta E_{F/Z}^{2}\right\rangle 
}$ associated with the interplay between $F$ and $Z$ oscillations $\sqrt{%
\left\langle \delta E_{F/Z}^{2}\right\rangle }\sim mg\hbar /\delta mc$, can
be viewed as a stored reactive energy. This indicates that it might be
possible, with an appropriate perturbative expansion, to identify a 
coupling $\widehat{V}_{F/Z}$ between CP eigenstates with a matrix element
such that $\left\langle M_{2/1}\right| \widehat{V}_{F/Z}\left|
M_{1/2}\right\rangle $ $\sim $ $\sqrt{\left\langle \delta
E_{F/Z}^{2}\right\rangle }$. With such a coupling the perturbation of the
eigenstates $\left| M_{1/2}\right\rangle $ is 
\begin{equation}
\frac{\left\langle M_{2/1}\right| \widehat{V}_{F/Z}\left|
M_{1/2}\right\rangle }{E_{1}-E_{2}}\approx \frac{\sqrt{\left\langle \delta
E_{F/Z}^{2}\right\rangle }}{\delta mc^{2}}\sim \varepsilon \text{.}
\end{equation}
This mixing amplitude $mg\hbar /\delta m^{2}c^{3}$ $\sim $ $O\left[
\varepsilon \right] $ displays the right order of magnitude for the well
documented $K^{0}/\overline{K}^{0}$ case. \ The identification of $\widehat{V%
}_{F/Z}$ is presented in section six. Beside CPV in the mixing, types ({\it %
ii}) and ({\it iii}) CPV have been shown to result from overlooked PCI
requirements resulting from this mixing. Note that within the framework of
GICPV all the CPV effects are associated with this type of gravity induced
mixing, mixing providing a regeneration when the decays are different.

Two interesting perspectives can be considered to extend and apply the new
results obtained in this paper.

First, a better modeling of interference type ({\it iii}) experiments is
needed to evaluate the impact of the non exponential regime of decay. 

The WW approximation describing the decay of an unstable state is a well
established model. However, an accurate analysis of unstable quantum states,
without the WW approximation, provides a universal three regimes picture 
\cite{30,31}. A generic decay displays three stages.

First an initial stage near $\tau =0$ where the decay is non exponential.
Then, a long exponential decay followed by a final non exponential regime
when $\tau \rightarrow \infty $. The transition between these last two
regimes displays oscillations that might be important in the context of type
({\it iii}) interference experiments. That the decay of an unstable quantum
state is not exponential can be simply established as follows. Consider a
CPV free evolution starting from one of the CP eigenstates $\left| M\left(
\tau =0\right) \right\rangle $ $=$ $\left| M_{1/2}\right\rangle $. The
survival amplitude of $M_{1/2}$, $A$, is defined by the relation

\begin{equation}
\left| M\right\rangle =\exp -j\widehat{H}\tau \left| M_{1/2}\right\rangle
=A\left( \tau \right) \left| M_{1/2}\right\rangle +\Sigma _{f}w_{f}\left(
\tau \right) \left| f\right\rangle \text{,}  \label{gamol}
\end{equation}
where $\widehat{H}$ is the full (CPV free) Hermitian Hamiltonian, before its
reduction through the WW approximation. The decay states are orthogonal to
the initial state $\sum_{f}w\left( f,\tau \right) \left\langle
M_{1/2}\right. \left| f\right\rangle $ = $0$ and $\left\langle f^{\prime
}\right. \left| f\right\rangle =0$. The survival probability is given by $%
\left| A\left( \tau \right) \right| ^{2}$ and the survival amplitude
fulfills the relation

\begin{equation}
A\left( \tau +\tau ^{\prime }\right) =A\left( \tau \right) A\left( \tau
^{\prime }\right) +\Sigma _{f}w_{f}\left\langle M_{1/2}\right| \exp -j%
\widehat{H}\tau ^{\prime }\left| f\right\rangle \text{,}
\end{equation}
obtained by the application of $\exp -j\widehat{H}\tau ^{\prime }$ on both
side of (\ref{gamol}). The occurrence of the last term on the right hand
side of this relation imply that the functional form of $A\left( \tau
\right) $ is not of the exponential type $A$ = $\exp \left( -\Gamma
_{1/2}\tau /2\right) $. The departure from an exponential decay has little
consequences on the analysis of type ({\it i}) and ({\it ii}) CPV
experiments, but the case is different for type ({\it iii}) interferences.

Second, GICPV not only explain ({\it i}, {\it ii} and {\it iii}) CPV
effects, and predict the associated observables, but it also renews, in
depth, the baryons asymmetry cosmological issue. In any environment where a
flavored neutral mesons $\left| M\right\rangle $ (mass $m$, mass spliting $%
\delta m$ and Compton wavelength $\lambda _{C}$) experiences a gravity ${\bf %
g}$, i.e. in any curved space-time environment, a CP eigenstates mixing with
amplitude $\sim $ $j\left( m/\delta m\right) ^{2}g\lambda _{C}/c^{2}$ will
be observed. For a massive spherical
object, with radius $R_{0}$ and Schwarzschild radius $R_{S}$, the scaling of
the mixing amplitude is given by $\sim j\left( m/\delta m\right) ^{2}\left(
R_{S}/R_{0}\right) (\lambda _{C}/R_{0})$. 

The proportionality to $g$ or 
$R_{S}$ indicates that this new CPV mechanism allows to set up cosmological
early evolution models predicting the strong asymmetry between the abundance
of matter and the abundance anti-matter in our present universe.

Among the three conditions identified by Sakharov \cite{4}, CP has to be
broken and an evolution displaying a strong departure from thermal
equilibrium is to be considered. This last conditions can be relaxed and
equilibrium asymmetric baryogenesis can be observed if CPT is broken. GICPV
offers a promising framework to set up both thermal equilibrium and
non-equilibrium asymmetric baryogenesis models during the early stage of the
cosmic evolution.

Beside the problem of early baryogenesis, neutrinos oscillations near a
spherical massive objects might be revisited to explore the impact of the
interplay between gravity and weak mixing. The previous calculations on the
impact of Earth gravity on neutral mesons oscillations can be extended to $%
D^{0}/\overline{D}^{0}$ $\sim $ $\left( c\overline{u}\right) /$ $\left( 
\overline{c}u\right) $ and $B_{s}^{0}/\overline{B_{s}}^{0}\sim $ $\left( s%
\overline{b}\right) /$ $\left( \overline{s}b\right) $. The parameters $%
m_{D}g\hbar /\delta m_{D}^{2}c^{3}$ and $m_{B_{s}}g\hbar /\delta
m_{B_{s}}^{2}c^{3}$ for both mesons systems are small, so a type ({\it i})
indirect violation will be extremely difficult to observe. However, type (%
{\it ii}) and type ({\it iii}) CPV can be analyzed on the basis of a clear
definition of PCI observables.

To summarize our findings the three types of CPV experiments are to be considered.

Type ({\it i}) CPV observed with $K^{0}/\overline{%
K}^{0}$ stems from a gravity-induced interplay between vertical
zitterbewegung oscillations at the velocity $\pm c$ on the one hand and the
strangeness oscillations $\Delta S=2$ on the other hand. 

Type ({\it ii})
small $K^{0}/\overline{K}^{0}$ direct CPV is associated with the CPT
invariant interpretation of the experiments and is elucidated with GICPV
through a careful analysis of the PCI definition of $\eta _{00}$. 

The large
type ({\it iii}) $B^{0}/\overline{B}^{0}$ CPV is associated with the CPT
invariant interpretation of the experiments and is elucidated with GICPV
through a careful analysis of the PCI definitions of $S_{f}$ and $C_{f}$.

Note that with GICPV the relation $\mathop{\rm Re}\varepsilon
_{PDG2024}\simeq \mathop{\rm Re}\left( \varepsilon ^{\prime }/\varepsilon
\right) _{PDG2024}$ is no longer {\it a coincidence}, as assumed in the
usual KM-CPV framework, but an identity resulting from GICPV theory and the
mandatory PCI requirement. 

Despite the operational success of the parametric KM phase modeling,
fundamental physics cannot rely on a parametrization that sidesteps the
question of the origin of the mechanism requiring the introduction of new
fundamental parameters. A relevant theory on the origin of CPV may be
described as one which proceeds logically from a few simple hypotheses, in
agreement with the well established principles of physics and the
experimental conditions, to conclusions which are in agreement with
observations, to within the limits of accuracy of these observations.

Moreover, the pertinence of a new theory is not simply validated by its
ability to explain and predict the measurement of existing observed physical
processes, but also by the absence of new predictions which are not
observed. The list of quantitative predictions supporting GICPV\ (\ref{t1})
must be supplemented with a list of open problems arising from the inclusion
of parametric KM phase modeling in the SM, problems which do not arise with
GICPV.

The main problems arising with KM-CPV are : ({\it i}) the asymmetric
baryogenesis during the early stage of the cosmic evolution, ({\it ii}) the
strong CP problem and ({\it iii}) the small TV observed in kaons experiments
under the CPT invariance assumption.

The GI-CPV theory does not suffer from these drawbacks and these three
issues are resolved within the framework of GICPV (the potentiality of GICPV
to solve the asymmetric baryogenesis problem it is to be confirmed with
explicit evolution models).

({\it i}) The very large matter-antimatter asymmetry observed in our
universe remains an open issue within the SM-KM framework of interpretation,
whereas GICPV displays the potential to set up cosmological evolution models
in agreement with the present state of our universe.

({\it ii}) GICPV restricts CPV to the dynamics of oscillating flavored
neutral mesons, such as $K^{0}/\overline{K}^{0}$, $B^{0}/\overline{B}^{0}$, $%
B_{s}^{0}/\overline{B_{s}}^{0}$ and $D^{0}/\overline{D}^{0}$, whose dynamics
is driven by weak interaction and gravity, so the {\it strong CP problem} is
not an issue and thus can be considered as a resolved issue given the lack
of experimental evidences. As there are no oscillations, driven by weak
interaction, between neutron and antineutron, GICPV predicts no CPV and its
dipole moment should be zero.

({\it iii}) With GICPV time reversal holds true and the confusion between
irreversibility and T violation is clarified. Without decay, when the mesons
are stable, the evolution is unitary and with GICPV there is CPV but no TV
at the fundamental level with the mass eigenstates $\left| K_{1/2}^{\oplus
}\right\rangle $. Then, when decays come into play, T is apparently broken
with the mass eigenstates $\left| K_{S/L}^{\oplus }\right\rangle $ as a
result of the irreversibility of the process $K_{2}\stackrel{\kappa }{%
\rightarrow }K_{1}\stackrel{\Gamma _{1}}{\rightrightarrows }f$, and CPT
appears restored if we assume T violation. CPV is due to the $\stackrel{%
\kappa }{\rightarrow }$ step where time reversal is not broken and the $%
\stackrel{\Gamma _{1}}{\rightrightarrows }$ step provides a dissipative
dressing, this dissipative dressing must not be
confused with TV. The apparent T violation with $\left| K_{S/L}^{\oplus
}\right\rangle $ and $\left| K_{S/L}\right\rangle $ is in fact an
irreversible evolution stemming from the very large classical phase space
offered to the forward transitions $K\rightarrow f$. As a result, the
backward transitions $f\rightarrow K$ can not be observed because of a large
Poincar\'{e} recurrence time. 

Within the framework of an assumed CPT
invariance, an over-interpretation of irreversibility as a TV is an example
of the confusion between (statistical) irreversibility and (dynamical) T
violation. The modelisation of experiments with CPT invariant states $\left|
K_{L/S}\right\rangle $ leads to the interpretation of the experimental data
as a T violation at the fundamental level of the interaction; the two step
GICPV mechanism ($\left| K_{1/2}^{\oplus }\right\rangle $ then $\left|
K_{S/L}\right\rangle $) clarify this issue of irreversibility versus TV, in
agreement with the experimental results.

The two steps GICPV mechanism is based on a conservative use of well
established first principles put at work to understand the overlooked
interplay between two low energy quantum oscillations and the irreversible
decays. From the accuracy and convergence of all the new results presented
in this paper, it is conjectured that Earth's gravity is the sole source of $%
\ \varepsilon $, $\varepsilon ^{\prime }$ and $\beta $ CPV effects observed
in $K^{0}/\overline{K}^{0}$ and $B^{0}/\overline{B}^{0}$ dedicated
experiments; the analysis and demonstrations of the previous sections,
summarized in (\ref{t1}), providing fairly convincing arguments in support
of this conjecture.

The two tables below summarize the difference between the two frameworks of
interpretations of CPV. 

With Gravity-Induced CPV, 
\begin{equation}
\begin{tabular}{cc}
\hline\hline
{\it CPV effects} & {\it GI-CPV} \\ \hline
Type ({\it i}), ({\it ii}), ({\it iii}) CPV & ($\varepsilon ,\varepsilon
^{\prime },\beta $) correctly \\ 
& predicted : (\ref{t1}) \\ \hline
Stable eigenstates and T & T conservation \\ \hline
Unstable states and CPT & CPT if irreversibility \\ 
& view as a T violation \\ \hline
SM strong CP problem & resolved issue \\ \hline
Non-equilib. asym. baryog. & possible, can be large \\ \hline
Equilib. asym. baryog. & possible, can be large \\ \hline\hline
\end{tabular}
\label{t2}
\end{equation}

With Kobayashi-Maskawa CPV, where the measured parameters are used to
parametrize the KM phase assuming CPT
invariance, 
\begin{equation}
\begin{tabular}{cc}
\hline\hline
{\it CPV effects} & {\it KM-CPV} \\ \hline
Type ({\it i}), ({\it ii}), ({\it iii}) CPV & ($\varepsilon ,\varepsilon
^{\prime },\beta $) measured to \\ 
& provide KM phase \\ \hline
Stable eigenstates and T & not considered \\ \hline
Unstable states and CPT & CPT assumed and \\ 
& T violation assumed \\ \hline
SM strong CP problem & unsolved issue \\ \hline
Non-equilib. asym. baryog. & far too small \\ \hline
Equilib. asym. baryog. & impossible \\ \hline\hline
\end{tabular}
\label{t3}
\end{equation}

To construct an effective SM Lagrangian on Earth, GICPV can be integrated in
an initially CPV free CKM matrix with an appropriate CPV phase adjusted to
recover $\varepsilon _{GICPV}$ and $\varepsilon ^{\prime }/\varepsilon
_{GICPV}$. This direct inclusion of GICPV in the initially CPV free quark
mixing matrix will provide the usual SM as the GICPV parameters, $%
\varepsilon _{GICPV}$ and $\varepsilon ^{\prime }/\varepsilon _{GICPV}$
agree with the experimental data $\varepsilon _{PDG2024}$ and $\varepsilon
^{\prime }/\varepsilon _{PDG2024}$  (\ref{t1}) used to parametrize the CKM
matrix. Such an integration display an obvious operational value but this
obviates the problem of the origin of CPV and creates misleading open issues
such as the {\it strong CP problem }and the small{\it \ time reversal
violation}. However, besides $\varepsilon $, $\varepsilon ^{\prime
}/\varepsilon $ and $\beta $, analyzed in this paper, such an integration of
GICPV as a KM phase extend the validity of GICPV to all the others CPV
effects observed {\it in agreement with the SM-KM model}.

\appendix

\section{ Parameters $\varepsilon $, $\beta $, $\eta _{f}$, $\lambda _{f}$, $%
S_{f}$, $C_{f}$}

The CPT invariance framework, reviewed in this Appendix, provides the usual
parametrization of CPV effects. For $K^{0}/\overline{K}^{0}$ type ({\it i})
CPV, the observed kaons mass eigenstates are parametrized as \cite{5,6}: 
\begin{equation}
\left| K_{L/H}\right\rangle =\left[ \left( 1+\varepsilon \right) \left|
K^{0}\right\rangle \pm \left( 1-\varepsilon \right) \left| \overline{K}%
^{0}\right\rangle \right] /\sqrt{2}\text{.}  \label{3}
\end{equation}
These superpositions ensure CPT invariance with CP and T violations. For $%
K^{0}/\overline{K}^{0}$ the observed mass eigenstates are usually indexed
with respect to the lifetime as the short-lived $S$ and the long-lived $L$
states with the correspondence $M_{H}\rightarrow K_{L}$ and $%
M_{L}\rightarrow K_{S}$. For $B^{0}/\overline{B}^{0}$ type ({\it iii}) CPV,
the angle $\beta $ is introduced to describe the observed mass eigenstates
under the assumption of CPT invariance as 
\begin{equation}
\left| B_{L/H}\right\rangle =\cos \beta \left| B_{1/2}\right\rangle +j\sin
\beta \left| B_{2/1}\right\rangle \text{.}  \label{9}
\end{equation}
The analysis of type ({\it ii}) and ({\it iii}) CPV are based the
measurement of the amplitudes ratios $\eta _{f}$ and $\lambda _{f}$
associated with decay ($M$ $\rightarrow f$ ) experiments : 
\begin{eqnarray}
\eta _{f} &=&\left\langle f\right| {\cal T}\left| M_{H}\right\rangle
/\left\langle f\right| {\cal T}\left| M_{L}\right\rangle \text{,}  \label{4}
\\
\lambda _{f} &=&\frac{\left\langle f\right| {\cal T}\left| \overline{M}%
^{0}\right\rangle \left\langle \overline{M}^{0}\right. \left|
M_{L}\right\rangle }{\left\langle f\right| {\cal T}\left| M^{0}\right\rangle
\left\langle M^{0}\right. \left| M_{L}\right\rangle }\text{.}  \label{5}
\end{eqnarray}
These parameters capture the informations on the CP asymmetry associated
with the decays to one final CP eigenstate $\left\langle f\right| $. With
CPT invariance these two parameters are not independent ($\eta =1-\lambda
/1+\lambda $). The definition and analysis of these various amplitudes
ratios $\eta $ and $\lambda $ must be guided by the fact that the use of PCI
observables is mandatory to set up a meaningful physical description. For
example the indirect CPV mixing parameter $\varepsilon $ in (\ref{3}) can be
defined as the PCI ratio 
\begin{equation}
\varepsilon =\frac{\left\langle K^{0}\right. \left| K_{1}\right\rangle }{%
\left\langle K^{0}\right. \left| K_{2}\right\rangle }\frac{\left\langle
K_{1}\right. \left| K_{H}\right\rangle }{\left\langle K_{2}\right. \left|
K_{H}\right\rangle }\text{.}  \label{a6}
\end{equation}
The{\it \ rephasing factor }(RF) $\left\langle K^{0}\right. \left|
K_{1}\right\rangle /$ $\left\langle K^{0}\right. \left| K_{2}\right\rangle =1
$ with the CPT invariant model (\ref{3}) and it offsets the $\left|
K_{1/2}\right\rangle $ phase-convention dependence of $\left\langle
K_{1}\right. \left| K_{H}\right\rangle /$ $\left\langle K_{2}\right. \left|
K_{H}\right\rangle $. Only PCI parameters are observable, and might be
measured in CPV type ({\it i}), ({\it ii}) and ({\it iii}) experiments. The
use of a RF is mandatory when PCI requirement is not fulfilled and RF is the
key to understand that type ({\it ii}) and ({\it iii}) CPV are GICPV-RF
effects.

If CPV was exclusively due to the CP mixing (\ref{3}), the amount of CPV in
any decay would be the same. For example, the neutral channel $%
K_{S/L}\rightarrow $ $\pi ^{0}\pi ^{0}$ and the charged channel $%
K_{S/L}\rightarrow $ $\pi ^{+}\pi ^{-}$ should display the same level of
CPV. This is not the case and experimentally

\begin{equation}
\eta _{00}=\frac{\left\langle \pi ^{0}\pi ^{0}\right| {\cal T}\left|
K_{L}\right\rangle }{\left\langle \pi ^{0}\pi ^{0}\right| {\cal T}\left|
K_{S}\right\rangle }\neq \eta _{+-}=\frac{\left\langle \pi ^{+}\pi
^{-}\right| {\cal T}\left| K_{L}\right\rangle }{\left\langle \pi ^{+}\pi
^{-}\right| {\cal T}\left| K_{S}\right\rangle }\text{.}  \label{10}
\end{equation}
However $\eta _{00}$ and $\eta _{+-}$ are not PCI and this argument requires
a deeper analysis. Type ({\it ii)} direct CPV is elucidated within the
framework of\ GICPV: it is due to Earth's gravity and the requirement to
define $\eta $ PCI observables with an appropriate RF. The usual definition
of $\eta _{00}$ in (\ref{10}) is invariant under rephasing of the pions
state $\left\langle \pi ^{0}\pi ^{0}\right| $, but not with respect to the
rephasing of the kaons states $\left| K_{L/S}\right\rangle $. A decay
amplitude ratio which is a PCI observable can be defined through the
multiplication of $\eta _{00}$ with the RF 
\begin{equation}
\varphi _{00}=\frac{\left\langle K^{0}\right. \left| K_{S}\right\rangle }{%
\left\langle K^{0}\right. \left| K_{L}\right\rangle }=\frac{\left\langle
K_{L}\right. \left| K^{0}\right\rangle }{\left\langle K_{S}\right. \left|
K^{0}\right\rangle }\text{.}  \label{12}
\end{equation}
With the CPT invariant parametrization (\ref{3}) this RF is equal to one, $%
\varphi _{00}=1$. Within the framework of GICPV $\left. \varphi _{00}\right|
_{GICPV}\neq 1$ and its value provides the experimental value of $%
\varepsilon ^{\prime }$ in remarkable agreement with {\it PDG 2024}. Thus
direct CPV is due to gravity induced CP mixing as demonstrated in section
nine.

Interferences experiments between a direct decay path $M_{0}\rightarrow f$
and a mixed path $M_{0}\rightarrow \overline{M}^{0}\rightarrow f$ $\ $have
been designed to measure $\lambda _{f}$ through the observables $S_{f}$ \
and $C_{f}$ \cite{22,23,24}, 
\begin{equation}
S_{f}=\frac{2\mathop{\rm Im}\lambda _{f}}{1+\left| \lambda _{f}\right| ^{2}}%
\text{, }C_{f}=\frac{1-\left| \lambda _{f}\right| ^{2}}{1+\left| \lambda
_{f}\right| ^{2}}\text{.}  \label{7}
\end{equation}
For example, the most pronounced CPV effects in the $B^{0}/\overline{B}^{0}$
system are displayed in interferences experiments dedicated to the study of
the phase difference between the decay path $B_{0}\rightarrow f$ $\ $and the
decay path $B_{0}\rightarrow \overline{B}^{0}\rightarrow f$. For an
appropriate choice of $f$, \ the relative phase between the two paths is
only due to the phase $-\exp \left( -2j\beta \right) $ between $B^{0}$ and $%
\overline{B}^{0}$and $S_{f}\approx \sin 2\beta $ and $C_{f}\approx 0$.
Gravity induced CPV\ provides a parametrization of $\left|
B_{L/H}\right\rangle $ different from (\ref{9}), however, $\left|
B_{H}^{\oplus }\right\rangle $ and $\left| B_{L}^{\oplus }\right\rangle $
gives the same observables $S_{f}$ and $C_{f}$ as the one predicted with the
CPT invariant parametrization $\left| B_{L/H}\right\rangle $ (\ref{9}).

\section{Dirac velocity matrix elements with $\left| \widetilde{M}%
_{0}\right\rangle $ and $\left| \widetilde{\overline{M}}_{0}\right\rangle $}

Two bounded spin 1/2 charged fermions can be described on a $4$ states
basis: the singlet and the triplets states. This requires $6$ real numbers
to define the global spin state. A neutral mesons is a singlet state and the
cancellation of the $4$ triplet coefficients can be achieved if the quark
model displays at least $4$ free parameters. The amplitudes of a generic $2$
components Pauli spinor $\left| \chi _{{\bf u}}\right\rangle $, projected on
the usual $\sigma _{z}$ eigenstates basis ($\left| \uparrow \right\rangle $ ,%
$\left| \downarrow \right\rangle $), are given by 
\begin{equation}
\left[ 
\begin{tabular}{c}
$\left\langle \uparrow \right. \left| \chi _{{\bf u}}\right\rangle $ \\ 
$\left\langle \downarrow \right. \left| \chi _{{\bf u}}\right\rangle $%
\end{tabular}
\right] =\left[ 
\begin{tabular}{c}
$\exp -j\varphi /2\cos \theta /2$ \\ 
$\exp j\varphi /2\sin \theta /2$%
\end{tabular}
\right] \text{,}  \label{zs7}
\end{equation}
where the polar angle $0\leqslant \theta \leqslant \pi $ and the azimuthal
angle $0\leqslant \varphi <2\pi $ define the direction of the vector ${\bf u}
$ with ${\bf u}\cdot {\bf u}$ = $1$. The {\it up} $\left( \left| \chi _{{\bf %
u}}\right\rangle \right) $ and {\it down} $\left( \left| \chi _{-{\bf u}%
}\right\rangle \right) $ spin states along ${\bf u}$ are orthogonal $%
\left\langle \chi _{-{\bf u}}\right. \left| \chi _{{\bf u}}\right\rangle $ = 
$0$. Beside the spin state, the wave functions in real space, $(X,Y,Z)$ and $%
(X^{\prime },Y^{\prime },Z^{\prime })$, is introduced with $\left|
n,l,k\right\rangle $ and $\left| n^{\prime },l^{\prime },k^{\prime
}\right\rangle $ : ${\bf n}$ $\equiv $ $\left( n,l,k\right) $, ${\bf n}%
^{\prime }$ $\equiv $ $\left( n^{\prime },l^{\prime },k^{\prime }\right) $
are integers as quarks are confined in a finite volume $Q$ $\ $such that $%
\int_{Q}\left| \left\langle X,Y,Z\right. \left| n,l,k\right\rangle \right|
^{2}$ $dXdYdZ$ = 1 (for example $\left\langle X^{\prime },Y^{\prime
},Z^{\prime }\right. \left| n^{\prime },l^{\prime },k^{\prime }\right\rangle 
$ $\sim $ $\sin \left( n^{\prime }\pi X^{\prime }/L_{X}\right) $ $\sin
\left( k^{\prime }\pi Y^{\prime }/L_{Y}\right) $ $\sin \left( l^{\prime }\pi
Z/L_{Z}\right) /$ $\sqrt{L_{X}L_{Y}L_{z}}$ describes the real space part of
a quark state confined in a rectangular box [$L_{X}$, $L_{Y}$, $L_{z}$]).
The generic quark states, $\left| q,{\bf u},{\bf n}\right\rangle $ and $%
\left| q^{\prime },{\bf u}^{\prime },{\bf n}^{\prime }\right\rangle $, with
spin parametrization $\left( \theta ,\varphi ,\theta ^{\prime },\varphi
^{\prime }\right) $ and confinement parametrization $\left( {\bf n},{\bf n}%
^{\prime }\right) $, are defined by the $4$ components Dirac spinors 
\begin{eqnarray}
\left| q,{\bf u},{\bf n}\right\rangle  &=&\left| n,l,k\right\rangle \left[ 
\begin{tabular}{c}
$\left| \chi _{{\bf u}}\right\rangle $ \\ 
$\left| 0\right\rangle $%
\end{tabular}
\right] \text{,  }  \label{qq1} \\
\left| q^{\prime },{\bf u}^{\prime },{\bf n}^{\prime }\right\rangle 
&=&\left| n^{\prime },l^{\prime },k^{\prime }\right\rangle \left[ 
\begin{tabular}{c}
$\left| \chi _{{\bf u}^{\prime }}\right\rangle $ \\ 
$\left| 0\right\rangle $%
\end{tabular}
\right] \text{ .}  \label{qq2}
\end{eqnarray}
They belong to 2 different Dirac Hilbert spaces whose tensorial product is $%
\widetilde{{\bf E}}_{Q}$. There are several ways to define the associated
antiquarks. The introduction of a (unitary) transformation $\widehat{\sigma }%
_{{\bf v}}$ defined on the Pauli matrices basis as 
\begin{equation}
\widehat{\sigma }_{{\bf v}}=\cos \varphi _{v}\sin \theta _{v}\widehat{\sigma 
}_{X}+\sin \varphi _{v}\sin \theta _{v}\widehat{\sigma }_{Y}+\cos \theta _{v}%
\widehat{\sigma }_{Z}\text{,}
\end{equation}
provides a general framework to define a generic set of antiquarks. The
polar angle $0\leqslant \theta _{v}\leqslant \pi $ and the azimuthal angle $%
0\leqslant \varphi _{v}<2\pi $ define the direction of the vector ${\bf v}$
with ${\bf v}\cdot {\bf v}$ = $1$. The antiquark states, $\left| \overline{q}%
,{\bf u},{\bf n}\right\rangle $ and $\left| \overline{q}^{\prime },{\bf u}%
^{\prime },{\bf n}^{\prime }\right\rangle $ are defined by the $4$
components Dirac spinors 
\begin{eqnarray}
\left| \overline{q},{\bf u},{\bf n}\right\rangle  &=&\left|
n,l,k\right\rangle \left[ 
\begin{tabular}{c}
$\left| 0\right\rangle $ \\ 
$\widehat{\sigma }_{{\bf v}}\cdot \left| \chi _{{\bf u}}\right\rangle $%
\end{tabular}
\right] \text{,}  \label{qq4} \\
\left| \overline{q}^{\prime },{\bf u}^{\prime },{\bf n}^{\prime
}\right\rangle  &=&\text{ }\left| n^{\prime },l^{\prime },k^{\prime
}\right\rangle \left[ 
\begin{tabular}{c}
$\left| 0\right\rangle $ \\ 
$\widehat{\sigma }_{{\bf v}}\cdot \left| \chi _{{\bf u}^{\prime
}}\right\rangle $%
\end{tabular}
\right] \text{ ,}  \label{qq5}
\end{eqnarray}
All these single quark and single antiquarks states (\ref{qq1}, \ref{qq2}, 
\ref{qq4}, \ref{qq5}) are normalized. These quarks and  antiquarks states
display $4$ free spin parameters ($\theta ,\theta ^{\prime },\varphi
,\varphi ^{\prime }$) as needed. The freedom on the remaining $2$ parameters
($\theta _{v}$,$\varphi _{v}$) will be fixed with a final mandatory
constraint on the model. The mesons $\left| \widetilde{M}_{0}\right\rangle
\in $ $\widetilde{{\bf E}}_{Q}$ and $\left| \widetilde{\overline{M}}%
_{0}\right\rangle $ $\in $ $\widetilde{{\bf E}}_{Q}$ are spin singlet
states. For the sake of generality a more general set of orthogonal diquark
states, encompassing the singlet case, is used below 
\begin{eqnarray}
\left| \widetilde{M}_{0}\right\rangle  &=&\frac{\left| q,{\bf u},{\bf n}%
\right\rangle \left| \overline{q}^{\prime },{\bf u}^{\prime },{\bf n}%
^{\prime }\right\rangle \pm \left| q,{\bf s},{\bf n}\right\rangle \left| 
\overline{q}^{\prime },{\bf s}^{\prime },{\bf n}^{\prime }\right\rangle }{%
\sqrt{2}}\text{,}  \label{zz124} \\
\left| \widetilde{\overline{M}}_{0}\right\rangle  &=&\frac{\left| \overline{q%
},{\bf u},{\bf n}\right\rangle \left| q^{\prime },{\bf u}^{\prime },{\bf n}%
^{\prime }\right\rangle \pm \left| \overline{q},{\bf s},{\bf n}\right\rangle
\left| q^{\prime },{\bf s}^{\prime },{\bf n}^{\prime }\right\rangle }{\exp -j%
\overline{\delta }\sqrt{2}}\text{.}  \label{zz125}
\end{eqnarray}
The restriction to singlet states is obtained with simple relations between $%
{\bf u}$, ${\bf u}^{\prime }$, ${\bf s}$, ${\bf s}^{\prime }$ and ${\bf v}$ (%
${\bf s}$ $=$ $-{\bf u}$, $\widehat{\sigma }_{{\bf v}}\cdot \left| \chi _{%
{\bf u}^{\prime }}\right\rangle $ $=\left| \chi _{-{\bf u}}\right\rangle $%
...). With this singlet choice normalization and orthogonality are ensured ($%
\left\langle q,{\bf u},{\bf n}\right. \left| q,{\bf u},{\bf n}\right\rangle $
= 1,  $\left\langle q,{\bf u},{\bf n}\right. \left| q,-{\bf u},{\bf n}%
\right\rangle $ = $0$...). Note that the spin state of $\left| \overline{q},%
{\bf u},{\bf n}\right\rangle $ is different from the spin state of $\left| q,%
{\bf u},{\bf n}\right\rangle $, or its reverse, because of the  $\widehat{%
\sigma }_{{\bf v}}$ factor between the definitions (\ref{qq4}) and (\ref{qq1}%
). Indeed, what matters to define $\left| \widetilde{\overline{M}}%
_{0}\right\rangle $ as the antiparticle of $\left| \widetilde{M}%
_{0}\right\rangle $ is that both are global spin singlet states with the
same spatial structure and reverse flavors. The arbitrary relative phase $%
\overline{\delta }$ is related to the arbitrary relative phase $\delta $
between a meson and its antiparticle discussed at the end of section two so $%
\exp j\overline{\delta }$ $=$ $\pm 1$ is to be considered {\it a priori.}

The issue of the reduction of a description in $\widetilde{{\bf E}}_{Q}$ to
a description in ${\bf E}_{LOY}\otimes {\bf E}_{\bot }$ is addressed now.
This requires a coherent definition of the instantaneous position operator
of the center of mass/barycenter in both spaces.

A conservative standpoint is adopted \ to define the barycenter of two
confined relativistic quarks inside $\left| \widetilde{M}_{0}\right\rangle $
or $\left| \widetilde{\overline{M}}_{0}\right\rangle $ defined by the
relations (\ref{zz124}, \ref{zz125}). To do so the canonical rule of quantum
mechanics is applied: the global probability amplitude is the sum of the
partial probability amplitudes. Four single quark states define a {\it meson}
of the type (\ref{zz124}, \ref{zz125}), thus the probability amplitude of
the position of each of these states is to be calculated. Then, given these
4 probability amplitudes, two (unknown) positive coefficients, $\lambda $
and $\lambda ^{\prime }$ such that $\lambda $ $+$ $\lambda ^{\prime }$ $=1$,
associated with the mass ponderation between $q/\overline{q}$ and $q^{\prime
}/\overline{q}^{\prime }$ ($\lambda /\lambda ^{\prime }\sim
m_{q}/m_{q^{\prime }}$), are used to set up a ponderation according to the
usual definition of a center of mass/barycenter. Following this usual point
of view, the amplitude ${\bf X}_{\widetilde{M}_{0}\widetilde{M}_{0}}$ of the
instantaneous barycentric position of $\ $(\ref{zz124}) is given by 
\begin{eqnarray}
{\bf X}_{\widetilde{M}_{0}\widetilde{M}_{0}} &=&\lambda \text{ (}%
\left\langle q,{\bf u},{\bf n}\right| \widehat{{\bf X}}\left| q,{\bf u},{\bf %
n}\right\rangle  \\
&&+\left\langle q,{\bf s},{\bf n}\right| \widehat{{\bf X}}\left| q,{\bf s},%
{\bf n}\right\rangle \text{)}/2  \nonumber \\
&&+\lambda ^{\prime }\text{ (}\left\langle \overline{q}^{\prime },{\bf u}%
^{\prime },{\bf n}^{\prime }\right| \widehat{{\bf X}}^{\prime }\left| 
\overline{q}^{\prime },{\bf u}^{\prime },{\bf n}^{\prime }\right\rangle  
\nonumber \\
&&+\left\langle \overline{q}^{\prime },{\bf s}^{\prime },{\bf n}^{\prime
}\right| \widehat{{\bf X}}^{\prime }\left| \overline{q}^{\prime },{\bf s}%
^{\prime },{\bf n}^{\prime }\right\rangle \text{)}/2\text{.}  \nonumber
\end{eqnarray}
This is one of the matrix element of the barycentric position operator $%
\widehat{{\bf X}}_{Q}\left( \tau \right) $ operating on (\ref{zz124}, \ref
{zz125}) in $\widetilde{{\bf E}}_{Q}$. The other diagonal matrix element $%
{\bf X}_{\widetilde{\overline{M}}_{0}\widetilde{\overline{M}}_{0}}$ is
obtained in a similar way with $q\rightarrow \overline{q}$ .... These two
diagonal matrix elements of the barycenter position are to be completed with
two others matrix elements to set up a complete definition of $\widehat{{\bf %
X}}_{Q}\left( \tau \right) $%
\begin{eqnarray}
{\bf X}_{\widetilde{M}_{0}\widetilde{\overline{M}}_{0}} &=&\lambda \exp j%
\overline{\delta }\text{(}\left\langle q,{\bf u},{\bf n}\right| \widehat{%
{\bf X}}\left| \overline{q},{\bf u},{\bf n}\right\rangle  \\
&&+\left\langle q,{\bf s},{\bf n}\right| \widehat{{\bf X}}\left| \overline{q}%
,{\bf s},{\bf n}\right\rangle \text{)}/2  \nonumber \\
&&+\lambda ^{\prime }\exp j\overline{\delta }\text{(}\left\langle \overline{q%
}^{\prime },{\bf u}^{\prime },{\bf n}^{\prime }\right| \widehat{{\bf X}}%
^{\prime }\left| q^{\prime },{\bf u}^{\prime },{\bf n}^{\prime
}\right\rangle   \nonumber \\
&&+\left\langle \overline{q}^{\prime },{\bf s}^{\prime },{\bf n}^{\prime
}\right| \widehat{{\bf X}}^{\prime }\left| q^{\prime },{\bf s}^{\prime },%
{\bf n}^{\prime }\right\rangle \text{)}/2\text{.}  \nonumber
\end{eqnarray}
The other matrix element ${\bf X}_{\widetilde{\overline{M}}_{0}\widetilde{M}%
_{0}}$ is obtained in a similar way with $q\rightarrow \overline{q}$ .... In
the operator $\widehat{{\bf X}}_{Q}$ the phase factors $\exp j\overline{%
\delta }$ in (\ref{zz125}) simplify and this ensure that 
\begin{eqnarray}
\widehat{{\bf X}}_{Q} &=&{\bf X}_{\widetilde{M}_{0}\widetilde{M}_{0}}\left| 
\widetilde{M}_{0}\right\rangle \left\langle \widetilde{M}_{0}\right| +{\bf X}%
_{\widetilde{\overline{M}}_{0}\widetilde{\overline{M}}_{0}}\left| \widetilde{%
\overline{M}}_{0}\right\rangle \left\langle \widetilde{\overline{M}}%
_{0}\right|  \\
&&+{\bf X}_{\widetilde{M}_{0}\widetilde{\overline{M}}_{0}}\left| \widetilde{M%
}_{0}\right\rangle \left\langle \widetilde{\overline{M}}_{0}\right| +{\bf X}%
_{\widetilde{\overline{M}}_{0}\widetilde{M}_{0}}\left| \widetilde{\overline{M%
}}_{0}\right\rangle \left\langle \widetilde{M}_{0}\right| \text{,}  \nonumber
\end{eqnarray}
is defined as a PCI operator. The barycenter position operator $\widehat{%
{\bf X}}_{Q}$ is meaningful in a restricted subspace of $\widetilde{{\bf E}}%
_{Q}$: for a fixed set (${\bf u}$, ${\bf u}^{\prime }$, ${\bf s}$, ${\bf s}%
^{\prime }$, ${\bf n}$, ${\bf n}^{\prime }$, ${\bf v}$) this subspace is
made of all the linear combinations of the two states $\left| \widetilde{M}%
_{0}\right\rangle $ and $\left| \widetilde{\overline{M}}_{0}\right\rangle $ (%
\ref{zz124}, \ref{zz125}), elsewhere in $\widetilde{{\bf E}}_{Q}$ it is
mathematically well defined but physically meaningless. The vertical
direction above Earth surface is taken along a direction ${\bf w}$ (${\bf %
w\cdot w}$ = $1$) such that the vertical position operator is defined as $%
\widehat{X}_{Q}\left( \tau \right) $ = ${\bf w}\cdot \widehat{{\bf X}}%
_{Q}\left( \tau \right) $. Following the definition (\ref{aver}), the
vertical fluctuations operator, $\widehat{x}_{Q}$, around the (time)
averaged barycenter position $\left\langle \widehat{X}_{Q}\right\rangle
_{\tau }$, is given by the difference between the instantaneous value minus
the time averaged value 
\begin{equation}
\widehat{x}_{Q}\left( \tau \right) =\widehat{X}_{Q}\left( \tau \right)
-\left\langle \widehat{X}_{Q}\left( \tau \right) \right\rangle _{\tau }\text{%
.}  \label{xxxc}
\end{equation}
The orders of magnitude of \ ({\it i}) $\left\langle \widetilde{M}%
_{0}\right| \widehat{x}_{Q}\left| \widetilde{M}_{0}\right\rangle $ is the
Compton wavelength $\hbar /mc$ and ({\it ii}) the typical frequency of $%
\left\langle \widetilde{M}_{0}\right| \widehat{x}_{Q}\left( \tau \right)
\left| \widetilde{M}_{0}\right\rangle $ is $mc^{2}/\hbar $. The operator $%
\widehat{x}_{Q}$ probes the internal structure of $\widetilde{M}_{0}$ and $%
\widetilde{\overline{M}}_{0}$ in $\widetilde{{\bf E}}_{Q}$. The hierarchy of
Hilbert spaces starts with the larger one $\widetilde{{\bf E}}_{Q}$. Then,
when most of the details of the quark structure are forgotten, the mesons are
described in ${\bf E}_{LOY}\otimes {\bf E}_{\bot }$.  The operator $\widehat{%
x}\left( \tau \right) $ in (\ref{31}, \ref{52}) operates in ${\bf E}%
_{LOY}\otimes {\bf E}_{\bot }$ thus an appropriate {\it vertical internal
wave function} $\left| \psi _{Q}\right\rangle \in {\bf E}_{\bot }$  is
introduced to complete $\left| M^{0}\right\rangle $ and $\left| \overline{M}%
^{0}\right\rangle $. If two given vectors $\left| M^{0}\right\rangle \left|
\psi _{Q}\right\rangle \in {\bf E}_{LOY}\otimes {\bf E}_{\bot }$ and $\left| 
\overline{M}^{0}\right\rangle \left| \psi _{Q}\right\rangle \in {\bf E}%
_{LOY}\otimes {\bf E}_{\bot }$ \ are assumed to describe the very same
physical objects as two vectors $\left| \widetilde{M}_{0}\right\rangle \in $ 
$\widetilde{{\bf E}}_{Q}$ and $\left| \widetilde{\overline{M}}%
_{0}\right\rangle $ $\in $ $\widetilde{{\bf E}}_{Q}$, but with fewer
details, and if $\widehat{x}_{Q}$ and $\widehat{x}$ are assumed to probe the
very same physical property on these objects: {\it the small fluctuations
around the average barycenter position}; then the matrix elements of $%
\widehat{x}\left( \tau \right) $ in (\ref{31}, \ref{52}) are 
\begin{eqnarray}
\left\langle \psi _{Q}\right| \left\langle M^{0}\right| \widehat{x}\left|
M^{0}\right\rangle \left| \psi _{Q}\right\rangle  &=&\left\langle \widetilde{%
M}_{0}\right| \widehat{x}_{Q}\left| \widetilde{M}_{0}\right\rangle \text{,}
\label{fg2} \\
\left\langle \psi _{Q}\right| \left\langle M^{0}\right| \widehat{x}\left| 
\overline{M}^{0}\right\rangle \left| \psi _{Q}\right\rangle  &=&\left\langle 
\widetilde{M}_{0}\right| \widehat{x}_{Q}\left| \widetilde{\overline{M}}%
_{0}\right\rangle \text{,}  \label{fg3}
\end{eqnarray}
$\left\langle \psi _{Q}\right| \left\langle \overline{M}^{0}\right| \widehat{%
x}\left| \overline{M}^{0}\right\rangle \left| \psi _{Q}\right\rangle $ and $%
\left\langle \psi _{Q}\right| \left\langle \overline{M}^{0}\right| \widehat{x%
}\left| M^{0}\right\rangle \left| \psi _{Q}\right\rangle $ are obtained in a
similar way. These four relations achieve the canonical reduction from $%
\widetilde{{\bf E}}_{Q}$ to ${\bf E}_{LOY}\otimes {\bf E}_{\bot }$. The
operator $d\widehat{x}/d\tau $ operates {\it a priori} in ${\bf E}%
_{LOY}\otimes {\bf E}_{\bot }$, however, it turns out that $d\widehat{x}%
/d\tau $ does not operate on the wave function $\left| \psi
_{Q}\right\rangle \in {\bf E}_{\bot }$ because $d\widehat{{\bf X}}/d\tau =c%
\widehat{\alpha }$ and $d\widehat{{\bf X}}^{\prime }/d\tau =c\widehat{\alpha 
}^{\prime }$ (\ref{zs3}, \ref{zs4}) do not operate on $\left|
n,l,k\right\rangle $ and $\left| n^{\prime },l^{\prime },k^{\prime
}\right\rangle $ as $\widehat{\alpha }$ and $\widehat{\alpha }^{\prime }$
are not functions of $\widehat{{\bf X}}$, $\widehat{{\bf P}}$ and $\widehat{%
{\bf X}}^{\prime }$, $\widehat{{\bf P}}^{\prime }$. Thus, the velocity
operator, $d\widehat{x}/d\tau $, describing the {\it zitterbewegung}
fluctuations velocity is given by
the matrix elements of $d\widehat{x}_{Q}/d\tau $ $=$ $d\widehat{X}_{Q}/d\tau 
$ as 
\begin{eqnarray}
\frac{d\widehat{x}}{d\tau } &=&\frac{dX_{\widetilde{M}_{0}\widetilde{M}_{0}}%
}{d\tau }\left| M^{0}\right\rangle \left\langle M^{0}\right|  \\
&&+\frac{dX_{\widetilde{\overline{M}}_{0}\widetilde{\overline{M}}_{0}}}{%
d\tau }\left| \overline{M}^{0}\right\rangle \left\langle \overline{M}%
^{0}\right|   \nonumber \\
&&+\frac{dX_{\widetilde{\overline{M}}_{0}\widetilde{M}_{0}}}{d\tau }\left|
M^{0}\right\rangle \left\langle \overline{M}^{0}\right| +\frac{dX_{%
\widetilde{M}_{0}\widetilde{\overline{M}}_{0}}}{d\tau }\left| \overline{M}%
^{0}\right\rangle \left\langle M^{0}\right| \text{.}  \nonumber
\end{eqnarray}
In order to evaluate these various matrix elements, the Dirac velocity
operators along the vertical direction ${\bf w}$, $c\widehat{\alpha }_{{\bf w%
}}$ and $c\widehat{\alpha }_{{\bf w}}^{\prime }$, are defined as

\begin{eqnarray*}
\widehat{\alpha }_{{\bf w}} &=&\cos \varphi _{w}\sin \theta _{w}\widehat{%
\alpha }_{X}+\sin \varphi _{w}\sin \theta _{w}\widehat{\alpha }_{Y}+\cos
\theta _{w}\widehat{\sigma }_{Z}\text{,} \\
\widehat{\alpha }_{{\bf w}}^{\prime } &=&\cos \varphi _{w}\sin \theta _{w}%
\widehat{\alpha }_{X}^{\prime }+\sin \varphi _{w}\sin \theta _{w}\widehat{%
\alpha }_{Y}^{\prime }+\cos \theta _{w}\widehat{\sigma }_{Z}^{\prime }\text{,%
}
\end{eqnarray*}
where the polar angle $0\leqslant \theta _{w}\leqslant \pi $ and the
azimuthal angle $0\leqslant \varphi _{w}<2\pi $ define the direction ${\bf w}
$ which is noted, without loss of generality, $x$. The diagonal element $%
\left\langle M^{0}\right| d\widehat{x}/d\tau \left| M^{0}\right\rangle $ is
given by the sum 
\begin{eqnarray}
c &&\lambda \text{ (}\left\langle q,{\bf u},{\bf n}\right| \widehat{\alpha }%
_{{\bf w}}\left| q,{\bf u},{\bf n}\right\rangle +\left\langle q,{\bf s},{\bf %
n}\right| \widehat{\alpha }_{{\bf w}}\left| q,{\bf s},{\bf n}\right\rangle 
\text{)}/2 \\
&&+c\lambda ^{\prime }\text{ (}\left\langle \overline{q}^{\prime },{\bf u}%
^{\prime },{\bf n}^{\prime }\right| \widehat{\alpha }_{{\bf w}}^{\prime
}\left| \overline{q}^{\prime },{\bf u}^{\prime },{\bf n}^{\prime
}\right\rangle +\left\langle \overline{q}^{\prime },{\bf s}^{\prime },{\bf n}%
^{\prime }\right| \widehat{\alpha }_{{\bf w}}^{\prime }\left| \overline{q}%
^{\prime },{\bf s}^{\prime },{\bf n}^{\prime }\right\rangle \text{)}/2\text{,%
}  \nonumber
\end{eqnarray}
and $\left\langle \overline{M}^{0}\right| d\widehat{x}/d\tau \left| 
\overline{M}^{0}\right\rangle $ is obtained in a similar way. The final
result is independent of $\ {\bf u},$ ${\bf n},$ ${\bf s},$ ${\bf u}^{\prime
},$ ${\bf n}^{\prime },$ ${\bf s}^{\prime }{\bf \ }$and is 
\begin{equation}
\left\langle M^{0}\right| d\widehat{x}/d\tau \left| M^{0}\right\rangle
=\left\langle \overline{M}^{0}\right| d\widehat{x}/d\tau \left| \overline{M}%
^{0}\right\rangle =0\text{.}
\end{equation}
This result can be interpreted with the definition of the probability
current ${\bf J}$ = $c\left\langle {}\right| \widehat{\alpha }\left|
{}\right\rangle $ associated with Dirac's equation \cite{8,9}. For a
localized confined ground state ${\bf J}$ = ${\bf 0}$ is expected. The non
diagonal matrix elements are given by 
\begin{eqnarray}
\left\langle M^{0}\right| \frac{d\widehat{x}}{d\tau }\left| \overline{M}%
^{0}\right\rangle  &=&c\frac{\lambda \exp j\overline{\delta }}{2}%
\left\langle q,{\bf u},{\bf n}\right| \widehat{\alpha }_{{\bf w}}\left| 
\overline{q},{\bf u},{\bf n}\right\rangle  \\
&&+c\frac{\lambda \exp j\overline{\delta }}{2}\left\langle q,{\bf s},{\bf n}%
\right| \widehat{\alpha }_{{\bf w}}\left| \overline{q},{\bf s},{\bf n}%
\right\rangle   \nonumber \\
&&+c\frac{\lambda ^{\prime }\exp j\overline{\delta }}{2}\left\langle 
\overline{q}^{\prime },{\bf u}^{\prime },{\bf n}^{\prime }\right| \widehat{%
\alpha }_{{\bf w}}^{\prime }\left| q^{\prime },{\bf u}^{\prime },{\bf n}%
^{\prime }\right\rangle   \nonumber \\
&&+c\frac{\lambda ^{\prime }\exp j\overline{\delta }}{2}\left\langle 
\overline{q}^{\prime },{\bf s}^{\prime },{\bf n}^{\prime }\right| \widehat{%
\alpha }_{{\bf w}}^{\prime }\left| q^{\prime },{\bf s}^{\prime },{\bf n}%
^{\prime }\right\rangle \text{.}  \nonumber
\end{eqnarray}
This requires the evaluation of four matrix elements of the type 
\begin{equation}
\left\langle q,{\bf u},{\bf n}\right| \widehat{\alpha }_{{\bf w}}\left| 
\overline{q},{\bf u},{\bf n}\right\rangle ={\bf w\cdot v}+j\text{ }%
\left\langle \chi _{{\bf u}}\right| \widehat{\sigma }_{{\bf w\times v}%
}\left| \chi _{{\bf u}}\right\rangle \text{,}
\end{equation}
where the relations: $\widehat{\sigma }_{{\bf w}}\cdot \widehat{\sigma }_{%
{\bf v}}$ = $\left( {\bf w\cdot v}\right) \widehat{\text{I}}$ $+$ $j$ $%
\widehat{\sigma }_{\left( {\bf w\times v}\right) }$, $\left\langle \chi _{%
{\bf u}}\right. \left| \chi _{{\bf u}}\right\rangle $ $=$ $1$ and $%
\left\langle n,l,k\right. \left| n,l,k\right\rangle $ $=$ $1$ have been used.

The results of section six demonstrate that, up to a real factor, the real
parts of $\left\langle \overline{M}^{0}\right| d\widehat{x}/d\tau \left|
M^{0}\right\rangle $ and $\left\langle M^{0}\right| d\widehat{x}/d\tau
\left| \overline{M}^{0}\right\rangle $ give the imaginary part of $%
\left\langle M_{1}\right. \left| M_{2}\right\rangle $ and the imaginary
parts of $\left\langle \overline{M}^{0}\right| d\widehat{x}/d\tau \left|
M^{0}\right\rangle $ and $\left\langle M^{0}\right| d\widehat{x}/d\tau
\left| \overline{M}^{0}\right\rangle $ give the real part of $\left\langle
M_{1}\right. \left| M_{2}\right\rangle $.

Following \cite{5} and \cite{10}, the analysis of the C, P and T properties
of the LOY model leads to two CPV possibilities: ({\it i}) $\left\langle
M_{1}\right. \left| M_{2}\right\rangle $ is real if CPT holds or ({\it ii}) $%
\left\langle M_{1}\right. \left| M_{2}\right\rangle $ is imaginary if\ T
holds. As there are no decays in the evolution described in section six, it
can be assumed that T holds without ambiguities and $\left\langle \overline{M%
}^{0}\right| d\widehat{x}/d\tau \left| M^{0}\right\rangle $ and $%
\left\langle M^{0}\right| d\widehat{x}/d\tau \left| \overline{M}%
^{0}\right\rangle $ must be real. Among the two remaining independent free
parameters $\overline{\delta }$ and ${\bf v}$ one is already given as $\exp j%
\overline{\delta }=\pm 1$ so that the constraint of time reversal invariance
can be fulfilled if and only if $\ {\bf v}=\pm {\bf w}$. With this time
reversal invariance prescription, the matrix elements of the fluctuating
velocity $d\widehat{x}/d\tau $ along a direction ${\bf w}$ are independent
of any particular choice of (${\bf u}$, ${\bf w}$) and (${\bf n}$, ${\bf n}%
^{\prime }$),
\begin{equation}
\left\langle \overline{M}^{0}\right| d\widehat{x}/d\tau \left|
M^{0}\right\rangle =\pm c\text{, }\left\langle M^{0}\right| d\widehat{x}%
/d\tau \left| \overline{M}^{0}\right\rangle =\pm c\text{.}
\end{equation}
With these results there are four possibilities to construct $d\widehat{x}%
/d\tau $, two of them are Hermitian and two of them are not. The operators $d%
\widehat{{\bf X}}_{Q}/d\tau $, $d\widehat{{\bf X}}/d\tau $ and $d\widehat{%
{\bf X}}^{\prime }/d\tau $ are not canonical observables like $\widehat{{\bf %
X}}$ and $\widehat{{\bf X}}^{\prime }$ because their definitions imply two
successive measurements of the positions with $d\tau \rightarrow 0$ such
that the uncertainty on the energy $\sim \hbar /d\tau $ becomes infinite. In
the operator $\widehat{x}_{Q}\left( \tau \right) $ $=$ $\widehat{X}%
_{Q}\left( \tau \right) $ $-$ $\left\langle \widehat{X}_{Q}\left( \tau
\right) \right\rangle _{\tau }$, the last term is not observable because it
is a sum $\int_{T\rightarrow +\infty }\widehat{X}_{Q}\left( \tau \right)
d\tau /T$ assuming an infinite set of successive measurements of the
position with $d\tau \rightarrow 0$ such that the uncertainty on the energy $%
\sim \hbar /d\tau $ becomes infinite. Effective operator which are not
observable are not necessarily Hermitian, $j\widehat{\gamma }$ provides such
an example. The operator $\widehat{x}_{Q}$ (\ref{xxxc}) is not an observable
and is not {\it a priori} Hermitian. Following the phase convention assumed
in section two, the definitions of the non diagonal matrix elements (\ref
{xxxc}) are given by 
\begin{eqnarray}
\left\langle \widetilde{M}_{0}\right| \widehat{x}_{Q}\left| \widetilde{%
\overline{M}}_{0}\right\rangle  &=&\left\langle \widetilde{M}_{0}\right| 
\widehat{x}_{Q}\cdot \widehat{CP}\left| \widetilde{M}_{0}\right\rangle \text{%
,} \\
\left\langle \widetilde{\overline{M}}_{0}\right| \widehat{x}_{Q}\left| 
\widetilde{M}_{0}\right\rangle  &=&\left\langle \widetilde{M}_{0}\right| 
\widehat{CP}\cdot \widehat{x}_{Q}\left| \widetilde{M}_{0}\right\rangle \text{%
.}
\end{eqnarray}
In these definitions $\widehat{P}$ in $\widehat{CP}$ is the {\it intrinsic }%
parity as $\widehat{x}_{Q}$ probes the {\it intrinsic} fermionic structure
of the mesons $\left| \widetilde{M}_{0}\right\rangle $ and $\left| 
\widetilde{\overline{M}}_{0}\right\rangle $ (\ref{zz124}, \ref{zz125}). The
parity operator $\widehat{P}$ anticommute with the {\it intrinsic} internal
fluctuations position operator $\widehat{x}_{Q}$ and commute with $\widehat{C%
}$ thus 
\begin{equation}
\left\langle \widetilde{M}_{0}\right| \widehat{x}_{Q}\left| \widetilde{%
\overline{M}}_{0}\right\rangle =-\left\langle \widetilde{\overline{M}}%
_{0}\right| \widehat{x}_{Q}\left| \widetilde{M}_{0}\right\rangle \text{.}
\label{pcv}
\end{equation}
The WW operator $j\widehat{\gamma }$ is also a non Hermitian effective
operator associated with a reduction from a large Hilbert space to a smaller
one. For the $j\widehat{\gamma }$ case hermicity can be restored in the
larger Hilbert space of all the final products $\left| f_{i}\right\rangle $
without the WW description. Note that the usual CPT\ parametrization of the
LOY Hamiltonian with CPV is also obtained on the basis of a non Hermitian
Hamiltonian \cite{5,6}. As a result of (\ref{pcv}), obtained with a parity
argument, we recover final velocity operator (\ref{ff}) obtained with the
Feynman prescription.


\begin{thebibliography}{500}

\bibitem{1}  J. H. Christenson, J. W. Cronin, V. L. Fitch and R. Turlay,
Evidence for the $2\pi $ decay of the $K_{2}^{0}$ meson, {\it Phys. Rev.
Lett.} {\bf 13}, 138-140 (1964).

\bibitem{2}  M. Kobayashi and  T. Maskawa, CP-violation in the
renormalizable theory of weak interaction. {\it Progress of Theoretical
Physics} {\bf 49}, 652-657 (1973).

\bibitem{3}  N. Cabibbo, Unitary symmetry and leptonic decays,  {\it Phys.
Rev. Lett.}  {\bf 10}, 531-533 (1963).

\bibitem{4}  A. D. Sakharov, Violation of CP invariance, C asymmetry, and
baryon asymmetry of the universe, Pisma Zh. Eksp. Teor. Fiz. 5, {\bf 32}
(1967).

\bibitem{5}  T. D. Lee, {\it Particle physics and introduction to field
theory} (Harwood Academic, New York, 1981).

\bibitem{6}  E. Paschos, {\it Electroweak Theory} (Cambridge University
Press, Cambridge, 2007).

\bibitem{7}  P. D. Group, Review of particle physics, Phys. Rev. D {\bf 110}
030001 (2024).

\bibitem{8}  J. Bjorken and S. Drell, {\it Relativistic Quantum Mechanics}
(McGraw-Hill, New York, 1964).

\bibitem{9}  A. Messiah, {\it Mecanique Quantique} (Dunod, Paris,1964).

\bibitem{10}  T. D. Lee, R. Oehme, and C. N. Yang, Remarks on possible
noninvariance under time reversal and charge conjugation, Phys. Rev. {\bf 106}, 340 (1957).

\bibitem{11}  T. T. Wu and C. N. Yang, Phenomenological analysis of
violation of CP invariance in decay of $K^{0}$ and $\overline{K}^{0}$, Phys.
Rev. Lett. {\bf 13}, 380 (1964).

\bibitem{12}  J-M. Rax, Zitterbewegung CP violation in a Schwarzschild
spacetime, arXiv 2403.07970 (03-2024).

\bibitem{13}  J-M. Rax, Gravity induced CP violation in K and B mixing,
decays and interferences experiments, Proceeding of Science 20263b, 346
(2026).

\bibitem{14} V. Weisskopf and E. Wigner, Berechnung der naturlichen linienbreite
linienbreite auf grund der diracschen lichttheorie, Zeitschrift fur Physik 
{\bf 63}, 54 (1930).  

\bibitem{15}  R. Penrose, {\it The Road to Reality} (Jonathan Cape, New
York, 2004).

\bibitem{16}  J. Bell and J. Steinberger, in {\it Proceedings of the Oxford
International Conference on Elementary Particles 1965}, edited by R. G.
Moorhouse, A. E. Taylor, and T. R. Walsh (Rutherford High Energy Laboratory,
1966) p. 194.

\bibitem{17}  H. Burkhardt et al. (NA31), First evidence for direct CP
violation, Phys. Lett. B {\bf 206}, 169 (1988).

\bibitem{18}  V. Fanti et al. (NA48), A new measurement of direct CP
violation in two pion decays of the neutral kaon, Phys. Lett. B {\bf 465},
335 (1999).

\bibitem{19}  A. Alavi-Harati et al. (KteV), Observation of direct CP
violation in $K_{S,L}\rightarrow \pi \pi $ decays, Phys. Rev. Lett. {\bf 83}, 22 (1999).

\bibitem{20}  B. Aubert et al. (BaBar), Observation of CP violation in the $%
B_{0}$ meson system, Phys. Rev. Lett. {\bf 87}, 091801 (2001).

\bibitem{21}  K. Abe et al. (Belle), Observation of large CP violation in
the neutral $B$ meson system, Phys. Rev. Lett. {\bf 87}, 091802 (2001).

\bibitem{22}  B. Aubert et al. (BaBar), Improved measurement of CP violation
in neutral $B$ decays to $c\overline{c}s$, Phys. Rev. Lett. {\bf 99}, 171803
(2007).

\bibitem{23}  I. Adachi et al. (Belle), Precise measurement of the CP
violation parameter, Phys. Rev. Lett. {\bf 108}, 171802 (2012).

\bibitem{24}  B. Aubert et al. (BaBar), Measurement of time-dependent
asymmetry in decays, Phys. Rev. D {\bf 79}, 072009 (2009).

\bibitem{25}  E. Fischbach, Test of general relativity at the quantum level,
in {\it Proceedings of the NATO Advanced Study Institute on Cosmology and
Gravitation}, edited by P. Bergmann and V. de Sabbata, NATO Scientific
Affairs Division (Plenum Press, New York and London, 1980) pp. 359-373.

\bibitem{26}  G. Chardin and J-M. Rax, CP violation. a matter of
(anti)gravity?, Phys. Lett. B {\bf 282}, 256 (1992).

\bibitem{27}  G. Chardin, CP violation and antigravity (revisited), Nuclear
Physics A {\bf 558}, 477 (1993).

\bibitem{28}  J-M. Rax, {\it Mecanique Analytique }(Dunod, Paris, 2020).

\bibitem{29}  N.J. Fisch, J-M. Rax, I.Y. Dodin, Current drive in a ponderomotive potential with sign reversal, 
Phys. Rev. Lett. {\bf 91}, 205004 (2003).

\bibitem{30}  K. Urbanowski, Decay law of relativistic particles: quantum
theory meets special relativity, Phys. Lett. B {\bf 737}, 346 (2014).

\bibitem{31}  L. Fonda, C. C. Ghirardi and A. Rimini, Decay theory of
unstable quantum systems, Rep. Prog. Phys. 41, 587 (1978).

\end{thebibliography}
\end{document}